\documentclass[nofootinbib,longbibliography,final,11pt]{revtex4-1}

\usepackage{graphicx}
\usepackage{hyperref}

\usepackage [latin1]{inputenc}
\usepackage{amsmath,amssymb}
\usepackage{bm} 
\usepackage{mathtools}
\usepackage[caption=false]{subfig}
\usepackage{booktabs}

\usepackage{color}
\usepackage{ulem}
\usepackage{comment}
\usepackage{cleveref}

\usepackage{natbib}

\graphicspath{{./fig/}}

\def\ie{\textit{i.e., }}

\begin{document}

\title{Mock-integrability and stable solitary vortices}

\author{Yukito Koike$^a$}
\email{r.yukitokoike@gmail.com}
\author{Atsushi Nakamula$^{b,g}$}
\email{nakamula@sci.kitasato-u.ac.jp}

\author{Akihiro Nishie$^c$}
\email{akihiro.a.nishie@gmail.com}

\author{Kiori Obuse$^d$}
\email{obuse@okayama-u.ac.jp}

\author{Nobuyuki Sawado$^{a,g}$}
\email{sawadoph@rs.tus.ac.jp}

\author{Yamato Suda$^a$}
\email{suda.y.ad@m.titech.ac.jp}

\author{Kouichi Toda$^{e,f,g}$}
\email{kouichi@yukawa.kyoto-u.ac.jp}

\affiliation{$^a$Department of Physics, Tokyo University of Science, Noda, Chiba 278-8510, Japan}
\affiliation{$^b$Department of Physics, School of Science, Kitasato University, Sagamihara, Kanagawa 252-0373, Japan}
\affiliation{$^c$Rakuten Group, Inc., Setagaya, Tokyo 158-0094, Japan}
\affiliation{$^d$Graduate School of Environmental and Life Science, Okayama University, Okayama 700-8530, Japan}
\affiliation{$^e$Department of Mathematical Physics, Toyama Prefectural University, Imizu, Toyama 939-0398, Japan}
\affiliation{$^f$Research and Education Center for Natural Sciences, Keio University, Hiyoshi 4-1-1, Yokohama, Kanagawa 223-8521, Japan}
\affiliation{$^g$Instituto de F\'isica de S\~ao Carlos; IFSC/USP, Universidade de S\~ao Paulo - USP,  
Caixa Postal 369, CEP 13560-970, S\~ao Carlos-SP, Brazil}

\begin{abstract}
Localized soliton-like solutions to a $(2+1)$-dimensional hydro-dynamical evolution equation are studied numerically. 
The equation is the so-called Williams-Yamagata-Flierl equation, which governs geostrophic fluid in a certain parameter range.
Although the equation does not have an integrable structure in the ordinary sense, we find there exist shape-keeping solutions with a very long life in a special background flow and an initial condition.
The stability of the localization at the fusion process of two soliton-like objects is also investigated. 
As for the indicator of the long-term stability of localization, we propose a concept of configurational entropy, which has been introduced in analysis for non-topological solitons in field theories.

\end{abstract}

\maketitle

\section{Introduction}

Nonlinear integrable systems have exceptional properties in that they possess a sufficient number of integrals of motion, 
Painl\'eve properties, reducibility into bilinear forms, and so on.
They admit as a consequence localized solutions that retains their identity even under collision, the solitons.
Solitons in nonlinear mathematical systems are eternally stable objects and simulate so many well-balanced phenomena in nature that a large amount of research has been done and is still underway \cite{Ablowitz-Segur,ablowitz_clarkson_1991,Miwa-Jimbo,hirota_2004}.

An archetypal example of integrable systems is the Korteweg-de Vries (KdV) equation, which describes the dynamics of surface waves propagating in one-dimensional shallow water.
The dynamics of the KdV equation consist of a balance between dispersion and nonlinearity,
and the system has an infinite number of integrals of motion and admits multi-soliton solutions.
What characterizes the solitons in the KdV equation is their stability at collisions and exponentially dumping tails of each soliton.
Several models of a generalization to the KdV equation into two-dimensional space as integrable systems are considered so far.
One of these models is the Kadomtsev-Petviashvili (KP) equation, which describes shallow water waves propagating in two-dimensional space \cite{Kadomtsev70}.
The soliton solutions to the KP equation have a structure that dumps rapidly in one spatial direction and does not localize in the other direction, \ie the KP solitons are extending in a two-dimensional plane with localization as a line. 
The system also has an infinite number of integrals of motion, the Lax formulation, the bilinear formulation, and the multi-soliton solutions~\cite{MANAKOV1977205}. 
For that reason, the class of non-linear systems such as KdV and KP is referred to as the completely integrable systems.

Alternatively, there have been found various localized solutions to some integrable systems with rapidly decaying in all spatial directions, 
so-called dromions~\cite{Davey74,FOKAS199099,HIETARINTA1990113,Nishinari94}, topological multi-vortices~\cite{Ishimori}, 
localized pulses in regularized-long-wave equations~\cite{KAWAHARA199279}, solitons in an extended Burgers equation~\cite{GAO2022112293},
and also several types of lump solitons~\cite{Kaup1981TheLS,Imai1996,Zhang2017,Zou2018}, so on. 
They could explain various nonlinear phenomena of fluid dynamics of the atmosphere or ocean.  
In the case of the ocean, the waves are generated by the wind and are the response made by the water under
the gravity or surface tension. 

Among them, the Davey-Stewartson I (DSI) equation~\cite{Davey74, Anker78} provides a
two-dimensional generalization of the non-linear Schr\"odinger equation and it possesses only one integral of motion in contrast to the completely integrable systems.
Nevertheless, the DSI system has the Lax and the bilinear formulation.
The lack of conserved quantities would be the characteristic that the system has lost, {\it i.e.}, the fundamental feature of integrability as one-dimensional systems~\cite{BOITI19952377}.
In consequence, the solutions to the initial value problem to the DSI equation cannot be defined uniquely without fixing suitable boundary conditions.
We notice here that the effects of the boundary conditions are the continuous supply of external ``flow".
In this sense, the DSI equation does not belong to the completely integrable systems in its strictly meaning.
Anyhow, the integrable systems in higher spatial dimensions that admit localized solutions in all directions are relatively rare.

In various ranges of mathematical physics, there are occasionally nonlinear systems that admit quasi-stable soliton-like solutions, 
although they do not have typical integrable structures,  
such as two-dimensional solitary Rossby-waves~\cite{Larichev76} or solitons in an  
inhomogeneous plasma~\cite{Hasegawa78, Hasegawa79}. Such drift-wave solitons are a typical example of a weak-integrable system where metastable 
solutions are observed though no concrete analytical methods exist. 
An illustration is the Zakharov-Kuznetsov (ZK) equation  \cite{Zakharov74} originally introduced to describe ion-acoustic waves in magnetized plasma.
The ZK equation is a three-dimensional generalization of the KdV equation and its two-dimensional analogue is given as
\begin{align}
\frac{\partial \phi}{\partial t}+\nabla^2\frac{\partial \phi}{\partial x}+2\phi\frac{\partial \phi}{\partial x}=0\,,\label{two-dimensional ZK}
\end{align}
where the Laplacian is $\nabla^2:=\partial_x^2+\partial_y^2$.
This equation also appears in some parts of physics such as a behavior of thin liquid film \cite{PETVIASHVILI1981329,Toh89,MELKONIAN1989255},
the solitary Rosby-waves~\cite{Petviashvilli80}, and vortex in the drift waves in the three-dimensional plasma~\cite{Nozaki81}, 
so it is continuously studied \cite{XU200521,Lannes2013,Klein21} with some extensions.
In the two-dimensional ZK equation \eqref{two-dimensional ZK}, the nonlinear term is similar to the KdV equation and we refer to this nonlinearity as scalar nonlinearity.
This equation has a rapidly decaying localized solution with cylindrical symmetry obtained numerically \cite{IWASAKI1990293}, whose behavior  under collision is shown to be soliton-like in certain conditions due to the scalar nonlinearity.
Despite the two-dimensional ZK equation \eqref{two-dimensional ZK}  possessing quasi-soliton solutions, the system has only a finite number of integrals of motion, thus it cannot be identified as a completely integrable system. 
Note that the single soliton exhibits a certain long-life (longevity) during the long-term evolution. 
Though the analysis of longevity seems critical to understand the stability of the solution, 
no rigorous research concerning the long-term dynamical evolution of (\ref{two-dimensional ZK}) exists so far.

In this paper, we draw attention to a nonlinear system in two spatial dimensions that are non-integrable and admit sufficiently localized 
and quasi-stable solutions. 
Collisions of traveling waves and isolated vortices are frequent phenomena in nature, 
observed both in the atmosphere and in the ocean. 
The dynamics of those familiar phenomena, and especially the behavior of long-lived, 
large-scale isolated vortices such as ``the Great Red Spot (GRS)'' of Jupiter in the atmosphere, 
``the Naruto whirlpools'' of Japan's coast in the ocean,  
have been of interest to many people. Therefore a lot of studies have been done 
so far~\cite{charney1981oceanic, bouchet2012statistical, constantin2009relevance, constantin2009solitons, constantin2011nonlinear,  
ibragimov2013invariant, ibragimov2013nonlinear, ibragimov2017nonlinear,  nezlin1989long,  BA2125604X}. 
We consider a similar non-linear system to the ZK equation~\eqref{two-dimensional ZK} that appears in the geophysical fluid dynamics of the atmosphere, 
or ocean, on a rotating planet, including additional non-linearity and non-autonomous terms.
It is an intermediate geostrophic equation for vorticity where the horizontal scales of the phenomena are smaller than the Rossby deformation radius. 
The equation is introduced first to explain the very long life of Jupiter's GRS and so-called Williams-Yamagata-Flierl (WYF) 
equation \cite{charney1981oceanic,Williams84}, or simply IG (Intermediate Geostrophic) equation, for the stream function $\eta$ of a fluid.
The equation in its normalized form is
\begin{align}
-\frac{\partial \eta}{\partial T}+\nabla^2\frac{\partial \eta}{\partial x}+2\eta\frac{\partial \eta}{\partial x}
-2y\frac{\partial \eta}{\partial x}+2J[\eta,\nabla^2\eta]=0\,,\label{eq:WYF intro}
\end{align}
which describes behavior of the vortex in a long time scale $T$. 
The last term, the Jacobian is defined as $J[A,B]:=\partial_x A\partial_y B-\partial_y A\partial_x B$ 
which represents the geostrophic advection of vorticity in the words of fluid dynamics, and we refer to it as the vector nonlinear term.
The WYF equation \eqref{eq:WYF intro} is derived from the shallow water $\beta$-plane model governing the stream function, 
or vorticity, of fluid under the influence of Coriolis force in a particular parameter range. The derivation is given in Appendix A.
The pioneer work \cite{Williams84} shows that a vortex, \ie a well-localized solution given by Gaussian initial shape, 
is sufficiently robust in a particular situation, and two distinct vortices merge into a single vortex, unlike soliton collisions.
The vector nonlinear term contributes to the shape-keeping of the localized solutions in not a similar manner to the soliton systems, 
so the merging phenomena of the multi-vortices are the consequence of the cooperation of the scalar and the vector nonlinearities.
We remark that this type of nonlinearity also appears in the Charney-Hasagawa-Mima equation \cite{Charney63, Hasegawa78, Hasegawa79}, 
which describes the drift waves in magnetized plasma, and also
the geophysical flows in the $\beta$-plane at a different parameter range to the WYF equation.
It is obvious that the varieties of nonlinearity increase as the spatial dimensions get larger, such as the vector nonlinearity appeared 
in the case of the WYF equation.
We expect that the interplay between those various types of nonlinearity plays significant role in the formation and the stability of 
higher-dimensional solitons and soliton-like objects.

We focus on here the origin of the longevity of the single vortex solutions to the equation
under the influence of this complicated nonlinearity together with external forces, \ie the background shear flow given with non-autonomous terms.
This paper investigates a modified WYF equation with, so-called the variable-coefficient form
~\cite{calogero1978exact,brugarino1980integration,joshi1987painleve,hlavaty1988painleve,brugarino1991painleve,
gao2001variable,kobayashi2005generalized,kobayashi2006painleve}
that implements the effect of background shear flows in the equation. 
The explicit form is given in Section II, Eq.\eqref{eq:WYF+shear}. 
We will find that the vortices secure eternal long-life in certain circumstances rather than expected from the previous works.
In the present work, although the WYF equation is not integrable in any sense, we demonstrate that a steady external force is crucial to the stability of the soliton-like objects, \ie vortices in two-dimensions.
Those effects are included in the dynamical equations as non-autonomous terms, which are absent in ordinary soliton systems.
We argue that the soliton-like stable, or very long-life objects appear occasionally as a consequence of those ``mock-integrability".   
In connection with this, we should remind the case of the DSI equation, in which there exist a suitable boundary condition for each soliton-like object.

The above non-integrable localized objects have some similarities with solitons in field theoretical models such as the $Q$-ball or the oscillon. 
The oscillon is a non-topological soliton in a scalar field theory with no explicit integrability. 
Typically, the oscillon has a bell shape that oscillates sinusoidally in time 
and has an extreme long life in some particular conditions~\cite{Honda:2001xg}.
The oscillon was discovered in the seventies of the last century by Bogolyubsky and Makhankov~\cite{Bogolyubsky:1976yu} and 
later revisited by~\cite{Gleiser:1993pt,Copeland:1995fq,Fodor:2008es,Fodor:2009xw}. 
A notable feature is that the lifetime of the oscillons is strongly affected by the size of the initial configuration which may be common in the several quasi-integrable, 
or mock-integrable systems.
We shall examine this issue: the initial condition dependency for longevity. 
For the analysis in the stability of localized object in field theory, an indirect method may be efficient for the estimation of the lifetime occasionally.
Among them, the so-called configurational entropy is a quite promising candidate for understanding the stable nature of the localized objects~\cite{Gleiser:2011di}. 
For an attempt of this concept into fluid mechanical systems, we apply it for the analysis of the longevity of the vortices in the WYF equation.
We expect that a criterion of the longevity of quasi-integrable vortices in the fluid dynamics would be given in the context of such information theoretical content.

This paper is organized as follows.
In Section II we shall describe the model, including a thorough discussion of the close relation with several quasi-integrable systems such like the 
Zakharov-Kuznetsov equation or the Davey-Stewartson equation. The conservation nature of the model is also given in this section. 
Section III presents the numerical solutions of the model. 
Conclusions and remarks are presented in the last Section.

\section{Non-integrable models in steady background  effects}

In this section, we introduce the ZK and the WYF equations in detail for the subsequent analysis.
As mentioned above, they are generalizations of the KdV equations into  two-spatial dimensions.

 \subsection{Zakharov-Kuznetsov equation}

The Zakharov-Kuznetsov equation originally was the model in three dimensions 
of plasma with a uniform magnetic field~\cite{Zakharov74}.
The majority of the subsequent works, however, 
have been done for the two-dimensional analogue of the model~\cite{PetYan82, IWASAKI1990293, Klein21}, 
which is defined as
\begin{align}
\frac{\partial \phi}{\partial t}+2\phi\frac{\partial \phi}{\partial x}+\frac{\partial}{\partial x}(\nabla^2\phi)=0\,.
\label{eq:ZK}
\end{align}
The model possesses meta-stable isolated solutions which 
enjoy the solitonic properties.
The single soliton is stable in dynamical sense, and the two solitons with same height scatter without merging as in the KdV solitons.
However, in the scattering of the two solitons with dissimilar height, the taller soliton becomes taller 
while the shorter one turns out to be shorter and radiates ripples \cite{IWASAKI1990293}. 
The stability of the solutions seem to rely on underlying KdV dynamics, especially the conserved quantities.

\begin{figure*}[t]
	\begin{center}

    \includegraphics[width=120mm]{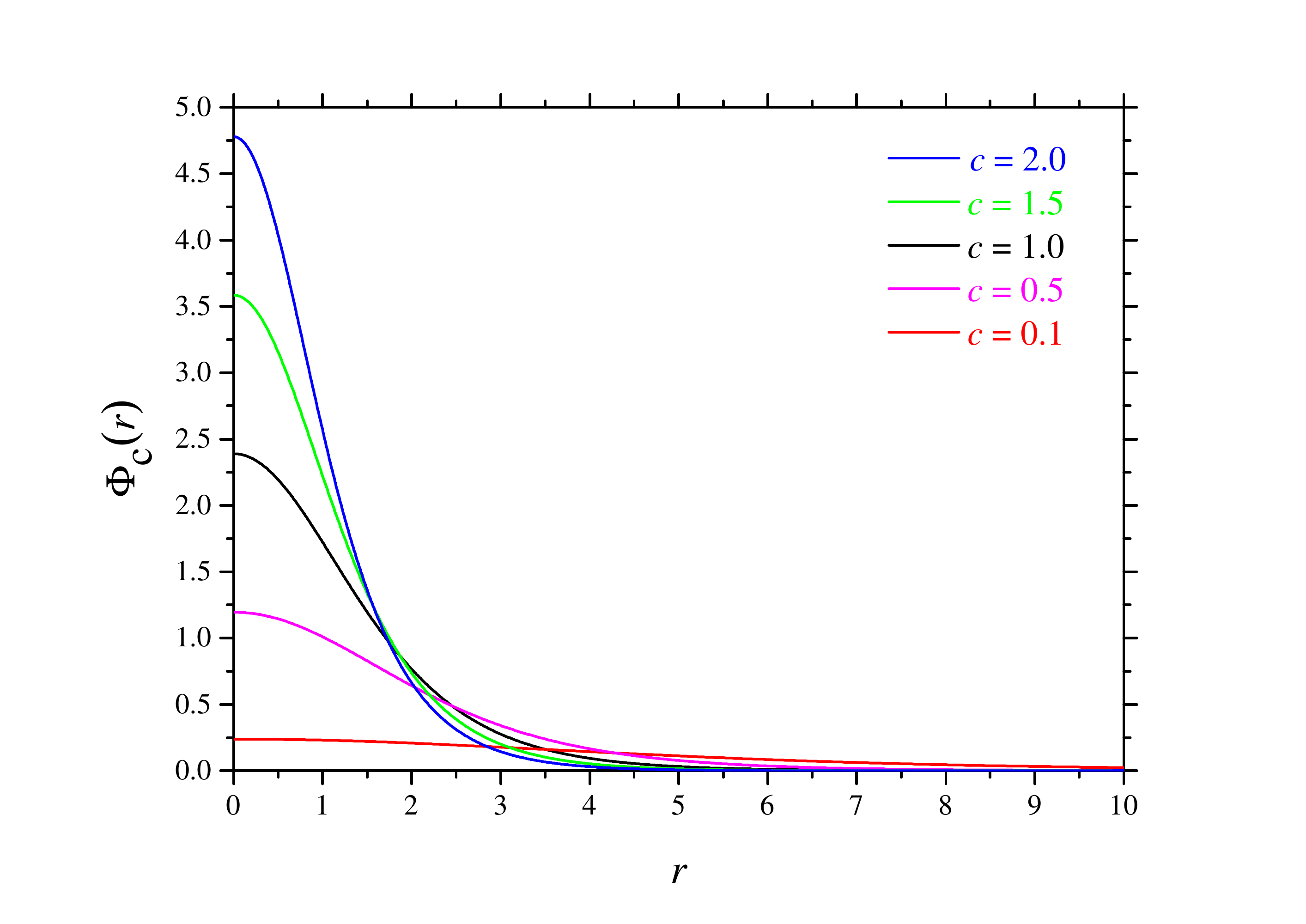}

	\caption{\label{fig:ZKcylind}The circularly symmetric solutions of the ZK equation (\ref{eq:ZK}), which
	are quasi solitary wave solutions with the wave velocity $c$.  }
  \end{center}
	\end{figure*}

The equation possesses the solutions propagating in a specific direction with uniform speeds. 
Here we set the direction in the positive $x$ orientation with the velocity $c$, namely assuming $\phi=\Phi(x-ct,y)$. 
Plugging it into (\ref{eq:ZK}) we obtain
\begin{align}
\nabla^2\Phi=c\Phi-\Phi^2\,,
\end{align}
where $\nabla^2=\partial_{\tilde{x}}^2+\partial_y^2$ and $\tilde{x}=x-ct$. 
A steady progressive exact wave solution is of the form
\begin{align}
\Phi_{c}^{\rm rec}=\frac{3c}{2}{\rm sech}^2\biggl[\frac{\sqrt{c}}{2}(\tilde{x}\cos\theta+y\sin\theta)\biggr],\label{eq:ZK wave}
\end{align}
where the $\theta$ is a given inclined angle of the solution. 
From this form, 
it is easy to see that the solution is just a trivial embedding of the KdV soliton into two spatial dimensions. 
Here we would like to find the solutions to \eqref{eq:ZK} keeping circular symmetry other than \eqref{eq:ZK wave}.
In order to find it,  we introduce cylindrical coordinate and rewrite the equation as \begin{align}
\frac{1}{r}\frac{d}{dr}\biggl(r\frac{d\Phi_c}{dr}\biggr)=c\Phi_c-\Phi_c^2\,,~~~~\Phi_c:=\Phi_c(r)\,,
\label{eq:ZKcylind}
\end{align}
where $r:=\sqrt{\tilde{x}^2+y^2}$. 
We are able to find a family of solutions to \eqref{eq:ZKcylind} with the boundary condition $\Phi_c\to 0$ as $r\to \infty$ in terms of a simple numerical study. 
The solutions form 
one parameter family of $c$ such as $\Phi_c(r):=cF(\sqrt{c}r)$. 
The result is shown in Fig.\ref{fig:ZKcylind}.
As is expected, the solution propagate in positive $x$ direction without any dissipation because of the character from the KdV like property.

Although the ZK equation \eqref{eq:ZK} has the soliton-like solutions given above,
the equation admits only a finite number of integrals of motion as shown in \cite{KUZNETSOV1986103}, which are
\begin{align}
&M:=\int M(y)dy= \int \phi dx dy,\ M(y):=\int \phi dx,
\label{CQ1}
\\
&P:=\int\frac{1}{2}\phi^2dxdy\,,
\label{CQ2}
\\
&H:=\int\biggl[\frac{1}{2}(\nabla\phi)^2-\frac{1}{6}\phi^3\biggr]dxdy\,,
\label{CQ3}
\\
&\bm{I}:=\int \bm{r}\phi dxdy-t\bm{e}_x\int\frac{1}{2}\phi^2dxdy\,,
\label{CQ4}
\end{align}
where $\boldsymbol{r}$ and $\boldsymbol{e}_x$ are the two-dimensional position vector and the unit vector in the $x$-direction.
Here $M$ is interpreted as the ``Mass" of the solution, and $M(y)$ itself is conserved similarly to the KdV equation, respectively.
We, therefore, conclude that the ZK equation is not an integrable system in the manner of ordinary soliton equations.

\subsection{Intermediate geostrophic regime and Williams-Yamagata-Flierl equation}

Several classes of geophysical fluid dynamics in two-dimensional approximation have been studied:
the quasi-geostrophic (QG; small-scale), the planetary geostrophic (PG; long-scale) and intermediate geostrophic (IG; medium-scale) motions. 
For the IG regime, Williams and Yamagata proposed the governing equation for the vortices on the planetary atmosphere with the zonal currents, which are obtained from the standard shallow water system on the $\beta$-plane of coordinates $x,y$ \cite{charney1981oceanic,Williams84}. 
The equation is defined as
\begin{align}
\frac{\partial \eta}{\partial T}-\frac{E}{S}\eta\frac{\partial\eta}{\partial x}-S\frac{\partial}{\partial x}(\nabla^2\eta)
+2y\frac{\partial \eta}{\partial x}+EJ[\nabla^2\eta,\eta]=0\,,~~~~\eta:=\eta(x,y,T)\,,
\label{eq:WYF}
\end{align}
where the Jacobian $J[A,~B]:=\partial_x A\partial_y B-\partial_y A\partial_x B$ and the Laplacian 
$\nabla^2:=\partial_x^2+\partial_y^2$.
Here, the long time scale $T$ is related with the standard time coordinate $t$ as $T:=\hat{\beta}t,~~\hat{\beta}\ll 1,$ 
and it denotes the slow time coordinate that describes the long-term evolution of the vortices, see Appendix A. 
As the authors of \cite{Williams84} numerically shown, 
the equation (\ref{eq:WYF}) have a Gaussian shaped vortex solutions supposed to simulate a candidate for Jupiter's red spot, which is considered to be an anti-cyclonic vortex. 
Though the results were sound, 
the discussions lack the mathematical discussions for the governing equation, \ie 
scope of how the solutions are to be stable in terms of effects of the nonlinearity and the background shear flow. 
In particular, we will analyze here the nonlinear evolution equation under the influence of numerous background, or external flows, given in (\ref{eq:WYF}).
We confirm that the equation (\ref{eq:WYF}) has well-localized  solutions with sufficient longevity like solitons in completely integrable systems as mentioned in Introduction.

	\begin{figure}[p]
	\begin{center}

	\includegraphics[width=120mm]{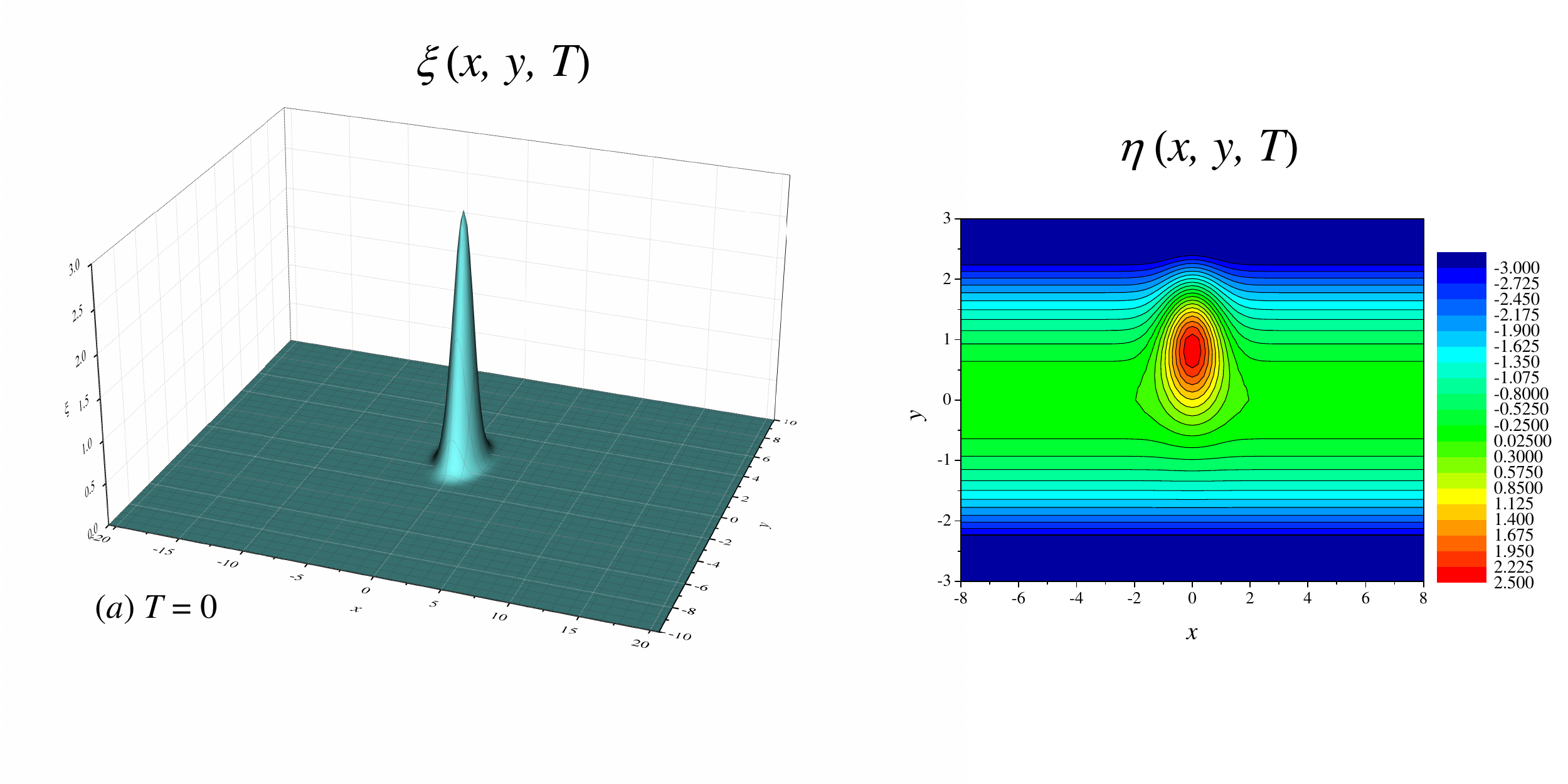}\\
	\vspace{-1cm}
	
	\includegraphics[width=120mm]{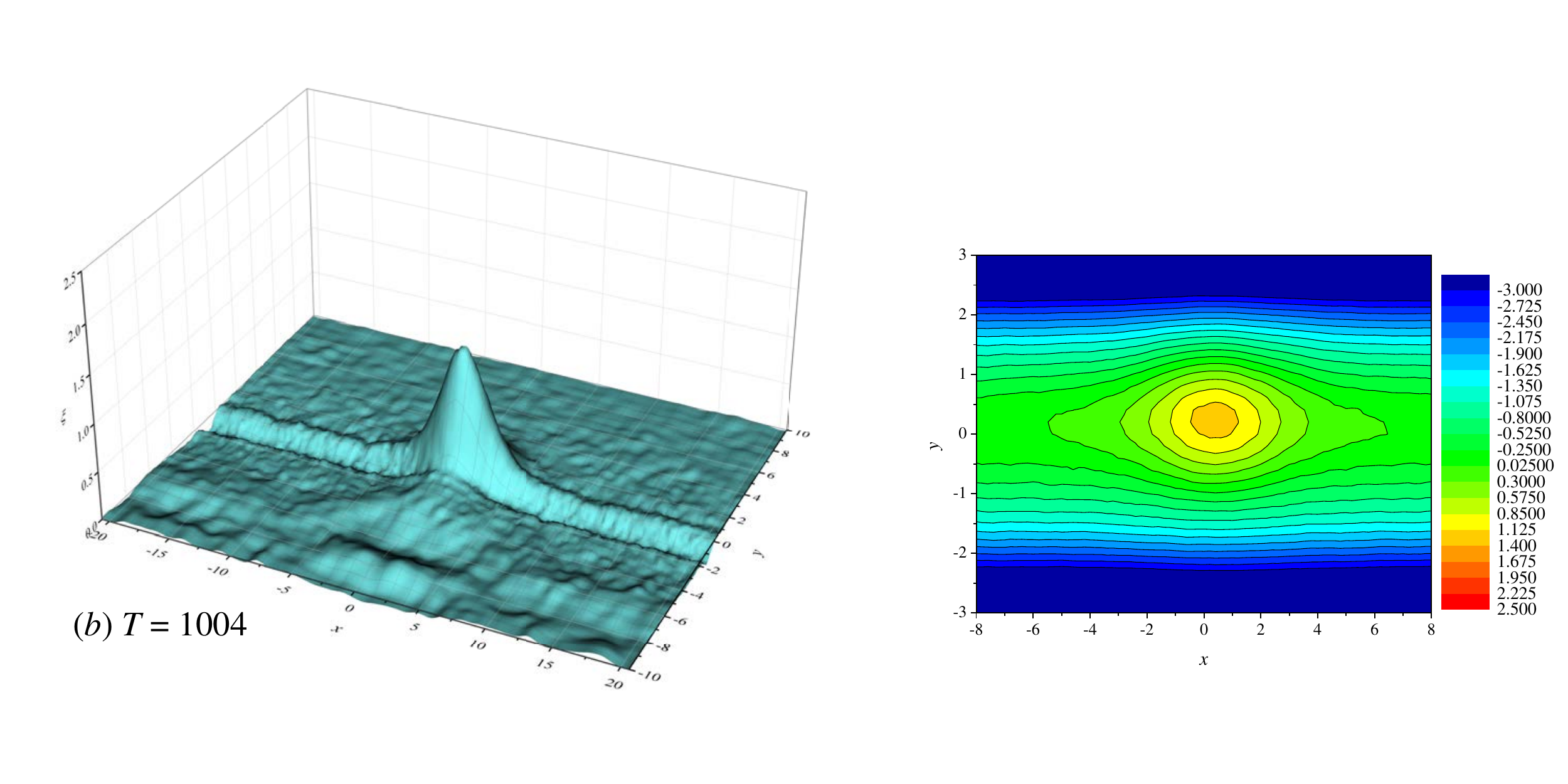}\\
	\vspace{-1cm}
	
	\includegraphics[width=120mm]{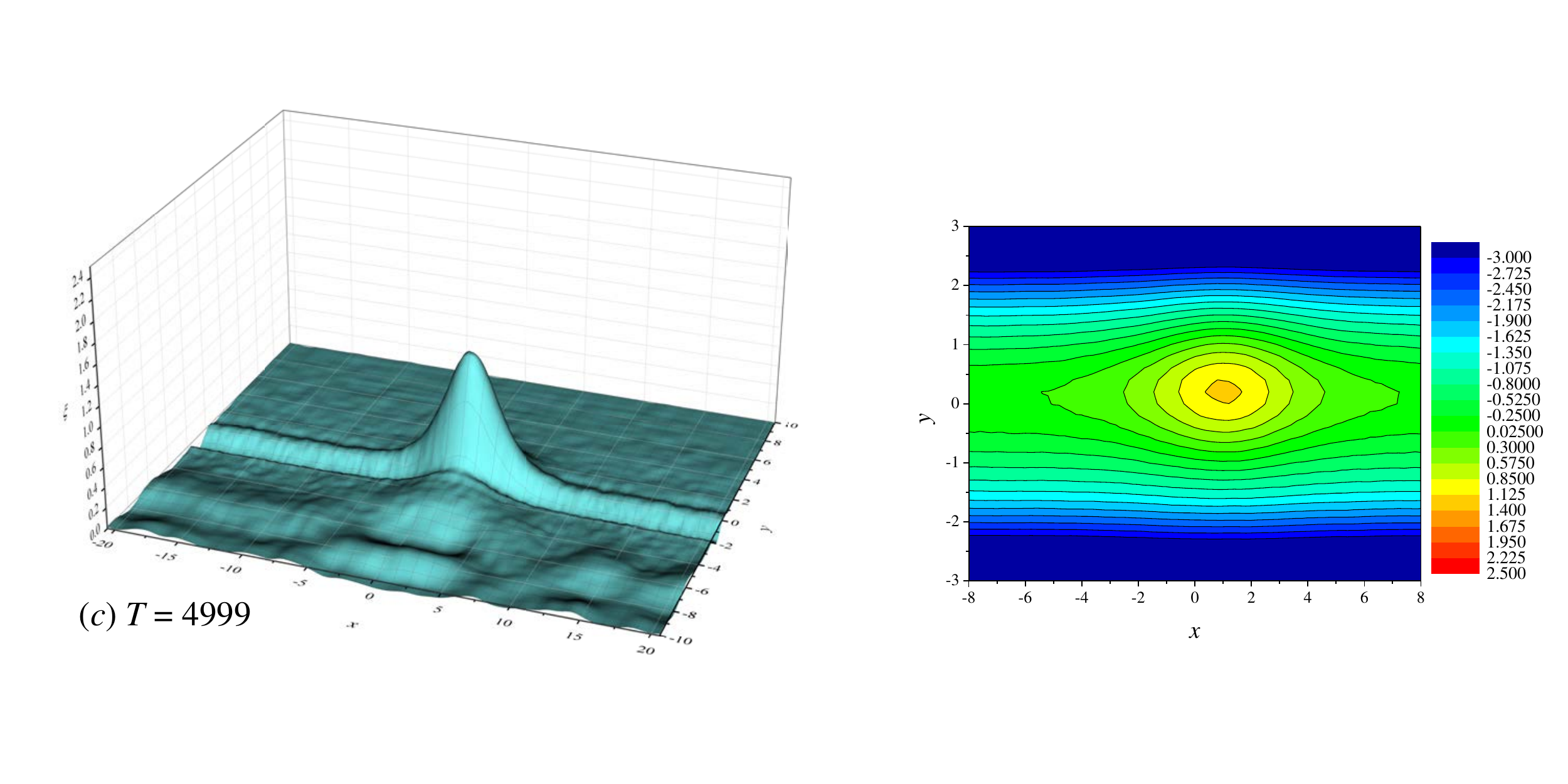}\\
	\vspace{-1cm}
	
	\includegraphics[width=120mm]{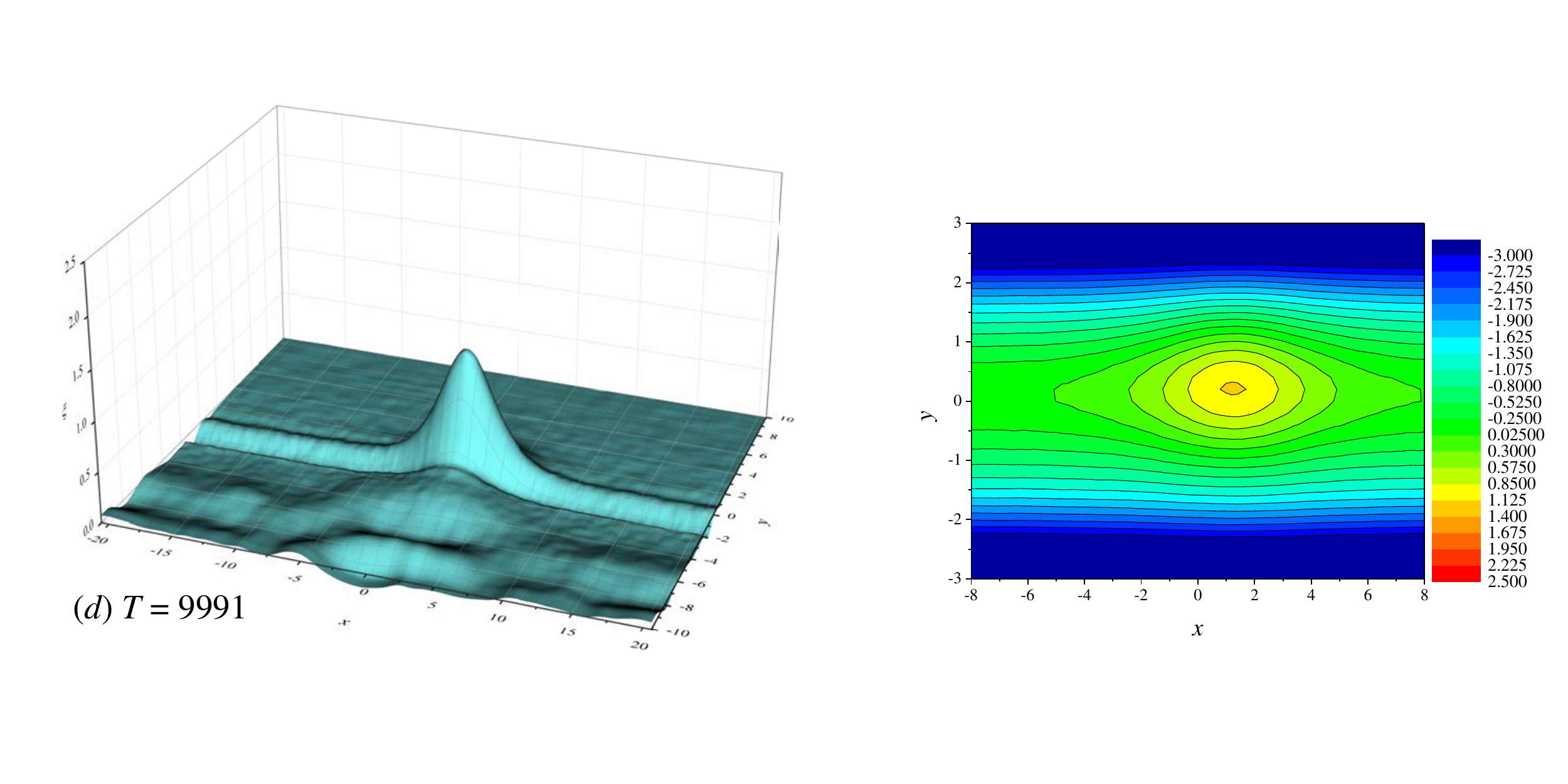}

	\caption{\label{fig:WYFprofile_gauss}~Long-term simulation of the profile of the vortex $\eta(x,y,T) $ in $(f_0,f_1)=(0.0,1.2)$
	for the iteration number: (a) $0$, (b)~$1004\times 10^{4}$, (c)~$4999\times 10^{4}$, (d)~$9991\times 10^{4}$, respectively. 
	We chose the time where the vortex is located at around the middle of the computational area. 
	The initial condition is the circular symmetric Gaussian: $\xi_{\rm init}=3.0\times\exp\{-x^2-(y-1)^2\}$.}
       \end{center}
	\end{figure}


	\begin{figure}[p]
	\begin{center}

	\includegraphics[width=120mm]{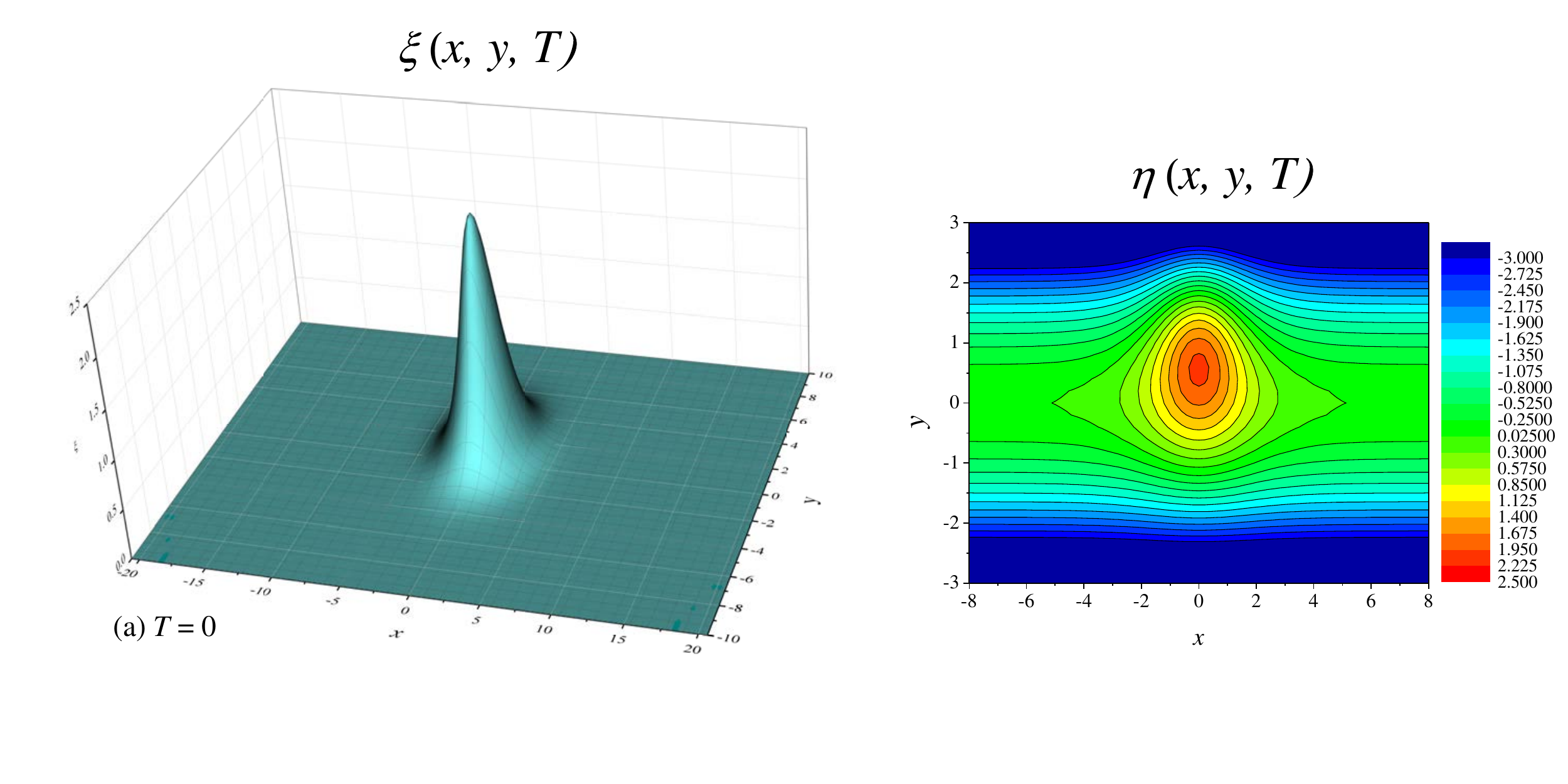}\\
	\vspace{-1.0cm}
	
	\includegraphics[width=120mm]{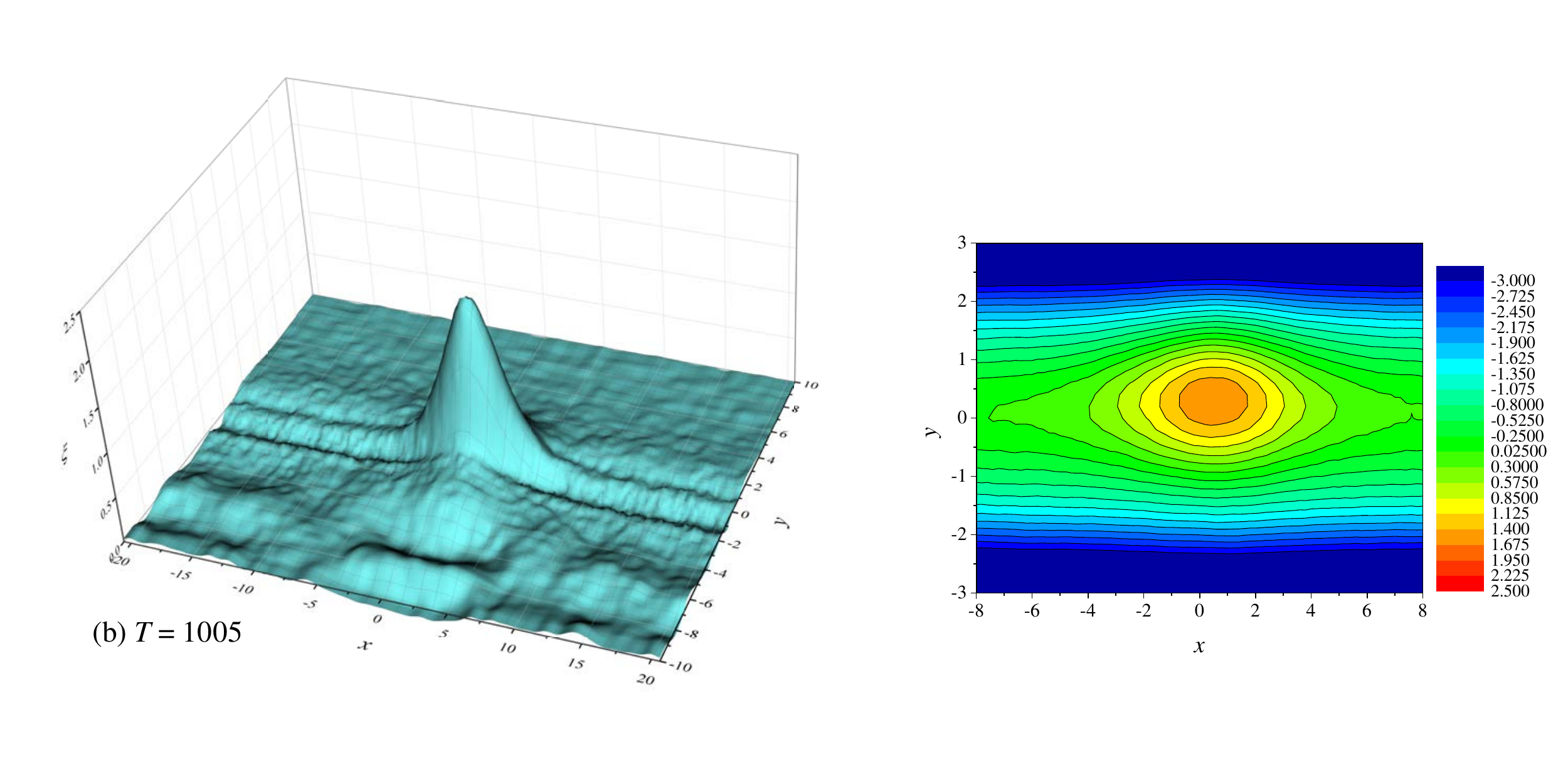}\\
	\vspace{-1.0cm}
	
	\includegraphics[width=120mm]{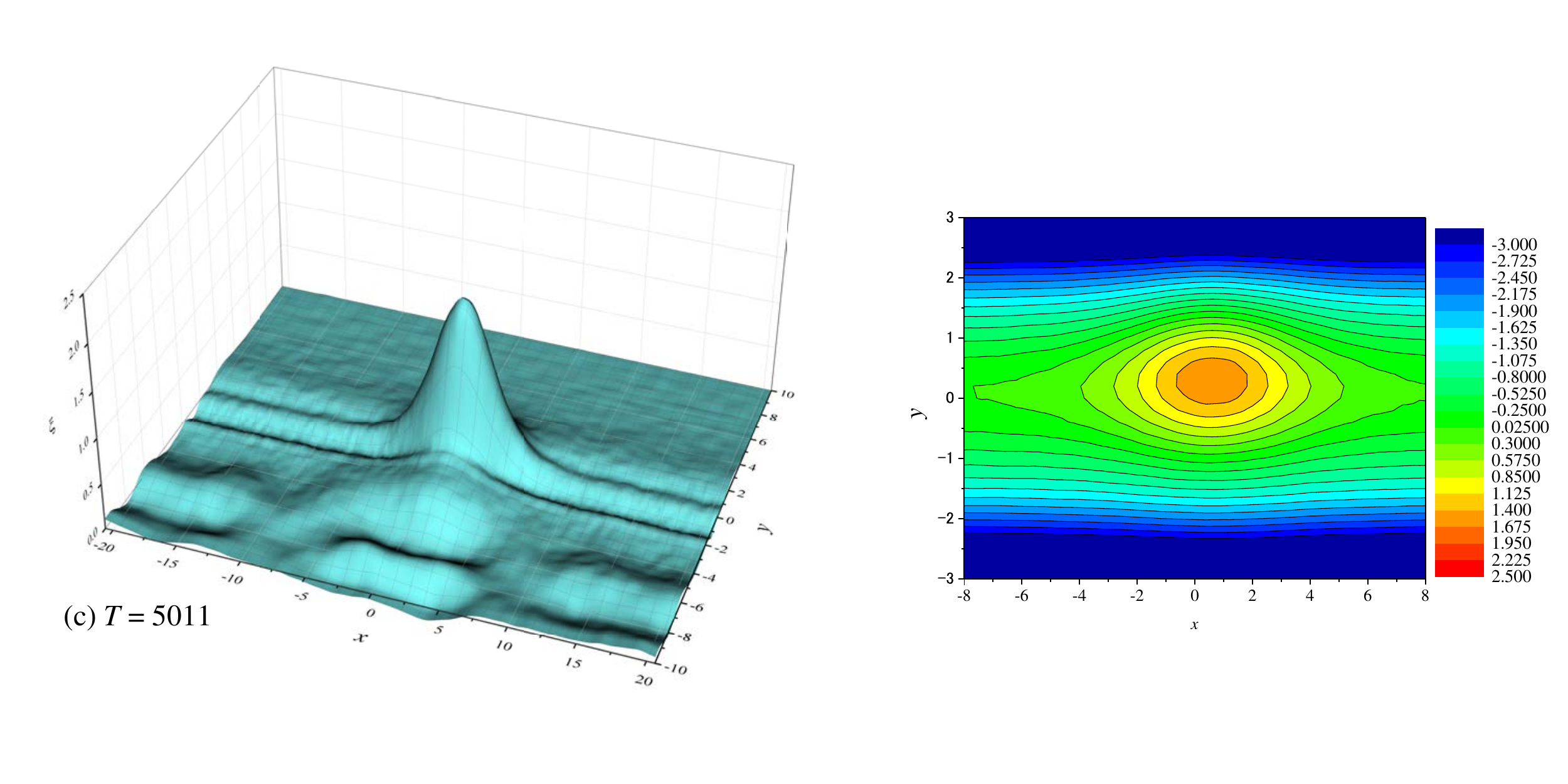}\\
	\vspace{-1.0cm}
	
	\includegraphics[width=120mm]{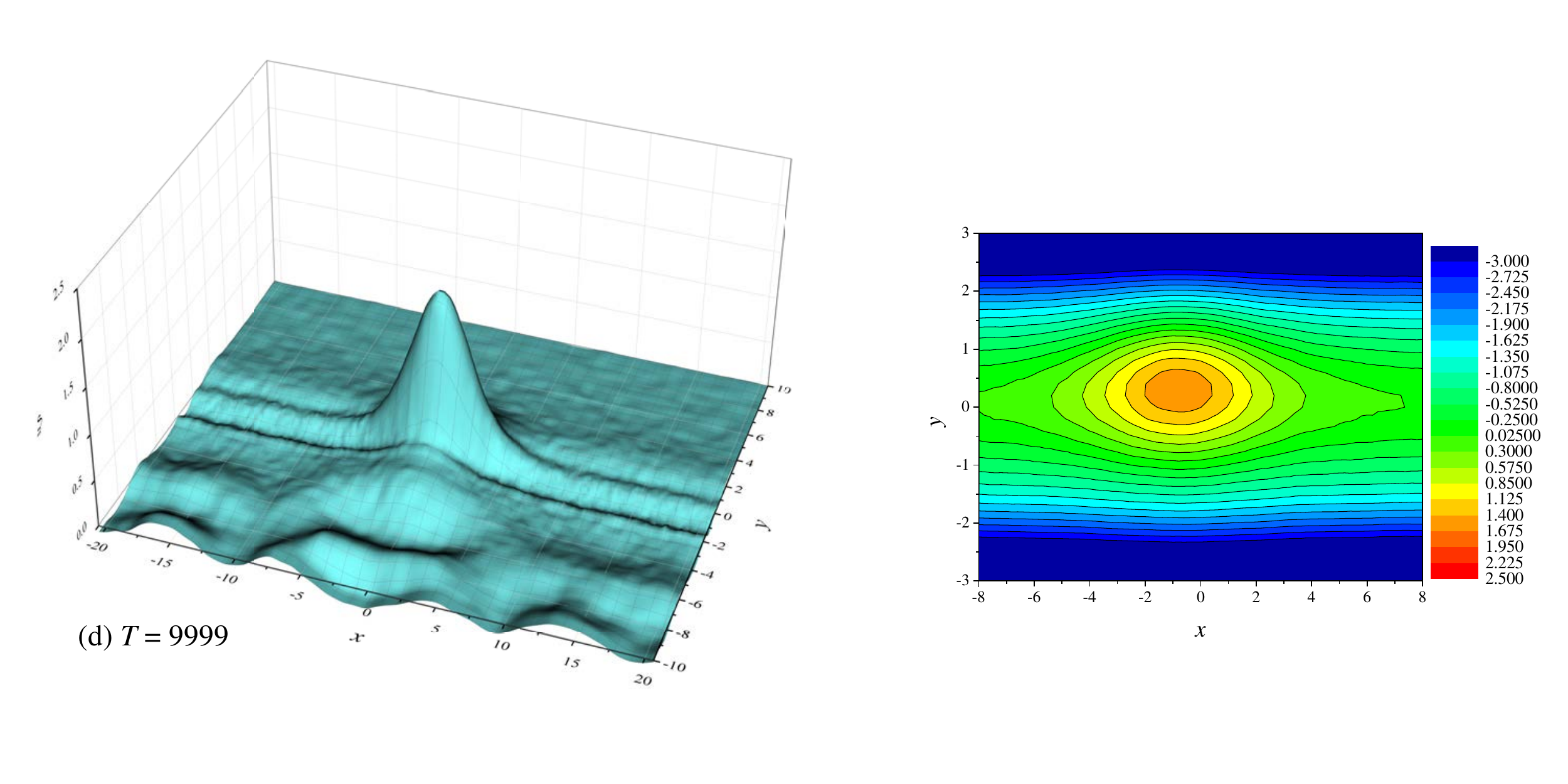}
	\caption{\label{fig:WYFprofile}~Long-term simulation of the profile of the vortex $\eta(x,y,T) $ in $(f_0,f_1)=(0.0,1.2)$ 
	for the iteration number: (a) $0$, (b)~$1005\times 10^{4}$, (c)~$5011\times 10^{4}$, (d)~$9999\times 10^{4}$, respectively. 
	The initial condition is the ZK profile of $c=1$~(see Fig.\ref{fig:ZKcylind}).}
       \end{center}
	\end{figure}


In (\ref{eq:WYF}), the parameters $E$ and $S$ are of $O(1)$, which defines the scales of the vortices. 
Mathematically they can be removed with redefinition of the field and the coordinate such as
\begin{align}
\eta\to \eta'=\frac{1}{2}S^{-2/3}E\eta\,,~~~~x \to x'=S^{-1/3}x\,.
\label{eq:redif}
\end{align} 
The equation (\ref{eq:WYF}) reduces to
\begin{align}
\frac{\partial \eta}{\partial T}-2\eta\frac{\partial\eta}{\partial x'}-\frac{\partial}{\partial x'}(\nabla'^2\eta)
+2y\frac{\partial \eta}{\partial x'}+2J[\nabla'^2\eta,\eta]=0\,.
\label{eq:WYF2}
\end{align}
Henceforth, we omit the prime for simplicity.
As shown in \cite{Williams84}, the WYF equation (\ref{eq:WYF}) has {\it anti-cyclonic} ($\eta>0$) Gaussian shaped vortex solutions when the appropriate background flow, or zonal currents, $u^0(y)$ are applied.
For implementing the effect, we introduce a new variable 
\begin{align}
\xi(x,y,T):=\eta(x,y,T)+\int^y_0 u^0(y')dy'\,.\label{stream function with background}
\end{align}
Substituting this into (\ref{eq:WYF2}), we obtain
\begin{subequations}\label{eq:WYF+shear} 
\begin{align}
\frac{\partial \xi}{\partial T}-2\xi\frac{\partial \xi}{\partial x}-P(y)\frac{\partial }{\partial x}(\nabla^2\xi)
+2Q(y)\frac{\partial \xi}{\partial x}+2J[\nabla^2\xi,\xi]=0\,, 
\\
P(y):=1+2u^0(y)\,,~~~~
Q(y):=y+\frac{\partial^2u^0}{\partial y^2}+\int^y_0u^0(y')dy'\,.
\end{align}
\end{subequations}
The $P,Q$ are regular functions providing the background shear flow or zonal currents.
(\ref{eq:WYF+shear}) is our central concern in this paper. 
Though the $u^0$ can be arbitrary chosen, we employ a linear approximation for the zonal current to simplify the discussion,
\begin{align}
u^0(y)=f_0+f_1 y,\qquad f_0\,,\;f_1\in \mathbb{R}\,.\label{shear flow parameters}
\end{align}
For the case $f_0=0$ and $f_1=1$, the equation reduces to the original WYF equation  \eqref{eq:WYF2}. 
On the other hand, when we choose $f_0=-1$ and $f_1=0$ and omit the Jacobian term, 
the equation (\ref{eq:WYF+shear}) reduces to the ZK type equation (\ref{eq:ZK}). 
Thus, it should have a stable {\it cyclonic} solution: $\xi(x,y):=-\phi(x,y),~(\phi>0)$.
In a general choice of $u^0(y)$, a large class of solutions would exist other than that we will search for in this paper.

At this stage, we observe the ``incomplete" integrable structure of the DSI equation, which is a coupled equation of 
a complex dynamical field $q(x,y,t)$ and a ``mean flow" or a potential $\varphi(x,y,t)$, 
\begin{subequations}\label{eq:DSI}
\begin{align}
    i\frac{\partial q}{\partial t}+\nabla^2 q-\left(\frac{\partial\varphi}{\partial x}+\frac{\partial\varphi}{\partial y}\mp|q|^2\right)q=0\,,\\
    2\frac{\partial^2\varphi}{\partial x\partial y}=\pm\left(\frac{\partial}{\partial x}+\frac{\partial}{\partial y}\right)|q|^2\,,
\end{align}
\end{subequations}
where $\nabla^2$ is the two-dimensional Laplacian as in the WYF equation.
The DSI equation \eqref{eq:DSI} is not a completely integrable system in the sense that it  does not have a sufficient number of conserved quantities \cite{Kaup_1993}. 
However, the DSI equation has numerous well-localized solutions in all spatial directions, the dromions \cite{FOKAS199099,HIETARINTA1990113,Nishinari94}, provided that the boundary conditions for the mean flow are appropriately fixed. 
The prescribed boundary conditions are, of course, not dynamical and considered to be a continuous supply of external flow in which the dromions can be alive stably.
Thus, we can regard the proper external flow, or force, as an ingredient of the integrability of the DSI equation, although the integrability is quite deficient. 
The DSI equation is an autonomous equation, meaning that the interaction with an external flow is incorporated only in the boundary conditions.
On the other hand, in the WYF equation \eqref{eq:WYF}, the effect of the external flow is given through the non-autonomous terms with 
the functions $P(y)$ and $Q(y)$ in \eqref{eq:WYF+shear} involving free parameters as \eqref{shear flow parameters}.
We will show that there exist the critical values of external flow, \ie the parameters $f_0$ and $f_1$, to stabilize a localized solution in the WYF equation.
This suggests that the interaction between the dynamical systems and the suitable supply of external flow gives rise to a stable soliton-like structure even if the dynamical system is not integrable.
We refer to such a system as a ``mock integrable system" hereafter.

We analyze the WYF equation in a region periodic in $x$-direction and bounded in $y$-direction, \ie
solving the equation in a rectangular domain $-L_x\leqq x\leqq L_x, -L_y\leqq y\leqq L_y$. 
We employ a stretched region $L_x=20, L_y=10$. 
Explicitly, the periodic boundary condition in $x$ and the free-slip condition for the $y$ boundary are written as
\begin{align}
&\xi(-L_x,y)=\xi(L_x,y),~~y\in [-L_y,L_y]\,,~~~~
\frac{\partial \xi(x,y)}{\partial y}\biggl|_{y=\pm L_y}=0,~~x\in [-L_x,L_x]\,.
\label{freeslip}
\end{align}
The velocity field is defined as
\begin{align}
v_x:=-\frac{\partial \eta}{\partial y}=-\frac{\partial \xi}{\partial y}+u^0(y)-u^0(0)\,,~~~~
v_y:=\frac{\partial \eta}{\partial x}=\frac{\partial \xi}{\partial x}\,.
\nonumber 
\end{align}
The boundary conditions (\ref{freeslip}) support that $x$ component of the velocity of the 
vortex smoothly connects to the external flow at the $y$-boundary.

	\begin{figure}[t]
	\begin{center}

    \includegraphics[width=130mm]{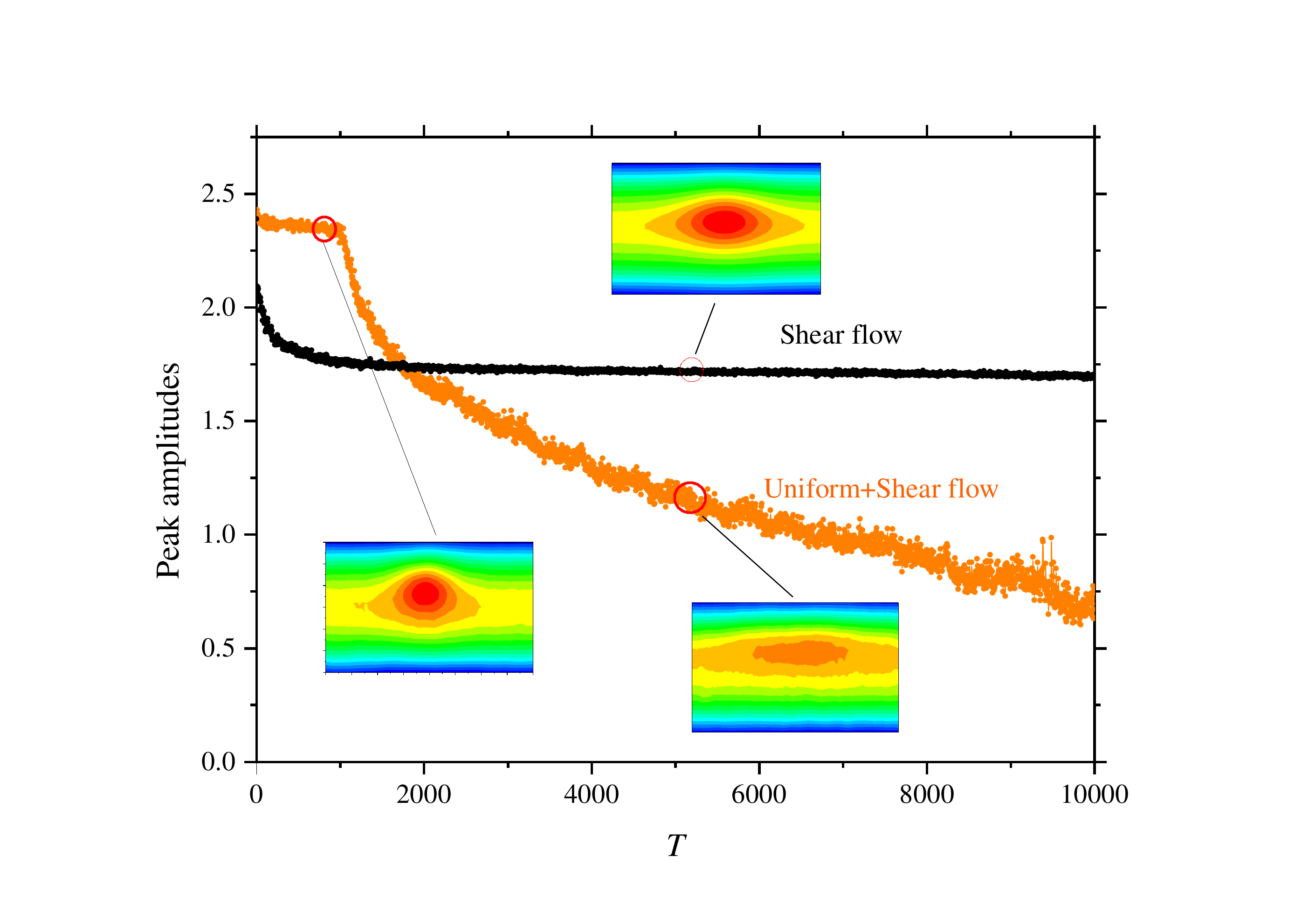}~~

	\caption{\label{fig:Peakvaluef_WYF} The peak amplitudes of the $\xi(x,y,T)$ of the shear flow $(f_0,f_1)=(0.0,1.2)$ 
	and also the shear flow with the uniform flow $(f_0,f_1)=(-1.0,1.2)$. We also plot the profiles at $T\sim 1000, 5000$. }
  \end{center}
	\end{figure}


	\begin{figure}[t]
	\begin{center}

	\includegraphics[width=130mm]{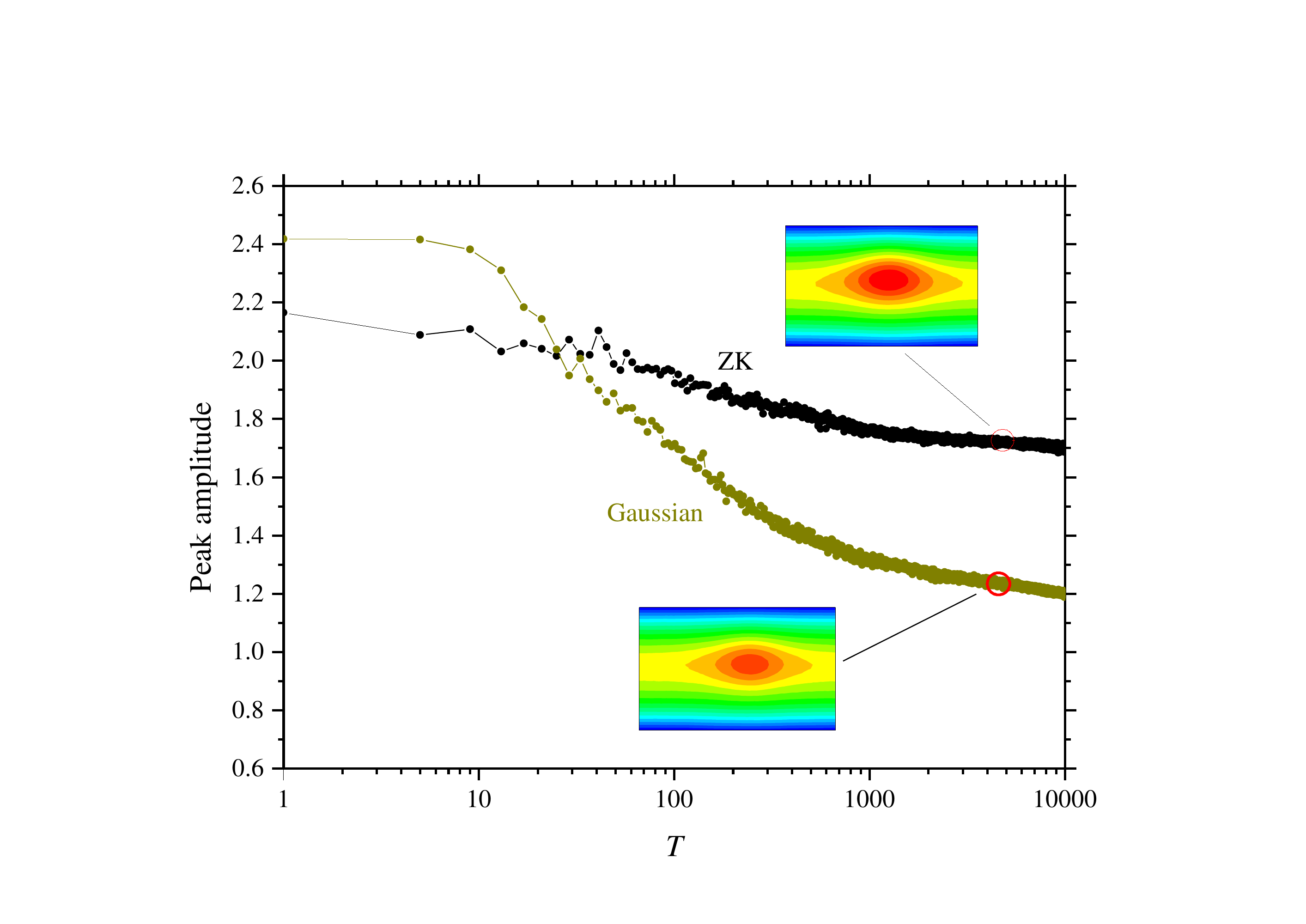}\hspace{-0.5cm}
		
	\caption{\label{fig:Peakvaluef_Gauss}~
The peak amplitudes of the $\xi(x,y,T)$ of the shear flow $(f_0,f_1)=(0.0,1.2)$ of the initial conditions
	of the ZK and the Gaussian $\xi_{\rm init}=3.0\times\exp\{-x^2-(y-1)^2\}$ profiles. 
	We also plot the profiles at $T\sim 5000$.}
       \end{center}
	\end{figure}


\subsection{A conservation nature of the integrals of motion in the WYF equation}

As in the case of the ZK equation, several known (semi-)integrable systems always possess a few (or infinite) 
conserved quantities and their existence strongly support the stability of the solutions. 
Although the WYF equation has a similar structure to the ZK equation in part, the existence of the 
Jacobian term breaks conservation laws that the ZK has. 
We briefly describe the situation.
For example, the conservation of the counterpart of the first quantity $M$ (\ref{CQ1}) in the equation~(\ref{eq:WYF+shear}) becomes
\begin{align}
\frac{d\tilde{M}}{dT} \nonumber
&= \frac{d}{dT}\int\xi dxdy \\
&= \int dy\biggl[\xi^2+P(y)\nabla^2\xi-2Q(y)\xi-2\nabla^2\partial_y\xi\biggr]^{L_x}_{-L_x}
+2\int dx\biggl[\nabla^2\xi\partial_x\xi\biggr]^{L_y}_{-L_y}\,.
\end{align}
The last term in the first integral and the second integral come from the Jacobian. 
All terms are the surface terms and then possibly to be zero caused by the boundary conditions.
The first integral becomes zero because of the periodic boundary condition of (\ref{freeslip}).
For the second term, we rewrite it by using the field $\eta$
\begin{align}
\int dx\biggl[\nabla^2\xi\partial_x\xi\biggr]^{L_y}_{-L_y}
=\int dx\biggl[\Bigl(\partial_x^2\eta+\partial_y^2\eta+f_1\Bigr)\partial_x\eta\biggr]^{L_y}_{-L_y}\,.
\label{CQerror}
\end{align}
There is no boundary condition which kills this term.
However, when the vortex is located at far from the boundary
and moving to $x$ direction, the notable modulation of $\eta$ at the $y$ boundary may be almost ignored, \ie
\begin{align}
\partial_x\eta=\partial_x^2\eta=\cdots =0\,,\qquad \partial^2_y\eta=\cdots =0\,,
\nonumber 
\end{align} 
which recovers the conservation of the quantity 
$\tilde{M}$ as the ZK does. 
The point is that, when the shear flow $f_1$ is finite, the value (\ref{CQerror}) may grows as 
the flow becomes stronger. 
Physically, we may say the shear flow supplies some influx to the system, which stabilize or destabilize the vortex. 

On the other hand, the counterpart of the second quantity $P$ (\ref{CQ2}) is apparently not conserved, unfortunately.
It becomes
\begin{align}
\frac{d\tilde{P}}{dT}&=\frac{d}{dT}\int\frac{1}{2}\xi^2dxdy
\nonumber \\
&=\int dy\biggl[\frac{2}{3}\xi^3+P(y)\xi\partial_x^2\xi-Q(y)\xi^2-\frac{1}{2}P(y)(\partial_x\xi)^2
+\frac{1}{2}P(y)\xi\partial_y^2\xi-2\xi\nabla^2\xi\partial_y\xi\biggr]_{-L_x}^{L_x}
\nonumber \\
&~~~
+\int dx\biggl[-\frac{1}{2}P(y)\partial_x\xi\partial_y\xi+\frac{1}{2}P(y)\xi\partial_x\partial_y\xi
+2\xi\nabla^2\xi\partial_x\xi\biggr]_{-L_y}^{L_y}
\nonumber \\
&~~~
+\int dxdy f_1(\partial_x\xi\partial_y\xi-\xi\partial_x\partial_y\xi)\,.
\end{align}
The last term of the right hand side is not the surface term thus unless the zero shear flow, $\tilde{P}$ is  not conserved.

\section{The numerical analysis}

The equation (\ref{eq:WYF+shear}) is highly non-linear and then is difficult to find any type of analytical 
solutions. Particularly, studies of the long-term simulation of the evolution of the vortex solutions, we need to 
rely on accurate and efficient numerical analysis. 
After describing our numerical setup, 
we shall show our successful numerical result where all the effects are fully incorporated. 

\subsection{The method}

There are several numerical methods to investigate the nonlinear evolution equations, {\it e.g.} KdV equations and others
~\cite{ZABUSKY1981195,TAHA1984203,TAHA1984231}. The semi-implicit method~\cite{LI1995121}, 
the pseudo-spectral method~\cite{Fornberg78,JAIN1997943,MUSLU2003503}, with discrete 
Galerkin methods for the spatial mesh~\cite{BONA1986859,XU200521} have been employed so far. 
For our numerical analysis, we use the simple explicit finite-difference method with the uniform mesh, 
which are known as the primer techniques for solving partial differential equations, see \cite{smith1985numerical}.
In order to gain accuracy, we have tried to implement the classical fourth-order Runge-Kutta method and also the Leap-frog method
for the time discretization, the former has an advantage for the numerical stability. To ensure the scheme fourth-order accurate, 
higher order discretization scheme for the spatial grid than the usual second-order must be implemented.    
The fourth-order finite differences method~\cite{Singer98} has been used for improving the numerical stability especially for solutions 
of several soliton models~\cite{HASSANIEN2005781,LEE201417,WANG2019310}. 
The derivatives of a field $u(x)$ can be expressed as follows by using the formula:
\begin{align}
&D_xu(x):=\frac{-u(x+2h)+8u(x+h)-8u(x-h)+u(x-2h)}{12h}\,,
\nonumber \\
&D_{xx}u(x):=\frac{-u(x+2h)+16u(x+h)-30u(x)+16u(x-h)-u(x-2h)}{12h^2}\,, 
\\
&D_{xxx}u(x):=\frac{u(x+2h)-2u(x+h)+2u(x-h)-u(x-2h)}{2h^3}\,,
\nonumber 
\end{align}
where $h$ is a grid spacing. 

Let the integers $(N_x,N_y)=(200,100)$ denote the number of spatial grids and $h := 2L_x/N_x = 2L_y/N_y=0.2$ be the uniform grid spacing. 
It is well-known that a special care is required with the Jacobian, the vector nonlinear term, for the numerical stability.
The Arakawa method is known for avoiding the numerical instability 
with conserving the several quantities such like the square of the vorticity and also the kinetic energy \cite{ARAKAWA1966119}.  
Since the Arakawa method is designed in the second-order finite difference for the spatial grid, 
we simply extend the method with the fourth-order. 
In Appendix~\ref{4thdifference}, we present an explicit form of the fourth-order difference scheme 
and also of the Arakawa scheme.

Most of our numerical calculations, the time difference $\Delta t=1.0\times 10^{-4}$ is employed, 
then the time $T$ is estimated via 
\begin{align}
T=\Delta t\times \textrm{(iteration number)}.
\end{align}
For the purpose of comparing our results with the previous one~\cite{Williams84}, we observe  
the relation between the real time scale $\tilde{T}$ and $T$ are defined 
in the derivation of the governing equation (\ref{eq:WYF}) from a shallow-water equation on a $\beta$ plane of planetary physics.
According to the prescription used in \cite{Williams84}, it is 
\begin{align}
\Tilde{T}:=\frac{1}{f\hat{s}\hat{\beta}^2}T;~~~~\hat{\beta}=\frac{\beta L}{f},~~\hat{s}=\frac{\lambda_R^2}{L^2}\,,~~
f=2\Omega\sin\phi,~~\beta=\frac{2\Omega}{R}\cos\phi\,,
\end{align} 
where $L$ is a typical size of the vortex, 
$R$ and $\Omega$ are a radius and a angular velocity of the planet and $\phi$ is a latitude where the vortex sits.
The Rossby deformation radius $\lambda_R:=\sqrt{gH}/f$ is defined in terms of the height $H$ of dynamics 
of our concern. 
For typical parameters of the Jupiter's Red spot, it gives the
rough estimation of the physical time $\tilde{T}\sim 0.27 \times T$.

\subsection{The initial profile of vortices}

In the actual simulation, a suitable initial profile has to be chosen for supporting stable behavior of the solution in the initial a few steps.
We have found that the choice is also essential for the long-term behavior (longevity), 
which means that genuine information of nonlinearity of the system is involved in the initial condition. 
In this paper, we have employed two types: the standard Gaussian profile and  
the ZK circular symmetric solution shown in Fig.\ref{fig:ZKcylind}. 
The Gaussian profile was used in \cite{Williams84} and this may be one of the reasons why their solutions were not so rigid. 
Instead, the ZK solution is more efficient for the stability because the equations share some basic features. 
Also, the zonal current $u^0(y)$ can be freely chosen and  
we study the two typical cases. We implement 
no uniform flow $f_0=0.0$ and a uniform flow $f_0=-1.0$, in which the equation has a common structure with the ZK, and in both cases with several values of positive shear flow $f_1$, which corresponds to the anti-cyclonic shear. 

On the initial configuration of the vortices, we notice the following aspect.
During the initial profiles eventually converging towards the stable configuration, some reformations or adjustments inevitably occur at the early stage. 
A small trick about an alignment of the initial profile by shifting upward point $y=1.0$ certainly mitigates the instability. 
The origin of this anisotropy in the $y$-direction should be investigated in future works.

Figures \ref{fig:WYFprofile_gauss} and \ref{fig:WYFprofile} show the single vortex profiles $\eta(x,y,T)$ of solutions of the 
Gaussian and the ZK profile initial conditions with the shear flow $f_1=1.2$ for several time steps: $T\sim 0, 1000, 5000, 10000$, respectively. 
In each case, the vortex moves to the right (positive $x$ direction) with roughly a constant speed.
In the Gaussian case Fig.\ref{fig:WYFprofile_gauss}, the shape is a typical oval and for later time it gradually dissipate 
and becomes small. 
On the other hand, in the ZK initial condition, the shape keeps for longer period than the Gaussian case (see Fig.\ref{fig:WYFprofile}).

In order to see the effects of the background uniform flow, we examine evolution of the peak behavior of the solutions. 
Figure \ref{fig:Peakvaluef_WYF} compares peak amplitude with/without the uniform flow over the long term simulation, 
which shows the uniform flow destabilizes the solutions unexpectedly.
Here, the initial profile is the ZK for both simulations. 
For $f_0=0.0$, after some small arrangements at the initial stage of the simulation,  
at $T\sim 2000$, the solution becomes more stable and this continues until the end of the simulation $T=10000$ with a very small dissipation.
On the other hand, with the uniform flow $f_0=-1.0$, the solution keeps the shape at the beginning and then, begins to decay after $T=1000$ and slowly goes to collapse. 
Figure \ref{fig:Peakvaluef_Gauss} shows the modulation of the peak amplitude of the vortex $\xi(x,y,T)$ for the initial conditions of the ZK and the Gaussian profile.  
In \cite{Williams84}, the authors employed the simple cylindrical Gaussian profile for initial condition 
and all of their results ended in short life.
Our result clearly indicates that the ZK profile has advantageous than the Gaussian. 
This suggests us that our solutions share some basic feature with that of the ZK equation because most of the terms in the equations are common and the WYF equation partially inherits the integrable structure of the ZK equation.

	\begin{figure*}[t]
	\begin{center}

	\includegraphics[width=60mm]{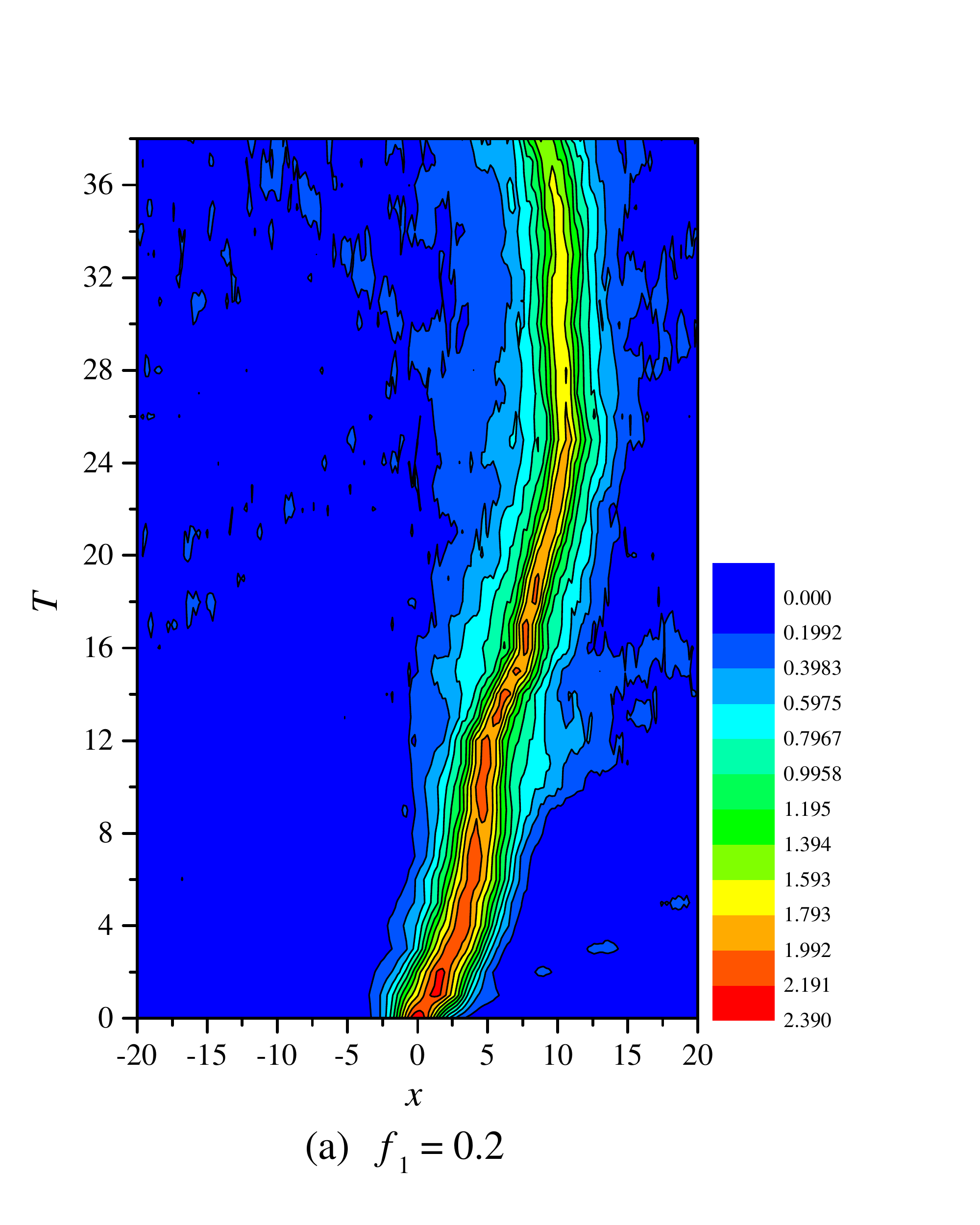}\hspace{-1.0cm}
	\includegraphics[width=60mm]{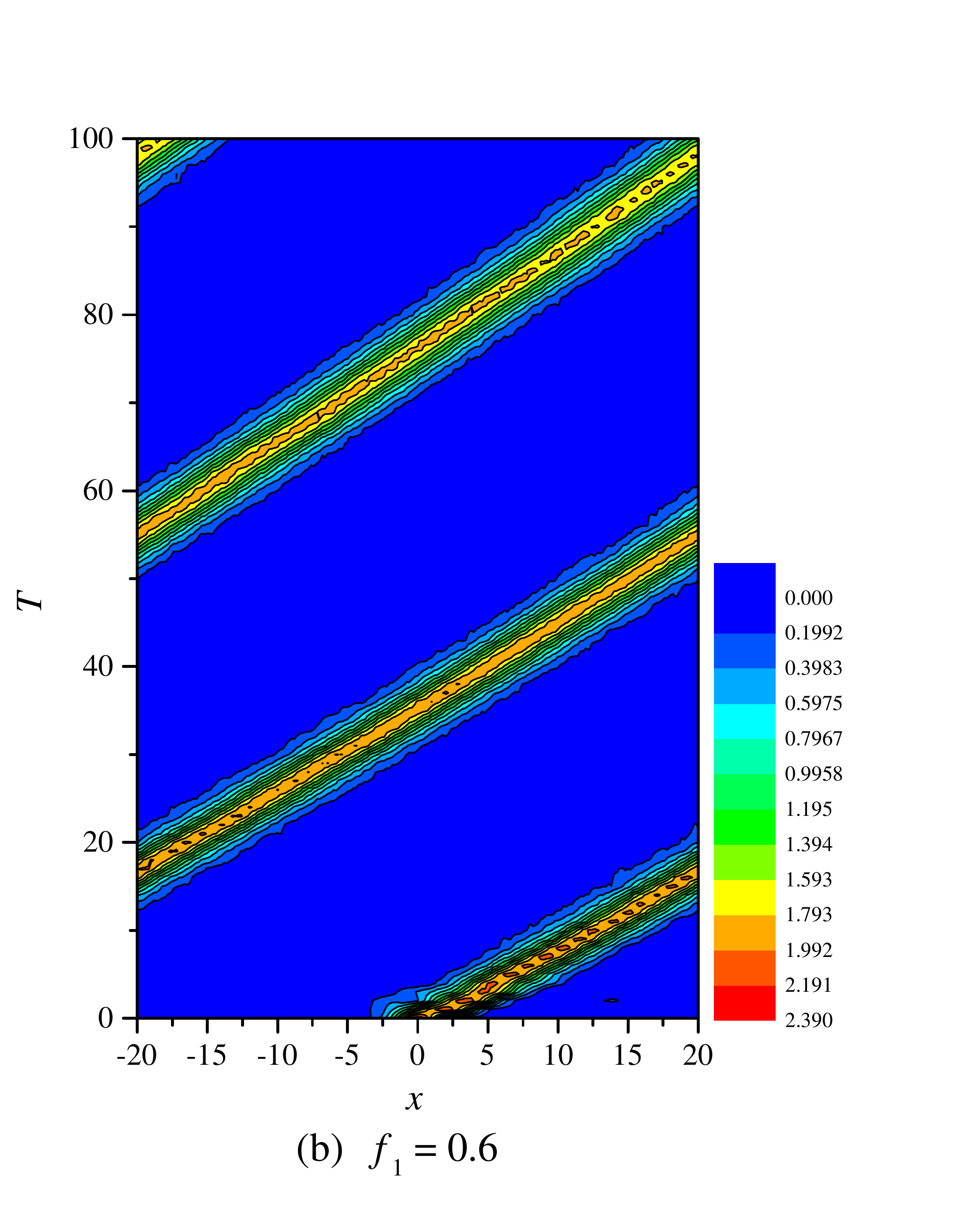}\hspace{-1.0cm}
	\includegraphics[width=60mm]{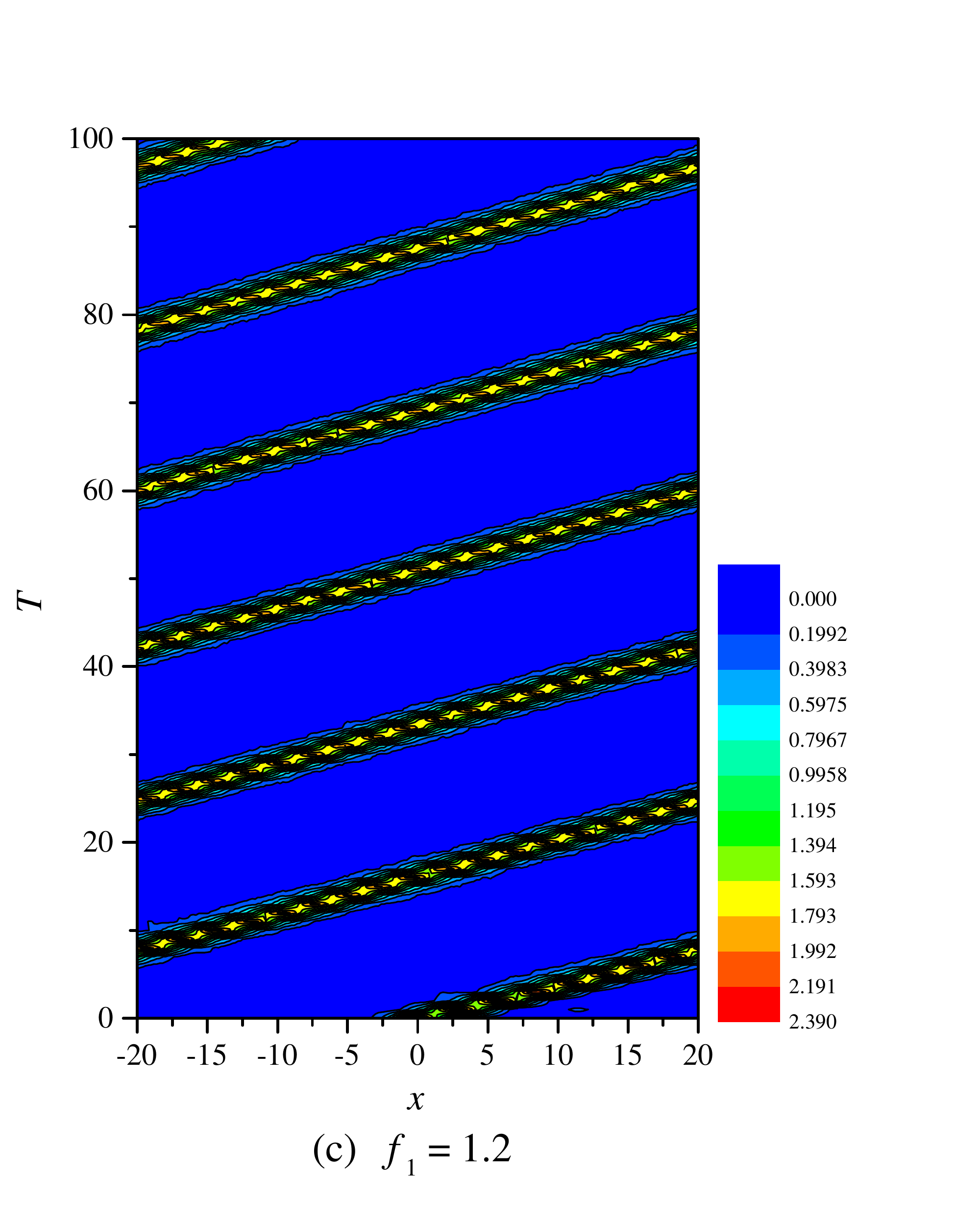}
	
	\caption{\label{fig:evolution0}~$\xi(x,y,T)$ in $f_0=0.0$ with several $f_1$
	in short period time evolution: $0\leq T \leq 100$. The $x$-slice solutions with $f_1=0.2,0.6,1.2$ are plotted. 
	In the case $f_1=0.2$, the solution collapses at $T\sim 38$. }
       \end{center}
	\end{figure*}


	\begin{figure}[h]
	\begin{center}

	\includegraphics[width=60mm]{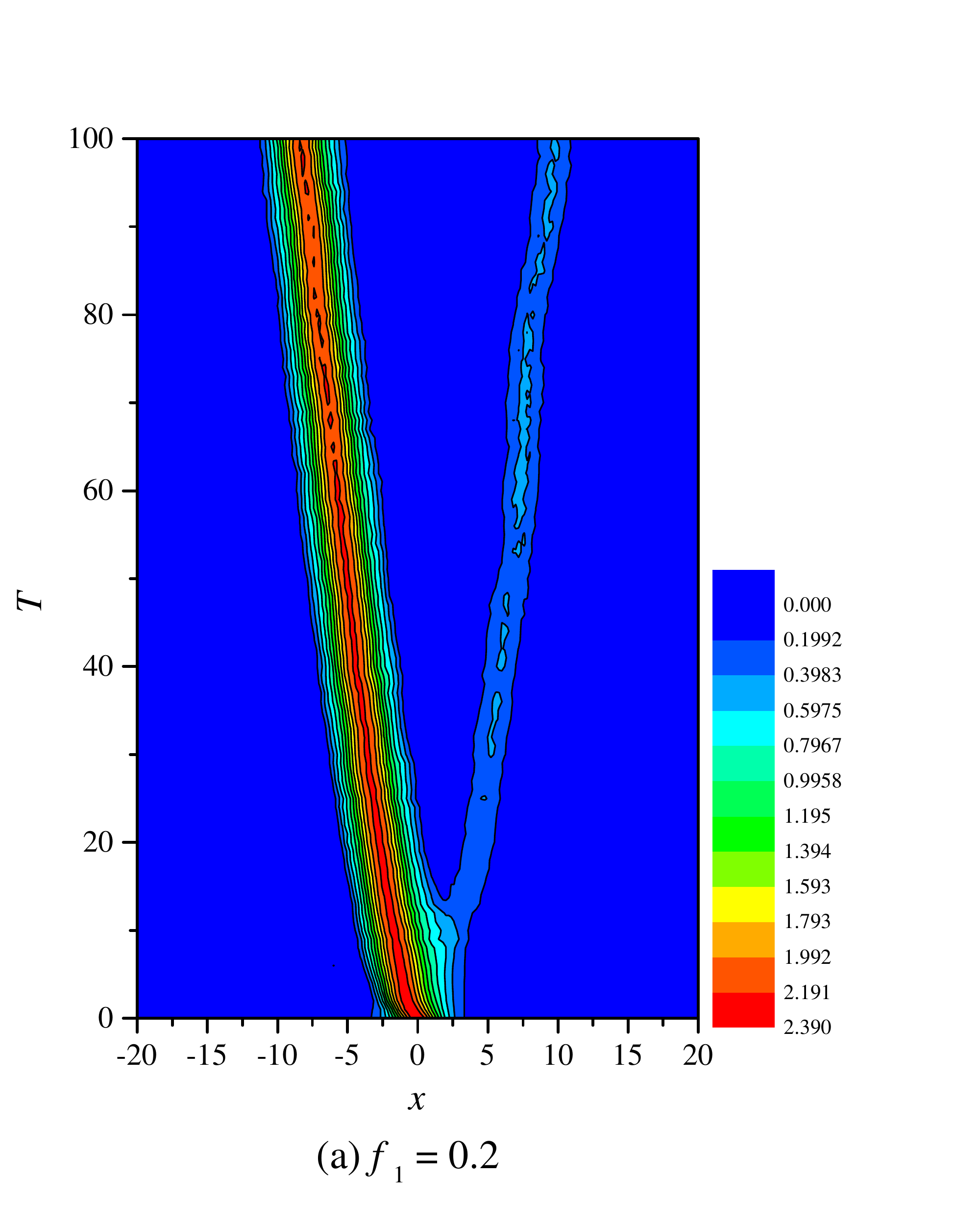}\hspace{-0.9cm}
	\includegraphics[width=60mm]{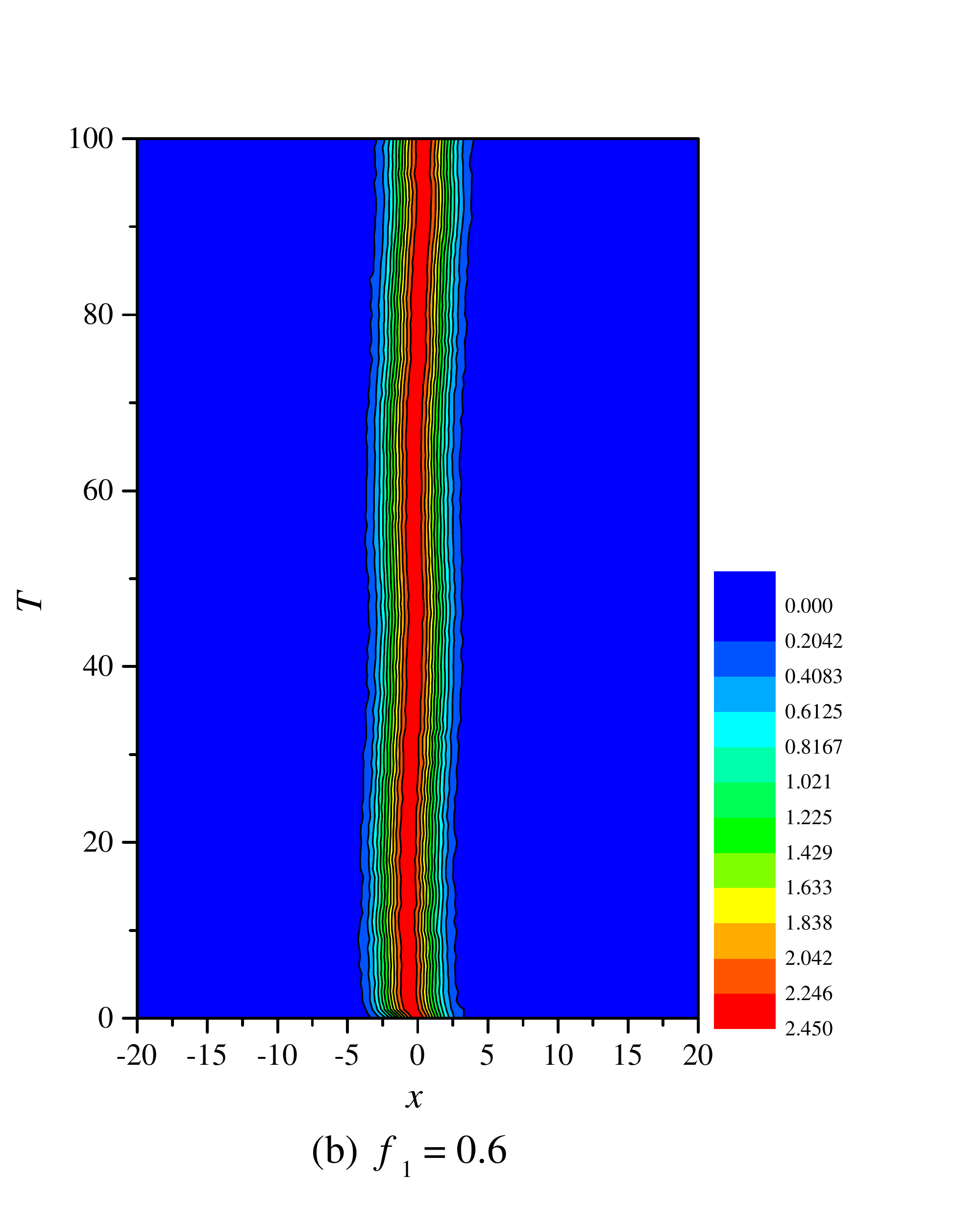}\hspace{-0.9cm}
	\includegraphics[width=60mm]{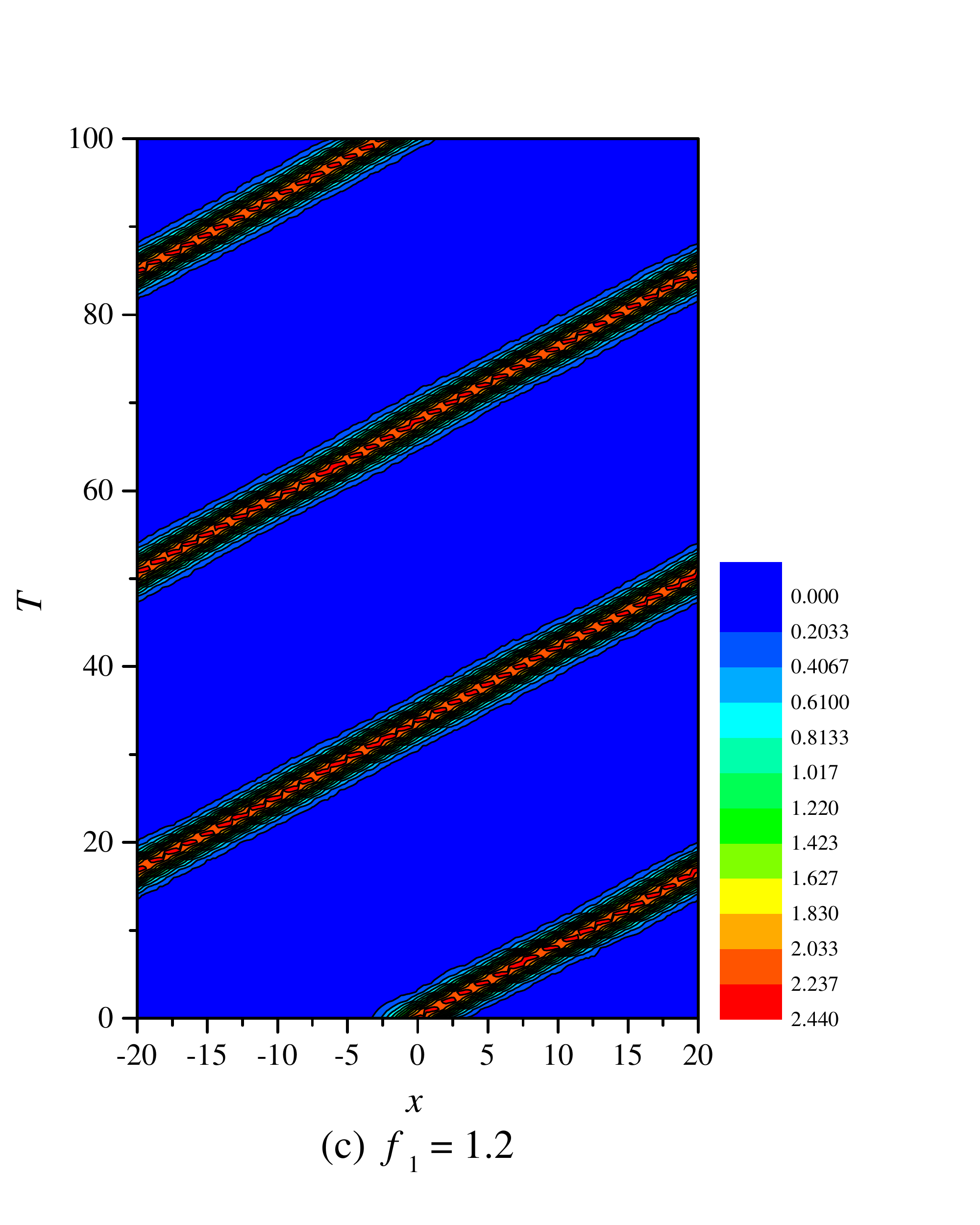}
	
	\caption{\label{fig:evolution-1}~$\xi(x,y,T)$ in $f_0=-1.0$ with several $f_1$ 
	in short period time evolution: $0\leq T \leq 100$. The $x$-slice solutions with $f_1=0.2,0.6,1.2$ are plotted. }
       \end{center}
	\end{figure}


	\begin{figure}[t]
	\begin{center}

	\includegraphics[width=130mm]{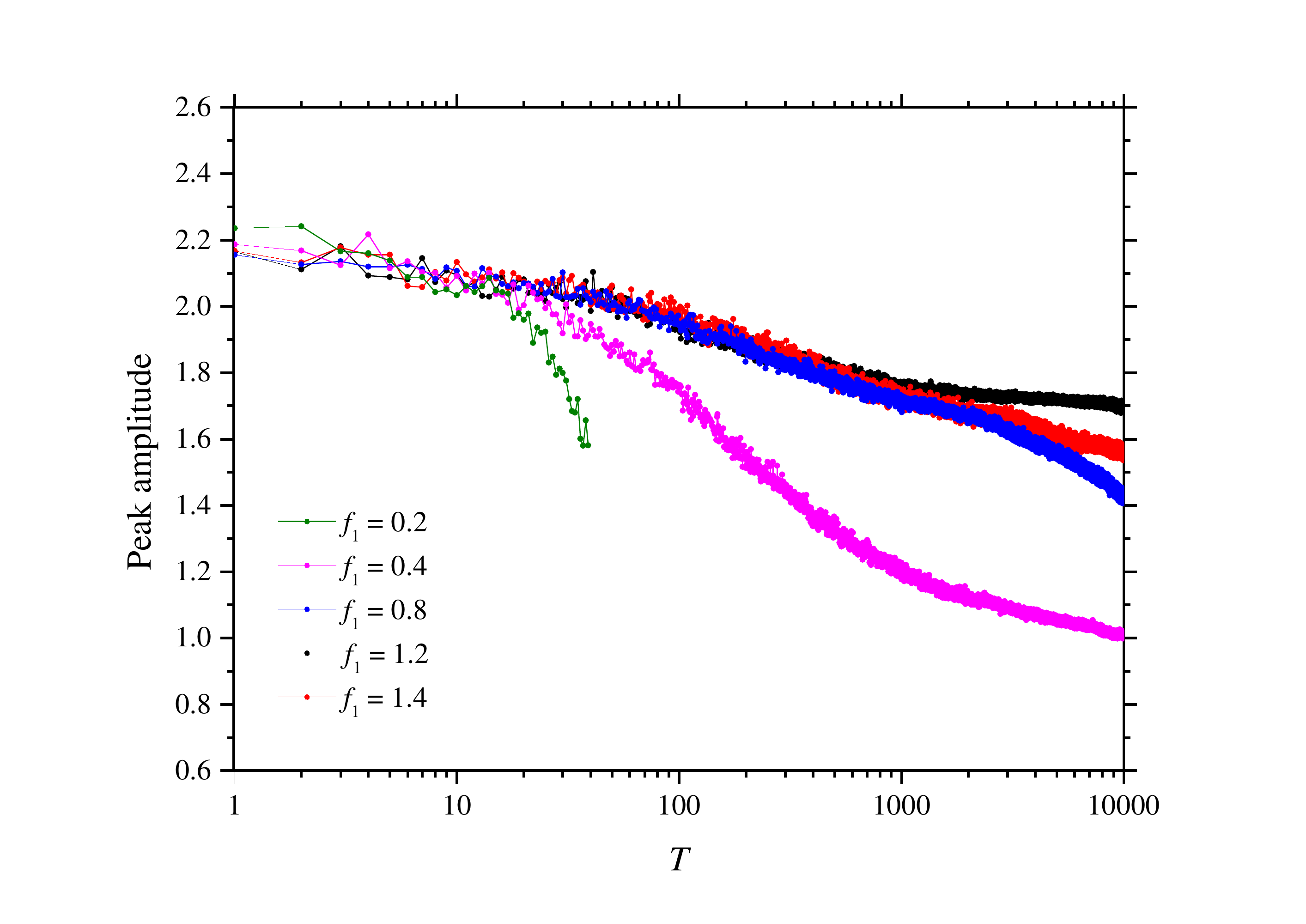}\hspace{-0.5cm}
		
	\caption{\label{fig:peakvalue}~The long-term behavior of the vortex in $f_0=0.0$ with $f_1=0.2,0.4,0.8,1.2,1.4$. 
	The solution with $f_1=1.2$ is stable and has a longevity. }
       \end{center}
	\end{figure}


	\begin{figure*}[t]
	\begin{center}
	
	\includegraphics[width=130mm]{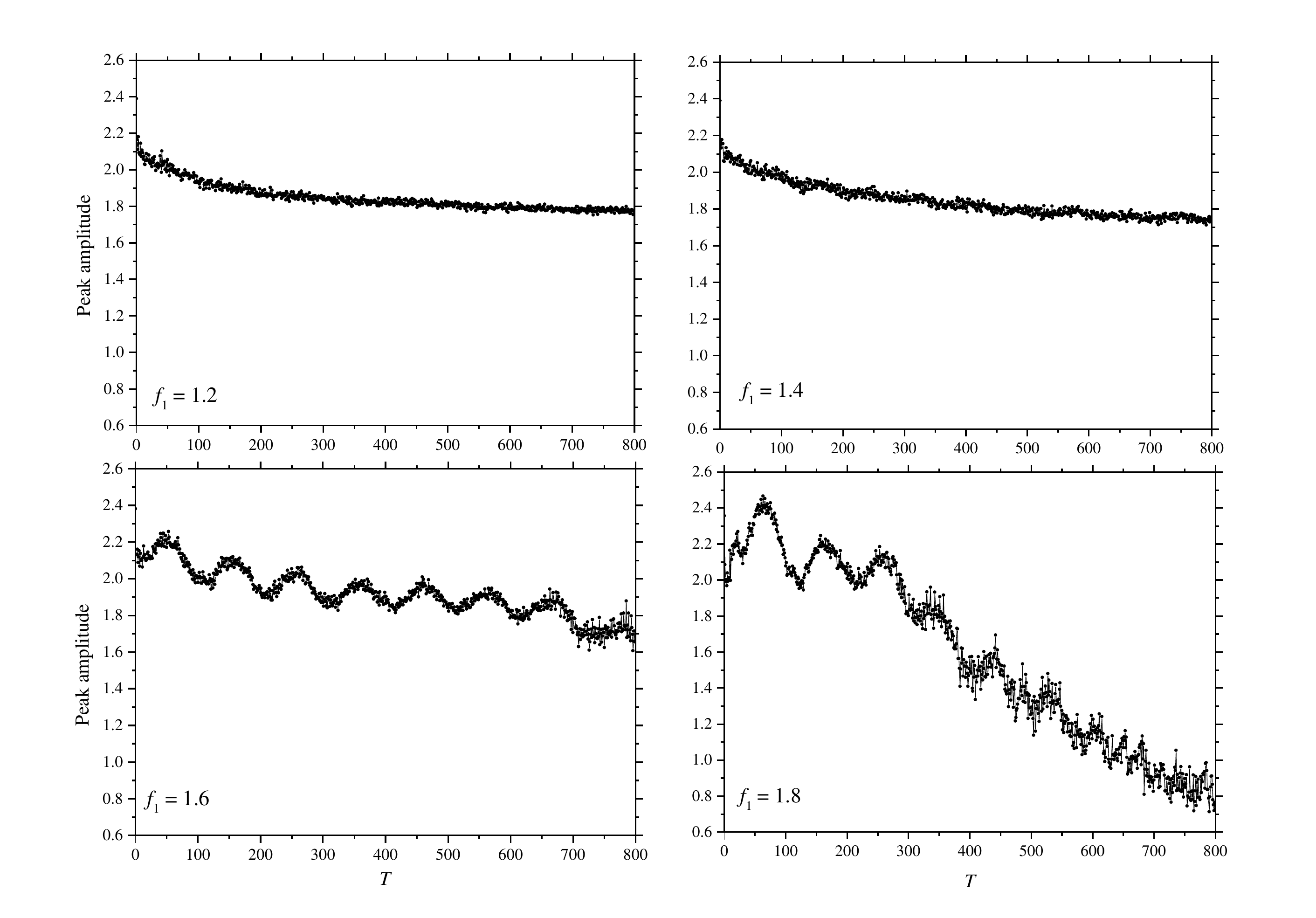}\hspace{-1.0cm}

	\caption{\label{fig:peakvaluef16}~Behaviors of the peak amplitude of the vortices in $f_0=0.0$ 
	with $f_1=1.2, 1.4, 1.6, 1.8$.  
	For $f_1\geqq 1.4$, excited vibrations is observed but is more apparent in the case of $f_1=1.6,1.8$.  
	Especially, case of $f_1=1.8$, the peak moves to the boundary area so we plot the value on $y\sim 1.0$. }
       \end{center}
	\end{figure*}


	\begin{figure*}[htbp]
	\begin{center}

	\includegraphics[width=130mm]{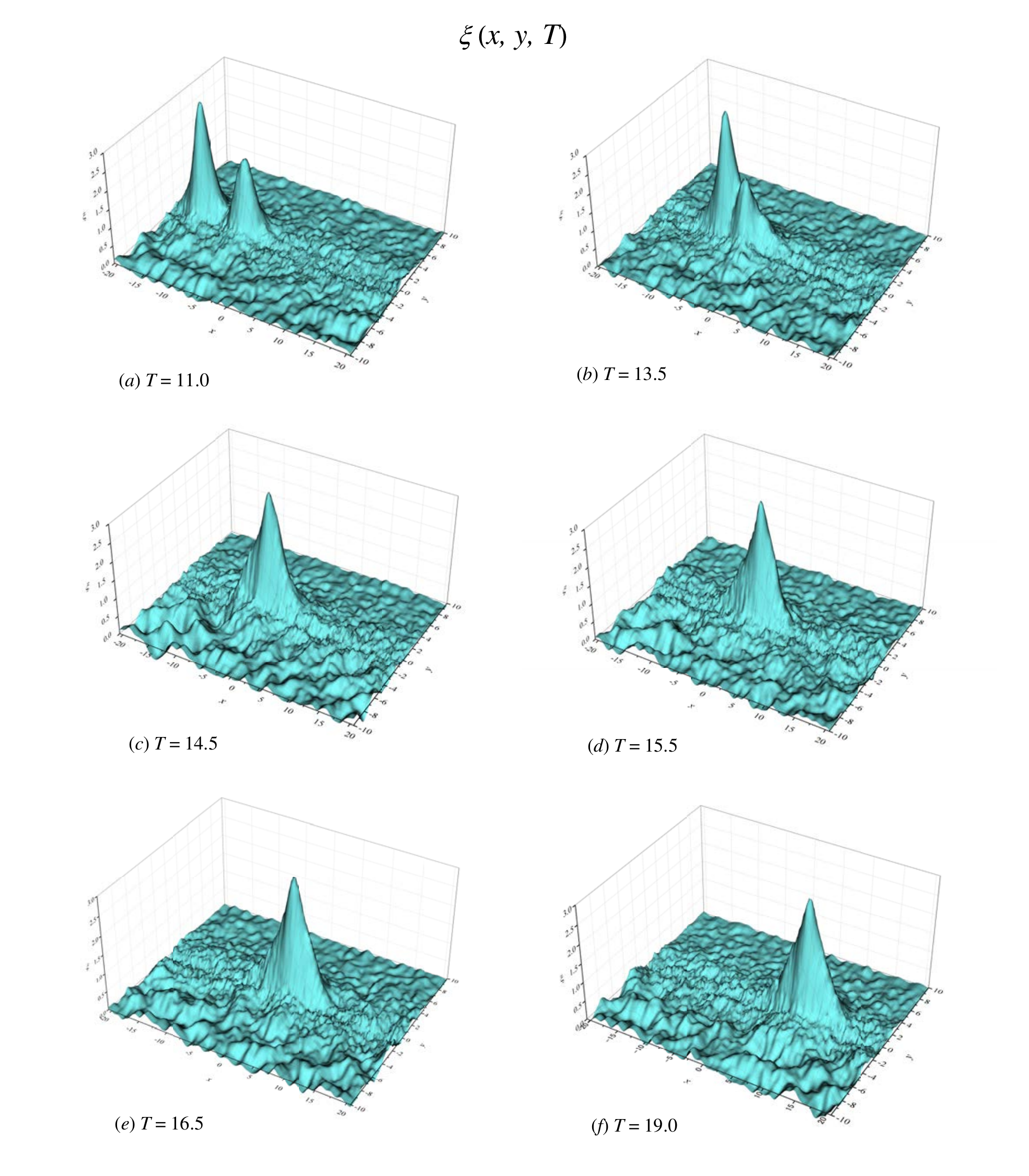}
	\caption{\label{fig:WYFprofile3d_2soliton}~Collision of the two vortices; 
	The profiles $\xi(x,y,T)$ in $(f_0.f_1)=(0.0,1.0)$. We employ the initial condition with ZK solutions
	with the taller (faster) vortex $c=1.5$ and the smaller (slower) one $c=1.0$.}
       \end{center}
	\end{figure*}


	\begin{figure*}[htbp]
	\begin{center}

	\includegraphics[width=130mm]{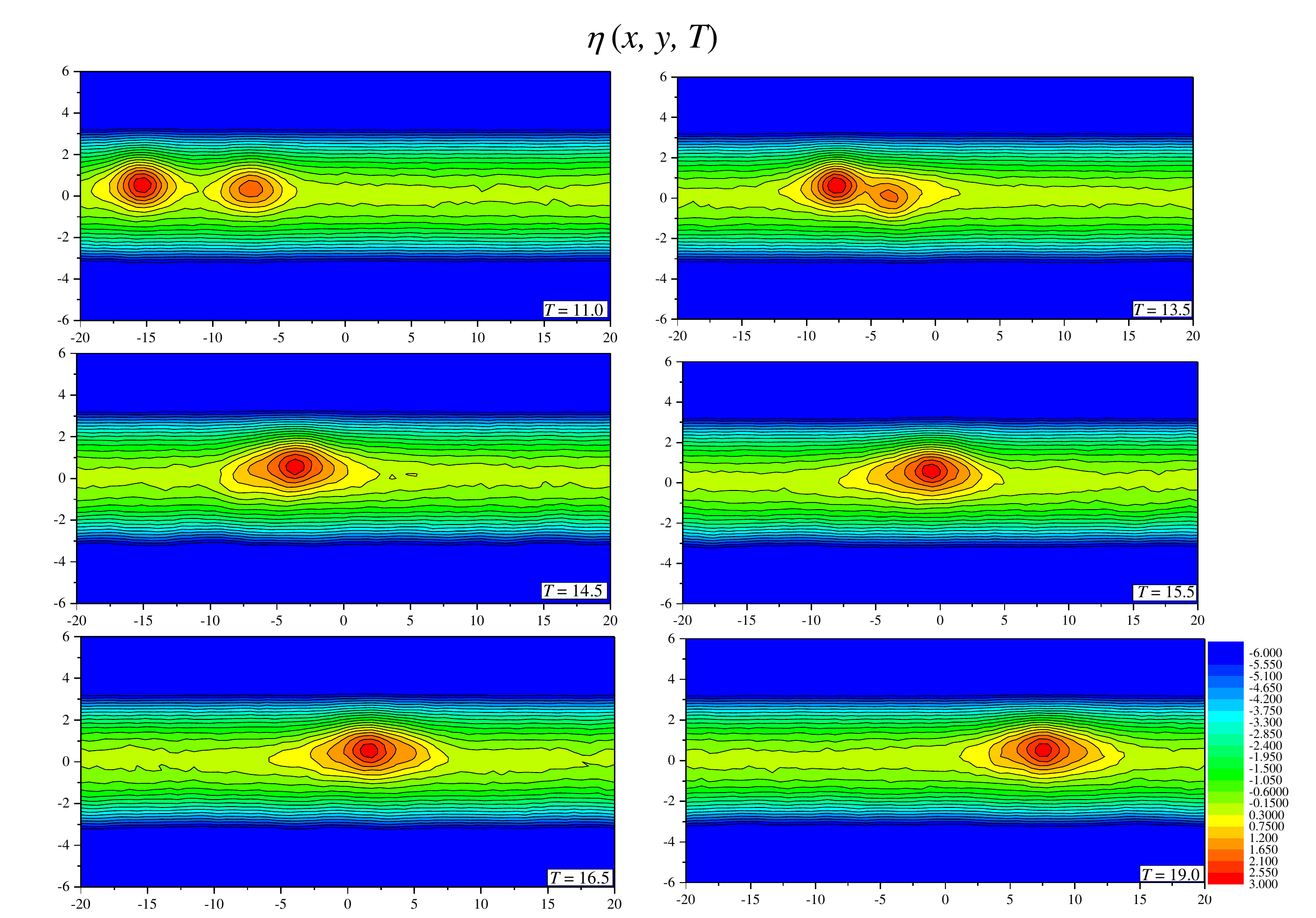}
	\caption{\label{fig:WYFprofile_2soliton}~Same as Fig.\ref{fig:WYFprofile3d_2soliton}, but
	here are the plots of the profiles $\eta (x,y,T)$. }
       \end{center}
	\end{figure*}


	\begin{figure}[htbp]
	\begin{center}

	\includegraphics[width=130mm]{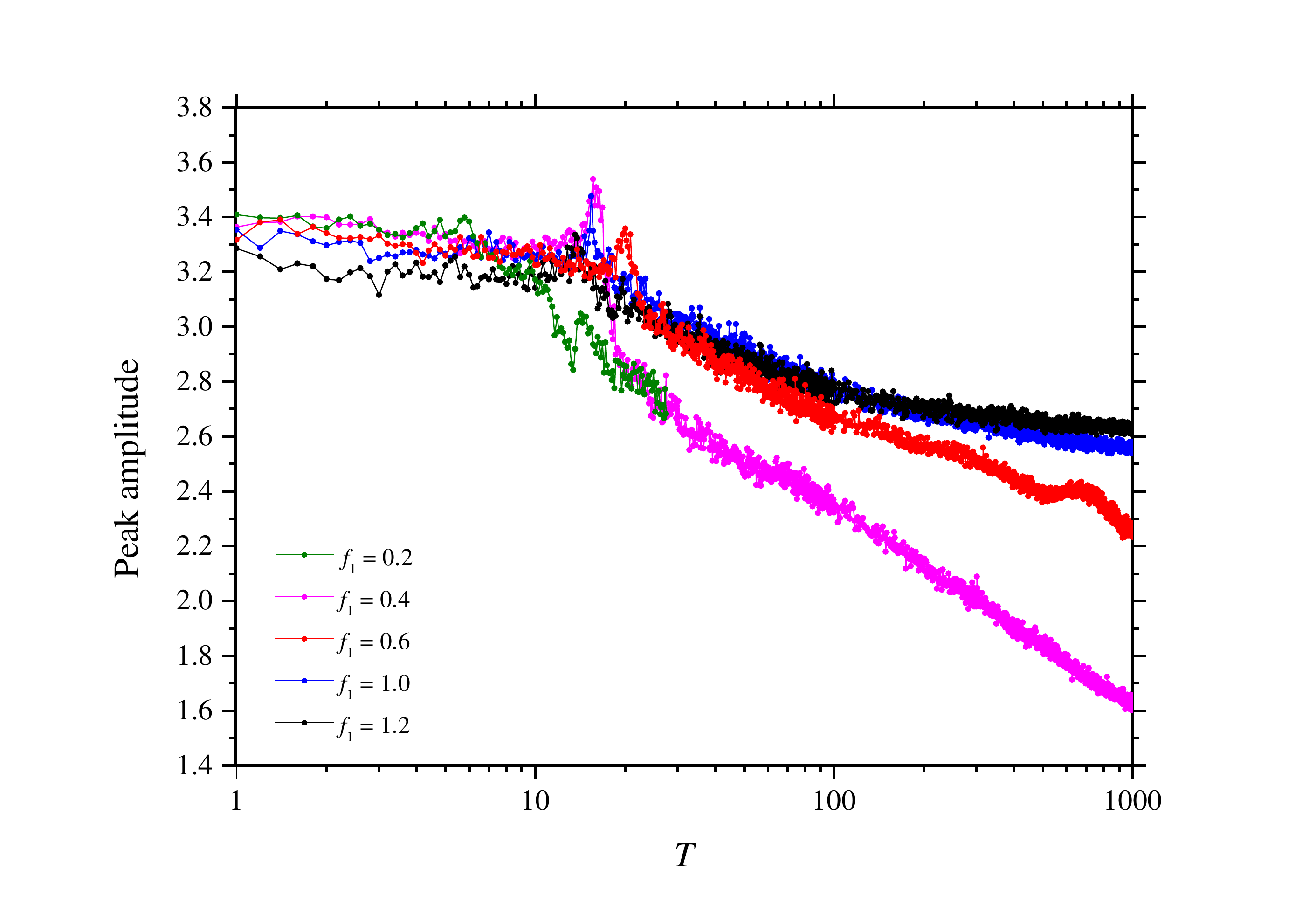}\hspace{-0.5cm}
		
	\caption{\label{fig:peakvalue_2soliton}~The long term behavior of collision of the two vortices 
	in $f_0=0.0$ with $f_1=0.2, 0.4, 0.6, 1.0, 1.2$. 
	The solution with $f_1=1.4$ rather oscillates and slightly is 
	attenuated (similar behavior is observed in the single vortex,  see Fig.\ref{fig:peakvalue}).  
	The peaks are observed at $T\sim 15$ where the two vortices collide and merge. 
	For solutions prior to the collisions, we plot the peak values of the larger vortex. 
	After the collision, the solutions behave as single large vortices without large dissipation.}
       \end{center}
	\end{figure}


	\begin{figure*}[t]
	\begin{center}

	\includegraphics[width=130mm]{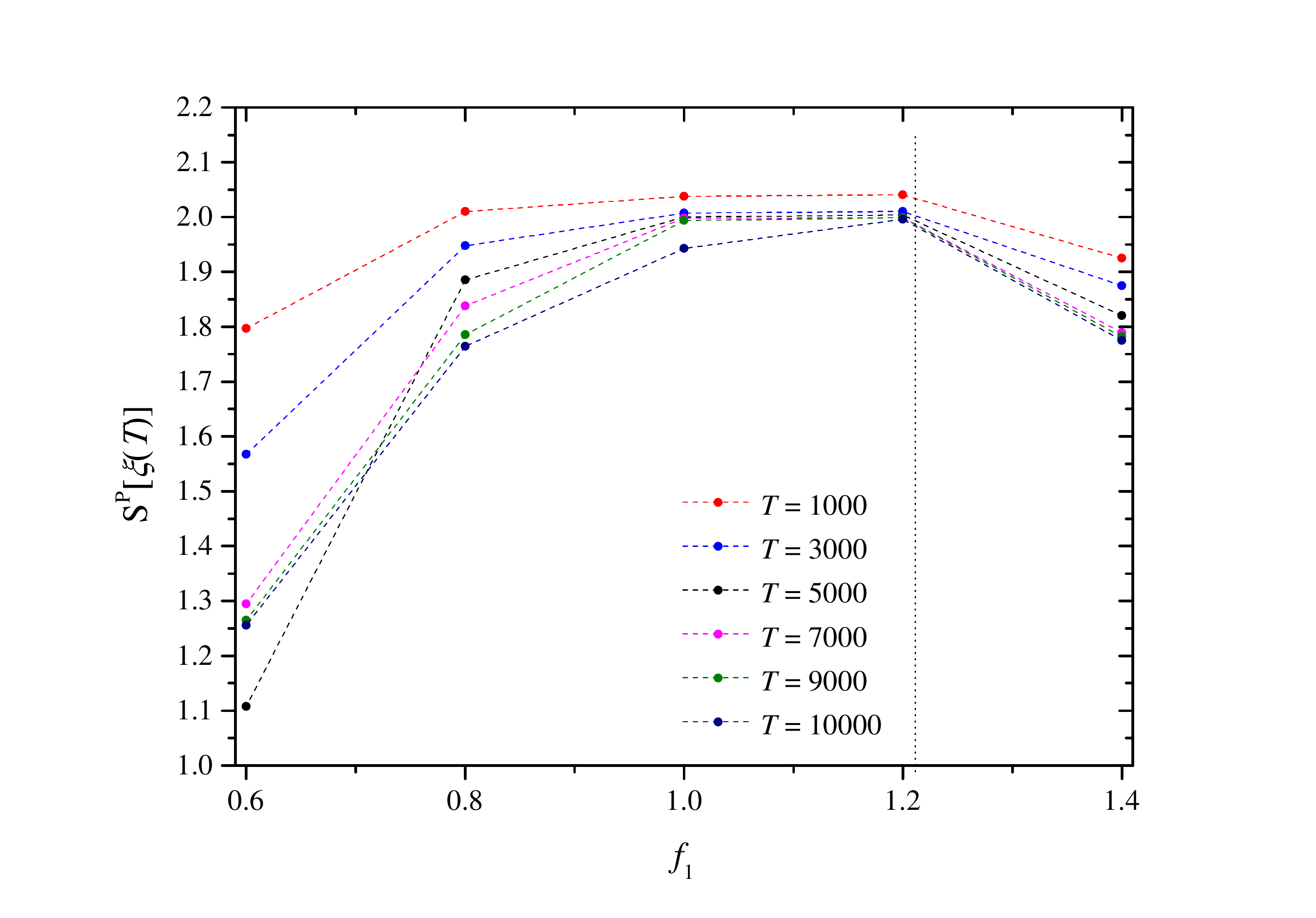}
	\caption{\label{fig:CEntropy}~The configurational entropy for several WYF 
	vortices $\xi(x,y,T)$ with the various shear flows $f_1$. The initial condition 
	is the ZK profile of $c=1$. 
	We plot of the time step $T=1000,3000,5000,7000,9000$ and 10000.  }
       \end{center}
	\end{figure*}


	\begin{figure*}[h]
	\begin{center}

	\includegraphics[width=85mm]{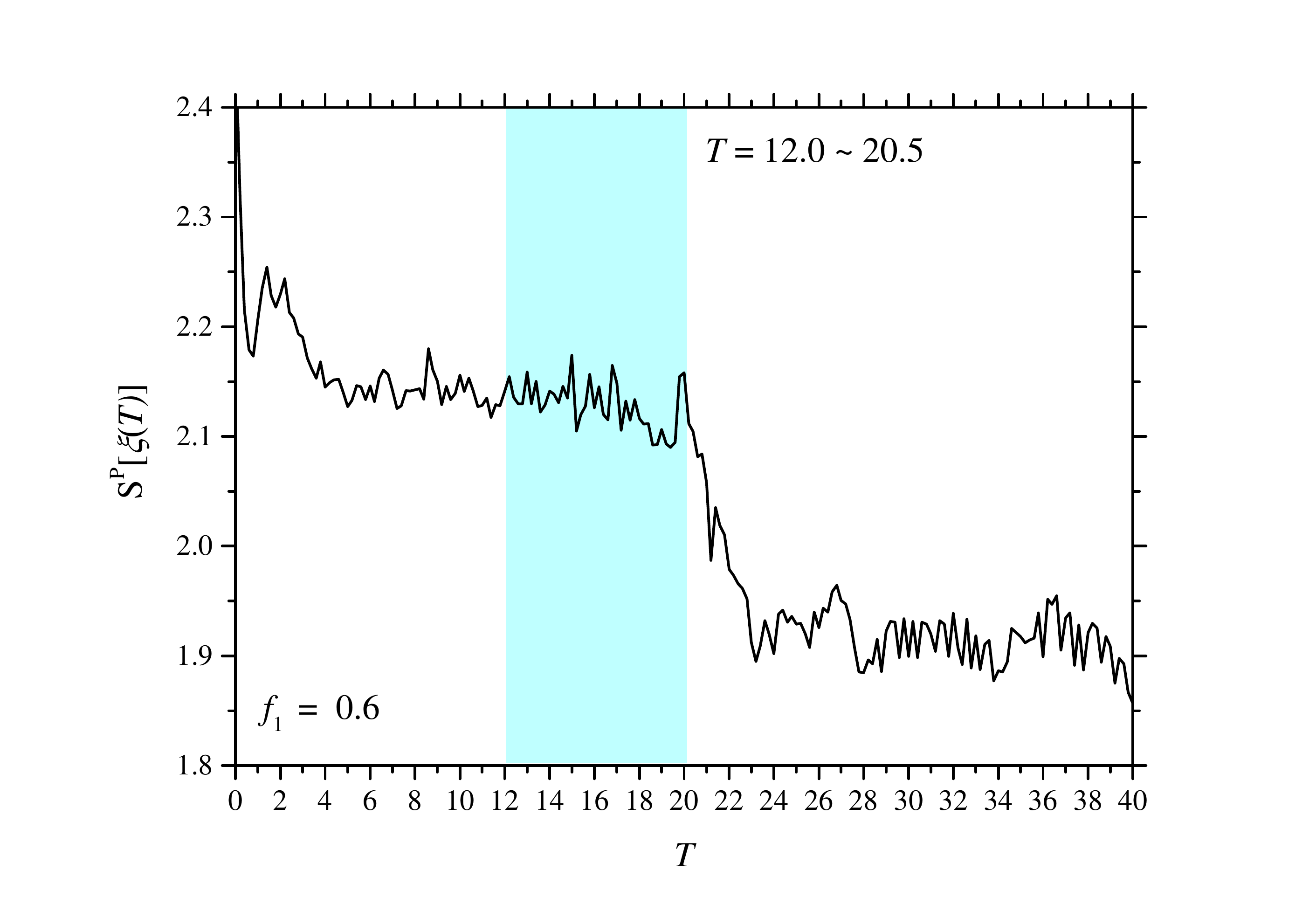}\hspace{-1.5cm}
	\includegraphics[width=85mm]{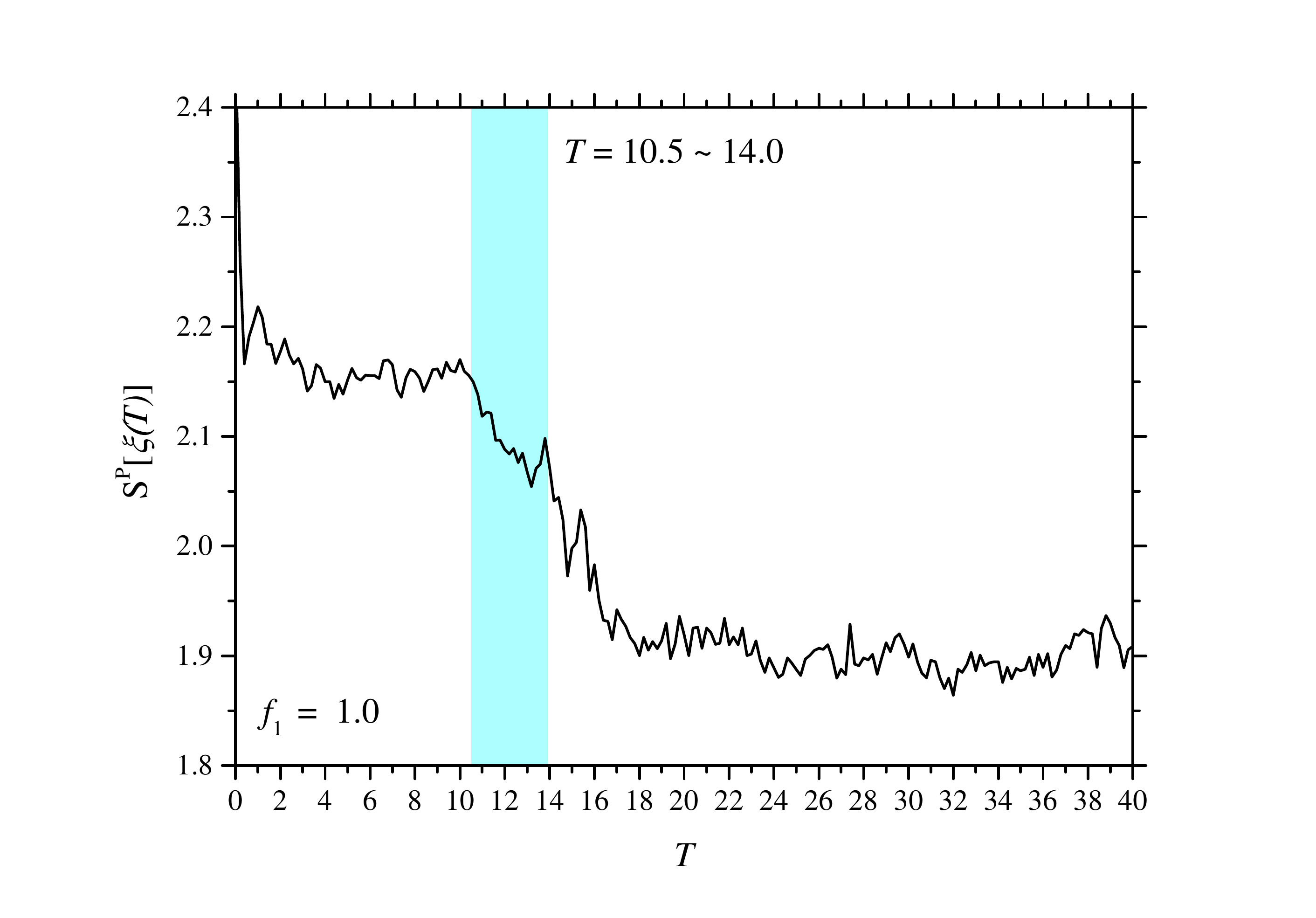}
	\caption{\label{fig:CEntropy2}~The configurational entropy for the two collision vortices $\xi(x,y,T)$ in
	$f_0=0.0$ with the shear flows $f_1=0.6, 1.0$. The $f_1=1.0$ corresponds to 
	Fig.\ref{fig:WYFprofile3d_2soliton}. The blue shaded areas indicate the time period where the two vortices collide.   }
       \end{center}
	\end{figure*}


\subsection{The strength of the background flows: uniform and shear}

On the stability of the vortices in the WYF equation, we find the intensity of the shear flow apparently plays a dominant role. 
Here, we examine the time evolution of the solution with different values of strength of the shear flow. 
In Fig.\ref{fig:evolution0}, we plot the time evolution of the solutions in $x$-direction at fixed $y$ value
for the shear flows $f_1$ of the strength $0.2, 0.6$ and $1.2$. 
The behavior of the solutions is always moderate and the vortices basically move positive $x$-direction in the effect of the shear flows. 
As for the speed of the vortex migration, it is higher in the strong shear flow. 
Then in the case of weak shear flow, $f_1=0.2$, the vortex becomes unstable and the life is short.
The vortex tends to be swept in negative $x$ direction then genuinely the vortex moves to that direction. 
In the case $f_1=0.6$, the stability improves but still it is inclined to dissipate. 
We find the vortex is more likely to stay stable in the case $f_1=1.2$. 
Interestingly, a short period oscillation of the peak amplitude emerges, which we shall mention later.

Shown in Fig.\ref{fig:evolution-1} is the cases including the uniform flow $f_0=-1.0$.
The equation is more like the ZK equation, and consequently the solutions may look stiffer. 
In the weak shear flows $f_1=0.2$, the behavior of the solution tends to be affected by a slight modulation of the backgrounds. 
The solution emits a small fraction at the beginning and then moves to the left. 
The medium case $f_1=0.6$, the vortex is rather stable and seems to be balanced in position. 
For the stronger $f_1$, the vortices are more stable and they move to the positive $x$ direction.
Vortex translation velocity is greater with stronger shear flow $f_1$ as in the cases without uniform flow.

In Fig.\ref{fig:peakvalue}, we present time evolution of the peak amplitude for the several values of the shear flow $f_1$ without the uniform flow.  
As $f_1$ increases, the ``decay rate" of the peaks gradually decreases and the solution reaches the maximum stability at $f_1\sim 1.2$.
Above this value, the decay rate grows again as the case $f_1=1.4$ shows.
The instability of the solutions at the larger $f_1$, where the decay of peak amplitude of the solution corresponds here, 
is indicated in Fig.\ref{fig:peakvaluef16}, in which we find the oscillation of the peak amplitude 
begins and the decay rate increases when $f_1\geq 1.4$.
There is nothing special at $f_1=1.2$ shown in Fig.\ref{fig:peakvaluef16} (upper left). 
For larger $f_1$, as in Figs.\ref{fig:peakvaluef16}, we observe 
the oscillating behaviors of the peak amplitude.
The solution of $f_1=1.8$ is unstable and the peak quickly moves to one of the $y$ boundaries, 
for this reason the lower right panel in Fig.\ref{fig:peakvaluef16} plots the peak at $y\sim 1$.  
These results suggest that the instability of the vortices is caused by the breakdown in the 
balance of the influx from the background flows and the dissipation of the energy.

We have thus confirmed that a constant shear flow with suitable value provides the longevity of vortices, and a uniform flow disturbs it.

\subsection{The vector nonlinearity and vortex fusion}

The stability of the vortices in the present model is due to several distinct effects: the strength of the background shear flows and the particular type of nonlinearities. 
In the equation \eqref{eq:WYF}, there are two different nonlinear terms: 
the KdV like term and the Jacobian term, are referred to as the scalar nonlinear term and the vector nonlinear term, respectively.
Both the nonlinear terms affect with the same order of intensity, \ie 
controlled by the $O(1)$ parameters $E$ and $S$.

Naive consideration from soliton theory indicates that the single vortices would collapse within a short time period when the nonlinear terms are absent due to the dispersive nature of the model, and the nonlinearity would stabilize them. 
For the models in two-dimensional space, however, the situation becomes considerably complicated at the collisions of multi-solitons because of their non-integrability. 
For example, the two solitons of the ZK equation \eqref{eq:ZK} collide inelastically such that the taller soliton gets more amplitude and the smaller one becomes smaller with ripples generated \cite{IWASAKI1990293}.


Although the instability at the soliton collisions is expected in those two-dimensional models, we now see another kind of stability in the WYF model
caused by vector nonlinearity.
In \cite{Williams84}, the authors extensively studied the two vortex collision process and observed that after the collision the two vortices are merged with each other. 
Here we reproduce the two vortex collision with the simple superposition of the two ZK initial profiles $\Phi_c(\tilde{r})$ of $c=1.5$ and $c=1.0$ 
\begin{align}
&\eta_{\rm init}(x,y,T=0)=\Phi_{c=1.5}(r_1)+\Phi_{c=1}(r_2)\,,
\nonumber \\
&\hspace{2cm}r_{1}:=\sqrt{(x+5)^2+(y-1)^2}\,,~~r_{2}:=\sqrt{(x-5)^2+(y-1)^2}\,,
\end{align}
and show that the merged vortex is sufficiently stable.
Fig.\ref{fig:WYFprofile_2soliton} and Fig. \ref{fig:WYFprofile3d_2soliton} show the behavior in the collision of the profiles in contour plot and 3-dimensional plot, respectively. 
We find the two individual vortices on the same $y$-location with different amplitude and $x$-velocity are merged into one single vortex.
The stability or longevity of the vortex after the fusion is depicted in
Fig. \ref{fig:peakvalue_2soliton} with several values of shear flow. 
The result shows the long-term stability of the single vortex after the fusion in the case of shear value $f_1=1.2$, which is consistent with the single vortex cases in the previous subsection. 
In fact, such a merging process has repeatedly been observed in the atmosphere of Jupiter.


\subsection{Fourier analysis: the configurational entropy}

Having seen the fusion phenomena of vortices in the WYF model, it is necessary to find a measure of the soliton excitation for further analysis on these non-linear systems.   
So far, we have mostly examined the peak amplitude of the vortices as the characteristic quantity of the soliton-like excitations.
In the rest of this section, we will try to define the qualitative indicator for the transmutation of the solutions.
To obtain further information on the dissipative behavior of the solutions, analysis using the Fourier transform may be efficient particularly. 
For this purpose, as an indicator we apply the so called configurational entropy (CE) \cite{Gleiser:2011di}, 
which is essentially a measure of spatial complexity of solitonic objects, \ie ``shape information of a soliton". 
The CE tells us many aspects of a soliton solution of a given physical system of our concern.  
It describes intrinsic structural change or it detects a bifurcation point of stable/unstable nature,
and also it is expected to bring us an knowledge of 
longevity of the object~\cite{Gleiser:2018kbq,Gleiser:2019rvw,Gleiser:2020zaj}.
The measure has been applied to many physical systems for the solitons in field theoretical context so far, 
and in this subsection, we employ it for the WYF vortices and see how the longevity is attained. 
Although this is an attempt to define an index of the soliton-like excitations and their lifetime at present, 
we hope the CE would be a promising measure for this kind of analysis.

We have studied the longevity of the WYF vortex with transition of the peak amplitude. 
It is worth investigating if property concerning the dissipation becomes more apparent by the use of Fourier analysis. 
Inspired by Shannon's information entropy $S_{\rm Shannon}=-\sum p_i\log p_i$, where 
$\{p_i\},i=1,\cdots,N$ gives a discrete probability distribution, the CE has been introduced in the following manner.
The Shannon's entropy represents an absolute limit on the best possible lossless compression of any communication. 
Gleiser and his collaborators extended the notion into configurations of the field theoretical models as providing their informational content.  
If our solutions are described by the set of square-integrable bounded functions $f(\bm{x})\in L^2(\mathbb{R}^d)$ and in terms of their Fourier transforms $F(\bm{k})$, we define the modal fraction 
\begin{align}
\mathfrak{f}(\bm{k})=\frac{F(\bm{k})}{\int F(\bm{k})d^d\bm{k}}\,.
\label{frac}
\end{align} 
The configurational entropy for the non-periodic solitonic object $S_C^\textrm{NP}$ 
is defined in terms of $f(\bm{k})$
\begin{align}
S^\textrm{NP}[f]=-\int \tilde{f}(\bm{k})\log [\tilde{f}(\bm{k})]d^dk\,,
\label{CENP}
\end{align}
where $\tilde{f}(\bm{k})=\mathfrak{f}(\bm{k})/\mathfrak{f}(\bm{k})_\textrm{max}$ and $\mathfrak{f}(\bm{k})_\textrm{max}$ is the maximum fraction. 
For the periodic functions where a Fourier series is defined, $\mathfrak{f}(\bm{k})$ turns out to be $f_n=|A_n|^2/\sum |A_n|^2$, where $A_n$ is the coefficient of the $n$-th Fourier mode.
In this case, the configurational entropy $S^\textrm{P}_C$ is defined as
\begin{align}
S^\textrm{P}_C[f]=-\sum_nf_n\log [f_n]\,.
\label{CEP}
\end{align}

We expect the configurational entropy to behave that: 
(i)~It promotes informational content of given configurations. 
(ii)~It detects bifurcation of stable/unstable branch, where the entropy takes the maximal value.  
(iii)~It reflects the structural change of the configuration. At the boundary of the domain, the entropy exhibits the extremum, often the minimum. 
(iv) It foresees the longevity of a configuration. 
In \cite{Gleiser:2011di}, the authors gave the case of a Gaussian in $d$ dimension $f(r)=N\exp (-\alpha r^2)$
as an example. Eq. (\ref{CENP}) gives 
\begin{align}
S^\textrm{NP}(\alpha)=\frac{d}{2}(2\pi\alpha)^{d/2}\,.
\end{align} 
For a very expanded Gaussian, $\alpha\to 0,~~S^\textrm{NP}\to 0$, while for a sharp peaked one $\alpha \to \infty,~~
S^\textrm{NP}\to\infty$. $S^\textrm{NP}$ estimates the information content required in $k$-space to build the function $f(r)$. 
It is a useful estimate because it gives a good insight into the CE concerning the dissipative property of a soliton. 
During a time evolution of a soliton, if it dissipates with releasing some ripples the CE should decreases, while if the soliton is stable against decay the CE tends to be constant.
From this point of view, we shall examine the CE of our vortex solutions.

In general, the evaluation is performed for an energy density or a charge density of a given field theoretical model 
in the analysis of CE, from which we extract the informational contents via the energy or the topological charge structure of the model.
Since our concern is how much the longevity of the soliton-like objects is achieved by both intrinsic and external effects,  
we evaluate the CE for the function $\xi(x,y,T)$ defined in \eqref{stream function with background} instead of the subject in the original analysis.
To analyze the CE, we should evaluate the modal fraction (\ref{frac}) and then the non-periodic function (\ref{CENP}), 
taking account of the boundary condition \eqref{freeslip} being periodic and free-slip on $x$ and $y$ boundaries, respectively.
However, it is not an easy task to implement them because of the property of this particular boundary conditions.  
At a glance at our numerical results, the dissipation in $y$ is relatively small and behavior of $x$ 
direction is almost responsible for the long-term stability so that   
we evaluate the CE for the $x$ section of the solutions in this paper.  
The Fourier series is directly obtained in the periodic boundary condition in $x$~(\ref{freeslip}), and the CE is computed using (\ref{CEP}).
Figure \ref{fig:CEntropy} is the CE of the WYF vortices with the ZK initial condition for varying shear flow $f_1$ for several time steps. 
The CE always takes the maximal value at $f_1=1.2$. 
As shown in \cite{Gleiser:2013mga,Gleiser:2015rwa}, in 
the cases of $Q$-balls/boson stars, the maximum of the CE often detects a bifurcation point of the stable/unstable branch. 
The present result seems to be consistent with the previous criterion,  
that is, $f_1= 1.2$ is the bifurcation point between the stable/unstable solutions.  
With increasing time, all the CE decay but the change is smallest at the case  $f_1=1.2$.
As a result, the solution of $f_1=1.2$ carries the largest informational content, which means the less dissipative, at the full-time scale.

We also examine the CE during the process of the two vortex collision in Fig.\ref{fig:CEntropy2} corresponding to Fig.\ref{fig:peakvalue_2soliton}.
Immediately after the collision, the CE rapidly decreases, which shows there is a release of the small fractions and slight dissipation occurs.
After that, the CE quickly ceased to decrease and in fact, it stays almost constant. 
It means the vortex remains stable after such a collision and a merging process. 

However, particularly for the tangled two-body process, the Fourier transformation in the entire two-dimension would be required for accomplishing the rigorous considerations. 
Such an analysis will be the next topic of this research.


\section{Conclusion}


We have presented two-dimensional vortex solutions in the modified Williams-Yamagata-Flierl equation~\eqref{eq:WYF+shear}. 
The equation was originally proposed in the context of geophysical fluid dynamics of the atmosphere of Jupiter and the solutions were the possible candidate for the great red spot. 
The model is not integrable in any sense: It has no conserved quantities and 
then the soliton-like stability of the solutions is not the consequence of such mathematical origin. 
We have found therefore the existence of solitonic objects is caused by the cooperation of several origins including external effects, 
\ie   the scalar/vector nonlinearities, and the shear flow. 
The scalar nonlinearity is equipped in Zakharov-Kuznetsov equation, which has relatively 
stable soliton solutions and the stability is realized by the underlying KdV dynamics. 
We have used the semi-analytical solutions of the Zakharov-Kusnetsov equation as an initial condition and it keeps the stability of the solution.
On the other hand, the vector nonlinearity, \ie Jacobian term has the role for the merging of two vortices upon collision,
 as well as the long-term stability of the single vortex of the model.


In this paper, we focused on the longevity of the vortex in terms of these effects.
We carried out extremely long-term simulations and clarified how the longevity of the solutions is attained. 
We compared the distinct initial conditions, the ZK vortex, and the standard Gaussian function.
The ZK is more stable with longevity, which means that the scalar nonlinearity has effects on the stability. 
Another important issue has been the background shear flows.
We have performed the long-term simulation for various shear flows.
The principal result of this paper is that we found the sweet spot of the value of the shear flow for stabilizing the vortex.
It remains a question, however, that what is the mechanism behind the determination of this special value. 
We will clarify it in the subsequent paper.
Another subject considered is the configurational entropy which detects the stable/unstable bifurcation point of a given solution. 
In terms of CE, we found that the sweet spot corresponds to the bifurcation point. 
It is consistent with the numerical observation. For larger $f_1$, apparently, the solutions become unstable. 

There would be many variants to nonlinear models in higher spatial dimensions which have quasi-stable localized objects.
They would have diverse types of nonlinear and external effects such as shear flows and other contents that make the solutions stable. 
We expect the study for such mock integrable systems will develop a fertile perspective of nonlinear dynamical systems.
We will inform the results in the subsequent papers in due course.


\section*{Acknowledgment}

The authors would like to thank Satoshi Horihata, Hiroshi Kakuhata, Ryu Sasaki, Yakov Shnir, Yves Brihaye and Pawe\l~Klimas 
for many useful advice and comments.  
N.S. deeply thanks Rafael Augusto Couceiro Correa for drawing our attention to the configurational entropy.  
A.N. and N.S. would like to thank Luiz Agostinho Ferreira for the kind hospitality 
at Instituto de F\'isica de S\~ao Carlos, Universidade de S\~ao Paulo.
Discussions during the YITP workshop YITP-W-20-03 on ``Strings and Fields 2020'' and 
YITP-W-21-04 on ``Strings and Fields 2021'' have been useful to complete this work. 
A.N., N.S. and K.T. were supported in part by JSPS KAKENHI Grant Number JP20K03278. 

\appendix

\section{\label{bplane}~The $\beta$-plane model and Williams-Yamagata-Flierl equation in the intermediate geostrophic regime}

\renewcommand{\theequation}{A.\arabic{equation} }
\setcounter{equation}{0}


We derive the Williams-Yamagata-Flierl equation \eqref{eq:WYF} for the stream function $\eta$ from the following  non-dimensional 
``shallow water $\beta$-plane model",
\begin{subequations}\label{beta-plane}
\begin{align}
&\hat{\varepsilon}\frac{D}{Dt}u-(1+\hat{\beta}y)v=-\frac{\partial \eta}{\partial x}\,,\label{beta-plane u}\\
&\hat{\varepsilon}\frac{D}{Dt}v+(1+\hat{\beta}y)u=-\frac{\partial \eta}{\partial y}\,,\label{beta-plane v}\\
&\frac{\hat{\varepsilon}}{\hat{s}}\frac{D}{Dt}\eta+\frac{\partial u}{\partial x}+\frac{\partial v}{\partial y}
+\frac{\hat{\varepsilon}}{\hat{s}}\eta\left(\frac{\partial u}{\partial x}+\frac{\partial v}{\partial y}\right)=0\,,
\label{beta-plane eta}
\end{align}
\end{subequations}
where the total-derivative, or Lagrange-derivative, is defined as
\begin{align}
\frac{D}{Dt}:=\frac{\partial }{\partial t}+u\frac{\partial }{\partial x}+v\frac{\partial }{\partial y}\,.\label{total derivative}
\end{align}
In this model, the dimensionless parameters $\hat{\beta},\; \hat{\varepsilon}$ and $\hat{s}$ are the so-called sphericity, Rossby and stratification parameters, respectively.
We apply the reductive perturbation method for the system \eqref{beta-plane} in terms of the sphericity parameter $\hat{\beta}\ll 1$.
For this purpose, we expand the dynamical variables $u,v$ and $\eta$  in $\hat{\beta}$,
\begin{subequations}
\begin{align}
u=u^{(0)}+\hat{\beta} u^{(1)}+\hat{\beta}^2 u^{(2)}+\cdots\,,\\
v=v^{(0)}+\hat{\beta} v^{(1)}+\hat{\beta}^2 v^{(2)}+\cdots\,,\\
\eta=\eta^{(0)}+\hat{\beta} \eta^{(1)}+\hat{\beta}^2 \eta^{(2)}+\cdots\,,
\end{align}
\end{subequations}
and substitute into \eqref{beta-plane}.
We now assume the order of the parameters as $\hat{\varepsilon}\sim \hat{\beta}^2$ and $\hat{s}\sim \hat{\beta}$.
This parameter range is known to characterize the dynamical length scale of phenomena between the smallest scale with respect to the radius of a planet, \ie the quasi-geostrophic (QG) scale, and the largest, \ie  the planetary-geostrophic (PG) scale, so that we refer to the scale as the intermediate-geostrophic (IG) scale.
Hereafter, we introduce $O(1)$ parameters $E$ and $S$ as $\hat{\varepsilon}=E\hat{\beta}^2$ and $\hat{s}=S\hat{\beta}$.

First of all,  we find the $O(1)$ equations in $\hat{\beta}$ of \eqref{beta-plane} give the relation
\begin{align}
\left\{\begin{array}{l}
\displaystyle{-v^{(0)}=-\frac{\partial \eta^{(0)}}{\partial x}}\\
\\
\displaystyle{u^{(0)}=-\frac{\partial \eta^{(0)}}{\partial y}}
\end{array}\right.\ \Rightarrow \ \frac{\partial u^{(0)}}{\partial x\ }+\frac{\partial v^{(0)}}{\partial y\ }=0,\label{IG 0}
\end{align}
known as the geostrophic balance. 
Next, the equations in $O(\hat{\beta})$ read
\begin{subequations}
\begin{align}
&-v^{(1)}-yv^{(0)}=-\frac{\partial \eta^{(1)}}{\partial x}\,,\label{IG 1-a}\\
&-u^{(1)}+yu^{(0)}=-\frac{\partial \eta^{(1)}}{\partial y}\,,\label{IG 1-b}\\
&\frac{E}{S}\frac{D^{(0)}}{Dt}\eta^{(0)}+\frac{\partial u^{(1)}}{\partial x\ }+\frac{\partial v^{(1)}}{\partial y\ }
+\frac{E}{S}\eta^{(0)}\left(  \frac{\partial u^{(0)}}{\partial x\ }+\frac{\partial v^{(0)}}{\partial y\ }\right)=0\,,\label{IG 1-c}
\end{align}
\end{subequations}
where
\begin{align}
\frac{D^{(0)}}{Dt}:=\frac{\partial}{\partial t}+u^{(0)}\frac{\partial }{\partial x}+v^{(0)}\frac{\partial }{\partial y}\,.
\end{align}
From  (\ref{IG 1-a}) and (\ref{IG 1-b}), we find
\begin{align}
&\frac{\partial v^{(1)}}{\partial y\ }+y\frac{\partial v^{(0)}}{\partial y\ }+v^{(0)}=\frac{\partial^2 \eta^{(1)}}{\partial x\partial y \ }\,,\\
&\frac{\partial u^{(1)}}{\partial x\ }+y\frac{\partial u^{(0)}}{\partial x\ }\qquad=-\frac{\partial^2 \eta^{(1)}}{\partial x\partial y \ }\,,
\end{align}
which lead to
\begin{align}
\frac{\partial u^{(1)}}{\partial x\ }+\frac{\partial v^{(1)}}{\partial y\ }=-v^{(0)}=-\frac{\partial \eta^{(0)}}{\partial x}.\label{IG 1-d}
\end{align}
Substituting \eqref{IG 1-d} into  (\ref{IG 1-c}), a linear wave equation for $\eta^{(0)}$
\begin{align}
\left(\frac{\partial}{\partial t}-\frac{S}{E}\frac{\partial}{\partial x}\right)\eta^{(0)}=0\,,\label{IG 1 final}
\end{align}
is obtained.
We therefore find that the dynamics at this order is given by a ``left-moving" stationary wave packet of velocity $S/E$, 
\begin{align}
\eta^{(0)}=f(t+(E/S)x)\,,\label{left-moving stationary wave}
\end{align}
where $f$ is an arbitrary function characterizing the shape of the wave packet.
We now assume that the dynamics can be decomposed into the stationary carrier wave of the form \eqref{left-moving stationary wave} and the other slowly moving motion. 
For this purpose, we introduce another time $T$ which describes the slow movement  by substituting the time derivative as
\begin{align}
\frac{\partial}{\partial t}\ \longrightarrow\ \frac{\partial}{\partial t}+\hat{\beta}\frac{S}{E}\frac{\partial}{\partial T}\,,
\end{align}
and suppose that the zeroth order stream function $\eta^{(0)}$ is a function of $\xi:=x+(S/E)t,~y$ and $ T$.

With these prescription, we find the equation in $O(\hat{\beta}^2)$ are
\begin{align}
&E\frac{D^{(0)}}{Dt}u^{(0)}-v^{(2)}-yv^{(1)}=-\frac{\partial \eta^{(2)}}{\partial x}\,,\label{IG 2-a}\\
&E\frac{D^{(0)}}{Dt}v^{(0)}+u^{(2)}+yu^{(1)}=-\frac{\partial \eta^{(2)}}{\partial y}\,,\label{IG 2-b}\\
&\frac{E}{S}\frac{D^{(0)}}{Dt}\eta^{(1)}+\frac{\partial}{\partial T}\eta^{(0)}
+\frac{\partial u^{(2)}}{\partial x}+\frac{\partial v^{(2)}}{\partial y}\nonumber\\
&+\frac{E}{S}\left\{\eta^{(0)}\left(\frac{\partial u^{(1)}}{\partial x}
+\frac{\partial v^{(1)}}{\partial y}\right)+\eta^{(1)}\left(\frac{\partial u^{(0)}}{\partial x}+\frac{\partial v^{(0)}}{\partial y}\right)
\right\}=0\,.
\label{IG 2-c}
\end{align}
Differentiating \eqref{IG 2-a} and  \eqref{IG 2-b} with respect to $y$ and $x$, respectively, and eliminating $\eta^{(2)}$, reads
\begin{align}
\frac{\partial u^{(2)}}{\partial x}+\frac{\partial v^{(2)}}{\partial y}=-E\left(
\frac{\partial }{\partial t}\nabla^2 \eta^{(0)}
+\frac{\partial \eta^{(0)}}{\partial x}\frac{\partial }{\partial y}\nabla^2 \eta^{(0)}
-\frac{\partial \eta^{(0)}}{\partial y}\frac{\partial }{\partial x}\nabla^2 \eta^{(0)}
\right)\nonumber\\
+y\frac{\partial \eta^{(0)}}{\partial x}-v^{(1)}\nonumber\\
=-E\left(\nabla^2\frac{\partial \eta^{(0)}}{\partial x}+J(\eta^{(0)},\nabla^2\eta^{(0)})\right)+2y\frac{\partial 
\eta^{(0)}}{\partial x}-\frac{\partial \eta^{(1)}}{\partial x}\,,\label{IG 2-ab}
\end{align}
where the zeroth and the first order equations (\ref{IG 0}) and \eqref{IG 1-a} are used, and the Laplacian is defined as
\begin{align}
\nabla^2:=\frac{\partial^2}{\partial x^2}+\frac{\partial^2}{\partial y^2}\,.
\end{align}
In the final expression, we have defined the Jacobian as
\begin{align}
J(a,b)&:=\frac{\partial a}{\partial x}\frac{\partial b}{\partial y}-\frac{\partial b}{\partial x}\frac{\partial a}{\partial y}\,.
\end{align}
Substituting \eqref{IG 2-ab}  into (\ref{IG 2-c}), we finally find the wave equation for $\eta^{(1)}$
\begin{align}
\frac{E}{S}\left(\frac{\partial}{\partial t}-\frac{S}{E}\frac{\partial}{\partial x}\right)\eta^{(1)}=
-\frac{\partial \eta^{(0)}}{\partial T}+S\nabla^2\frac{\partial \eta^{(0)}}{\partial x}+
EJ(\eta^{(0)},\nabla^2\eta^{(0)})\nonumber\\
-2y\frac{\partial \eta^{(0)}}{\partial x}+\frac{E}{S}\eta^{(0)}\frac{\partial \eta^{(0)}}{\partial x}\,.\label{IG 2 final}
\end{align}
This is a wave equation for the $O(\hat{\beta})$ correction to the stream function $\eta^{(1)}$ with right-hand-side as a ``forced oscillation" given by $\eta^{(0)}$.
From \eqref{IG 1 final}, we find that the dispersion relation of $\eta^{(1)}$ is exactly the same as that of $\eta^{(0)}$, if we consider the homogeneous equation.
Thus, there would be occurring a resonance for $\eta^{(1)}$ at the same frequency of the $O(1)$ wave $\eta^{(0)}$ concerning for the original time $t$, which strongly be a conflict with the assumption of the perturbation expansion.
In order to avoid this, the right-hand-side, namely the secular term, have to vanish.
We therefore find that the IG equation for $\eta^{(0)}$ with respect to the ``slow time" $T$ is
\begin{align}
-\frac{\partial \eta^{(0)}}{\partial T}+S\nabla^2\frac{\partial \eta^{(0)}}{\partial x}+
EJ(\eta^{(0)},\nabla^2\eta^{(0)})+\frac{E}{S}\eta^{(0)}\frac{\partial \eta^{(0)}}{\partial x}
-2y\frac{\partial \eta^{(0)}}{\partial x}=0\,,\label{IG eqn}
\end{align}
which is exactly the Williams-Yamagata-Flierl equation \eqref{eq:WYF}.

\section{\label{4thdifference}~The fourth order finite difference scheme}

\renewcommand{\theequation}{B.\arabic{equation} }
\setcounter{equation}{0}


The explicit form of the fourth order differentials for (\ref{eq:WYF+shear}) are defined as follows:
for the field $ \xi_{i,j}\equiv \xi(x,y)$ with the grid spacing $h\equiv \Delta x,~\ell\equiv \Delta y$, 
\begin{align}
&\biggl(\frac{\partial \xi}{\partial x}\biggr)_{ij}=\frac{1}{12h}\Bigl(-\xi_{i+2j}+8\xi_{i+1j}-8\xi_{i-1j}+\xi_{i-2j}\Bigr)+O(h^5)\,,
\nonumber \\
&\biggl(\frac{\partial^2 \xi}{\partial x^2}\biggr)_{ij}=\frac{1}{12h^2}\Bigl(-\xi_{i+2j}+16\xi_{i+1j}-30\xi_{ij}+16\xi_{i-1j}-\xi_{i-2j}\Bigr)+O(h^5)\,,
\\
&\biggl(\frac{\partial^3 \xi}{\partial x^3}\biggr)_{ij}=\frac{1}{2h^3}\Bigl(\xi_{i+2j}-2\xi_{i+1j}+2\xi_{i-1j}-\xi_{i-2j}\Bigr)+O(h^5)\,.
\nonumber 
\end{align}
From these, we are able to write down the mixed integral
\begin{align}
\biggl(\frac{\partial^3\xi}{\partial x\partial^2y}\biggr)_{ij}
=\frac{1}{144h\ell^2}
\Bigl[
&-\Bigl(-\xi_{i+2j+2}+16\xi_{i+2j+1}-30\xi_{i+2j}+16\xi_{i+2j-1}-\xi_{i+2j-2}\Bigr)
\nonumber \\
&+8\Bigl(-\xi_{i+1j+2}+16\xi_{i+1j+1}-30\xi_{i+1j}+16\xi_{i+1j-1}-\xi_{i+1j-2}\Bigr)
\nonumber \\
&-8\Bigl(-\xi_{i-1j+2}+16\xi_{i-1j+1}-30\xi_{i-1j}+16\xi_{i-1j-1}-\xi_{i-1j-2}\Bigr)
\nonumber \\
&+\Bigl(-\xi_{i-2j+2}+16\xi_{i-22j+1}-30\xi_{i-2j}+16\xi_{i-2j-1}-\xi_{i-2j-2}\Bigr)
\Bigr]\,.
\end{align}

For the advection term, we employ the Arakawa form which is given in the following finite difference form
~\cite{ARAKAWA1966119}
\begin{align}
&J^{DD}_{ij}(\zeta,\xi):=\frac{1}{4h\ell}\Bigl[(\zeta_{i+1j}-\zeta_{i-1j})(\xi_{ij+1}-\xi_{ij-1})-
(\zeta_{ij+1}-\zeta_{ij-1})(\xi_{i+1j}-\xi_{i-1j})\Bigr]\,,
\nonumber \\
&J^{DC}_{ij}(\zeta,\xi):=\frac{1}{4h\ell}\Bigl[\zeta_{i+1j}(\xi_{i+1j+1}-\xi_{i+1j-1})-\zeta_{i-1j}(\xi_{i-1j+1}-\xi_{i-1j-1})
\nonumber \\
&\hspace{4cm}-\zeta_{ij+1}(\xi_{i+1j+1}-\xi_{i-1j+1})+\zeta_{ij-1}(\xi_{i+1j-1}-\xi_{i-1j-1})
\Bigr]\,,
\\
&J^{CD}_{ij}(\zeta,\xi):= -J^{DC}(\xi,\zeta):=\frac{1}{4h\ell}\Bigl[\zeta_{i+1j+1}(\xi_{ij+1}-\xi_{i+1j})-\zeta_{i-1j-1}(\xi_{i-1j}-\xi_{ij-1})
\nonumber \\
&\hspace{4cm}-\zeta_{i-1j+1}(\xi_{ij+1}-\xi_{i-1j})+\zeta_{i+1j-1}(\xi_{i+1j}-\xi_{ij-1})
\Bigr]\,,
\nonumber \\
&\zeta:=\nabla^2\xi\,,
\nonumber \\
&\textrm{and} \nonumber \\
&J(\zeta,\xi)_{ij}:=\frac{1}{3}\Bigl[J^{DD}_{ij}(\zeta,\xi)+J^{DC}_{ij}(\zeta,\xi)+J^{CD}_{ij}(\zeta,\xi)\Bigr]\,.
\end{align}
Therefore, the fourth order differential's version can be extended via
\begin{align}
&J^{DD}_{ij}(\zeta,\xi):=\frac{1}{144h\ell}\Bigl[(-\zeta_{i+2j}+8\zeta_{i+1j}-8\zeta_{i-1j}+\zeta_{i-2j})
(-\xi_{ij+2}+8\xi_{ij+1}-8\xi_{ij-1}+\xi_{ij-2})
\nonumber \\
&\hspace{2cm}-(-\zeta_{ij+2}+8\zeta_{ij+1}-8\zeta_{ij-1}+\zeta_{ij-2})
(-\xi_{i+2j}+8\xi_{i+1j}-8\xi_{i-1j}+\xi_{i-2j})\Bigr]\,,
\nonumber \\
&J^{DC}_{ij}(\zeta,\xi):=
\frac{1}{144h\ell}\Bigl[-\zeta_{i+2j}(-\xi_{i+2j+2}+8\xi_{i+2j+1}-8\xi_{i+2j-1}+\xi_{i+2j-2})
\nonumber \\
&\hspace{4cm}+8\zeta_{i+1j}(-\xi_{i+1j+2}+8\xi_{i+1j+1}-8\xi_{i+1j-1}+\xi_{i+1j-2})
\nonumber \\
&\hspace{4cm}-8\zeta_{i-1j}(-\xi_{i-1j+2}+8\xi_{i-1j+1}-8\xi_{i-1j-1}+\xi_{i-1j-2})
\nonumber \\
&\hspace{4cm}+\zeta_{i-2j}(-\xi_{i-2j+2}+8\xi_{i-2j+1}-8\xi_{i-2j-1}+\xi_{i-2j-2})
\nonumber \\
&\hspace{4cm}+\zeta_{ij+2}(-\xi_{i+2j+2}+8\xi_{i+1j+2}-8\xi_{i-1j+2}+\xi_{i-2j+2})
\nonumber \\
&\hspace{4cm}-8\zeta_{ij+1}(-\xi_{i+2j+1}+8\xi_{i+1j+1}-8\xi_{i-1j+1}+\xi_{i-2j+1})
\nonumber \\
&\hspace{4cm}+8\zeta_{ij-1}(-\xi_{i+2j-1}+8\xi_{i+1j-1}-8\xi_{i-1j-1}+\xi_{i-2j-1})
\nonumber \\
&\hspace{4cm}-\zeta_{ij-2}(-\xi_{i+2j-2}+8\xi_{i+1j-2}-8\xi_{i-1j-2}+\xi_{i-2j-2})\,,
\\
&J^{CD}_{ij}(\zeta,\xi):= -J^{DC}_{ij}(\xi,\zeta)\,.
\nonumber 
\end{align}  

\bibliography{redspot}

\begin{thebibliography}{81}%
\makeatletter
\providecommand \@ifxundefined [1]{%
 \@ifx{#1\undefined}
}%
\providecommand \@ifnum [1]{%
 \ifnum #1\expandafter \@firstoftwo
 \else \expandafter \@secondoftwo
 \fi
}%
\providecommand \@ifx [1]{%
 \ifx #1\expandafter \@firstoftwo
 \else \expandafter \@secondoftwo
 \fi
}%
\providecommand \natexlab [1]{#1}%
\providecommand \enquote  [1]{``#1''}%
\providecommand \bibnamefont  [1]{#1}%
\providecommand \bibfnamefont [1]{#1}%
\providecommand \citenamefont [1]{#1}%
\providecommand \href@noop [0]{\@secondoftwo}%
\providecommand \href [0]{\begingroup \@sanitize@url \@href}%
\providecommand \@href[1]{\@@startlink{#1}\@@href}%
\providecommand \@@href[1]{\endgroup#1\@@endlink}%
\providecommand \@sanitize@url [0]{\catcode `\\12\catcode `\$12\catcode
  `\&12\catcode `\#12\catcode `\^12\catcode `\_12\catcode `\%12\relax}%
\providecommand \@@startlink[1]{}%
\providecommand \@@endlink[0]{}%
\providecommand \url  [0]{\begingroup\@sanitize@url \@url }%
\providecommand \@url [1]{\endgroup\@href {#1}{\urlprefix }}%
\providecommand \urlprefix  [0]{URL }%
\providecommand \Eprint [0]{\href }%
\providecommand \doibase [0]{http://dx.doi.org/}%
\providecommand \selectlanguage [0]{\@gobble}%
\providecommand \bibinfo  [0]{\@secondoftwo}%
\providecommand \bibfield  [0]{\@secondoftwo}%
\providecommand \translation [1]{[#1]}%
\providecommand \BibitemOpen [0]{}%
\providecommand \bibitemStop [0]{}%
\providecommand \bibitemNoStop [0]{.\EOS\space}%
\providecommand \EOS [0]{\spacefactor3000\relax}%
\providecommand \BibitemShut  [1]{\csname bibitem#1\endcsname}%
\let\auto@bib@innerbib\@empty
\bibitem [{\citenamefont {Ablowitz}\ and\ \citenamefont
  {Segur}(1981)}]{Ablowitz-Segur}%
  \BibitemOpen
  \bibfield  {author} {\bibinfo {author} {\bibfnamefont {Mark~J.}\ \bibnamefont
  {Ablowitz}}\ and\ \bibinfo {author} {\bibfnamefont {Harvey}\ \bibnamefont
  {Segur}},\ }\href {\doibase 10.1137/1.9781611970883} {\emph {\bibinfo {title}
  {Solitons and the Inverse Scattering Transform}}}\ (\bibinfo  {publisher}
  {Society for Industrial and Applied Mathematics},\ \bibinfo {year} {1981})\
  \Eprint
  {http://arxiv.org/abs/https://epubs.siam.org/doi/pdf/10.1137/1.9781611970883}
  {https://epubs.siam.org/doi/pdf/10.1137/1.9781611970883} \BibitemShut
  {NoStop}%
\bibitem [{\citenamefont {Ablowitz}\ and\ \citenamefont
  {Clarkson}(1991)}]{ablowitz_clarkson_1991}%
  \BibitemOpen
  \bibfield  {author} {\bibinfo {author} {\bibfnamefont {M.~A.}\ \bibnamefont
  {Ablowitz}}\ and\ \bibinfo {author} {\bibfnamefont {P.~A.}\ \bibnamefont
  {Clarkson}},\ }\href {\doibase 10.1017/CBO9780511623998} {\emph {\bibinfo
  {title} {Solitons, Nonlinear Evolution Equations and Inverse Scattering}}},\
  London Mathematical Society Lecture Note Series\ (\bibinfo  {publisher}
  {Cambridge University Press},\ \bibinfo {year} {1991})\BibitemShut {NoStop}%
\bibitem [{\citenamefont {T.Miwa}(2000)}]{Miwa-Jimbo}%
  \BibitemOpen
  \bibfield  {author} {\bibinfo {author} {\bibfnamefont {E.Date}\ \bibnamefont
  {T.Miwa}, \bibfnamefont {M.Jimbo}},\ }\href
  {https://ci.nii.ac.jp/ncid/BA44542015} {\emph {\bibinfo {title} {Solitons:
  differential equations, symmetries and infinite dimensional algebras}}},\
  \bibinfo {series} {Cambridge tracts in mathematics}\ No.\ \bibinfo {number}
  {135}\ (\bibinfo  {publisher} {Cambridge University Press},\ \bibinfo {year}
  {2000})\BibitemShut {NoStop}%
\bibitem [{\citenamefont {Hirota}(2004)}]{hirota_2004}%
  \BibitemOpen
  \bibfield  {author} {\bibinfo {author} {\bibfnamefont {Ryogo}\ \bibnamefont
  {Hirota}},\ }\href {\doibase 10.1017/CBO9780511543043} {\emph {\bibinfo
  {title} {The Direct Method in Soliton Theory}}},\ edited by\ \bibinfo
  {editor} {\bibfnamefont {Atsushi}\ \bibnamefont {Nagai}}, \bibinfo {editor}
  {\bibfnamefont {Jon}\ \bibnamefont {Nimmo}}, \ and\ \bibinfo {editor}
  {\bibfnamefont {Claire}\ \bibnamefont {Gilson}},\ Cambridge Tracts in
  Mathematics\ (\bibinfo  {publisher} {Cambridge University Press},\ \bibinfo
  {year} {2004})\BibitemShut {NoStop}%
\bibitem [{\citenamefont {B.B.Kadomtsev}\ and\ \citenamefont
  {V.I.Petviashivilli}(1970)}]{Kadomtsev70}%
  \BibitemOpen
  \bibfield  {author} {\bibinfo {author} {\bibnamefont {B.B.Kadomtsev}}\ and\
  \bibinfo {author} {\bibnamefont {V.I.Petviashivilli}},\ }\bibfield  {title}
  {\enquote {\bibinfo {title} {{On the stability of solitary waves in weakly
  dispersive media}},}\ }\href@noop {} {\bibfield  {journal} {\bibinfo
  {journal} {Sov. Phys. Dokl}\ }\textbf {\bibinfo {volume} {15}},\ \bibinfo
  {pages} {539--541} (\bibinfo {year} {1970})}\BibitemShut {NoStop}%
\bibitem [{\citenamefont {Manakov}\ \emph {et~al.}(1977)\citenamefont
  {Manakov}, \citenamefont {Zakharov}, \citenamefont {Bordag}, \citenamefont
  {Its},\ and\ \citenamefont {Matveev}}]{MANAKOV1977205}%
  \BibitemOpen
  \bibfield  {author} {\bibinfo {author} {\bibfnamefont {S.V.}\ \bibnamefont
  {Manakov}}, \bibinfo {author} {\bibfnamefont {V.E.}\ \bibnamefont
  {Zakharov}}, \bibinfo {author} {\bibfnamefont {L.A.}\ \bibnamefont {Bordag}},
  \bibinfo {author} {\bibfnamefont {A.R.}\ \bibnamefont {Its}}, \ and\ \bibinfo
  {author} {\bibfnamefont {V.B.}\ \bibnamefont {Matveev}},\ }\bibfield  {title}
  {\enquote {\bibinfo {title} {{Two-dimensional solitons of the
  Kadomtsev-Petviashvili equation and their interaction}},}\ }\href {\doibase
  https://doi.org/10.1016/0375-9601(77)90875-1} {\bibfield  {journal} {\bibinfo
   {journal} {Physics Letters A}\ }\textbf {\bibinfo {volume} {63}},\ \bibinfo
  {pages} {205--206} (\bibinfo {year} {1977})}\BibitemShut {NoStop}%
\bibitem [{\citenamefont {Davey}\ and\ \citenamefont
  {Stewartson}(1974)}]{Davey74}%
  \BibitemOpen
  \bibfield  {author} {\bibinfo {author} {\bibfnamefont {A.}~\bibnamefont
  {Davey}}\ and\ \bibinfo {author} {\bibfnamefont {Keith}\ \bibnamefont
  {Stewartson}},\ }\bibfield  {title} {\enquote {\bibinfo {title} {{On
  three-dimensional packets of surface waves}},}\ }\href {\doibase
  10.1098/rspa.1974.0076} {\bibfield  {journal} {\bibinfo  {journal}
  {Proceedings of the Royal Society of London. A. Mathematical and Physical
  Sciences}\ }\textbf {\bibinfo {volume} {338}},\ \bibinfo {pages} {101--110}
  (\bibinfo {year} {1974})}\BibitemShut {NoStop}%
\bibitem [{\citenamefont {Fokas}\ and\ \citenamefont
  {Santini}(1990)}]{FOKAS199099}%
  \BibitemOpen
  \bibfield  {author} {\bibinfo {author} {\bibfnamefont {A.S.}\ \bibnamefont
  {Fokas}}\ and\ \bibinfo {author} {\bibfnamefont {P.M.}\ \bibnamefont
  {Santini}},\ }\bibfield  {title} {\enquote {\bibinfo {title} {Dromions and a
  boundary value problem for the davey-stewartson 1 equation},}\ }\href
  {\doibase https://doi.org/10.1016/0167-2789(90)90050-Y} {\bibfield  {journal}
  {\bibinfo  {journal} {Physica D: Nonlinear Phenomena}\ }\textbf {\bibinfo
  {volume} {44}},\ \bibinfo {pages} {99--130} (\bibinfo {year}
  {1990})}\BibitemShut {NoStop}%
\bibitem [{\citenamefont {Hietarinta}(1990)}]{HIETARINTA1990113}%
  \BibitemOpen
  \bibfield  {author} {\bibinfo {author} {\bibfnamefont {Jarmo}\ \bibnamefont
  {Hietarinta}},\ }\bibfield  {title} {\enquote {\bibinfo {title} {One-dromion
  solutions for genetic classes of equations},}\ }\href {\doibase
  https://doi.org/10.1016/0375-9601(90)90535-V} {\bibfield  {journal} {\bibinfo
   {journal} {Physics Letters A}\ }\textbf {\bibinfo {volume} {149}},\ \bibinfo
  {pages} {113--118} (\bibinfo {year} {1990})}\BibitemShut {NoStop}%
\bibitem [{\citenamefont {Nishinari}\ and\ \citenamefont
  {Yajima}(1994)}]{Nishinari94}%
  \BibitemOpen
  \bibfield  {author} {\bibinfo {author} {\bibfnamefont {Katsuhiro}\
  \bibnamefont {Nishinari}}\ and\ \bibinfo {author} {\bibfnamefont {Tetsu}\
  \bibnamefont {Yajima}},\ }\bibfield  {title} {\enquote {\bibinfo {title}
  {Numerical studies on stability of dromion and its collisions},}\ }\href
  {\doibase 10.1143/JPSJ.63.3538} {\bibfield  {journal} {\bibinfo  {journal}
  {Journal of the Physical Society of Japan}\ }\textbf {\bibinfo {volume}
  {63}},\ \bibinfo {pages} {3538--3541} (\bibinfo {year} {1994})},\ \Eprint
  {http://arxiv.org/abs/https://doi.org/10.1143/JPSJ.63.3538}
  {https://doi.org/10.1143/JPSJ.63.3538} \BibitemShut {NoStop}%
\bibitem [{\citenamefont {Ishimori}(1984)}]{Ishimori}%
  \BibitemOpen
  \bibfield  {author} {\bibinfo {author} {\bibfnamefont {Yuji}\ \bibnamefont
  {Ishimori}},\ }\bibfield  {title} {\enquote {\bibinfo {title} {{Multi-Vortex
  Solutions of a Two-Dimensional Nonlinear Wave Equation}},}\ }\href {\doibase
  10.1143/PTP.72.33} {\bibfield  {journal} {\bibinfo  {journal} {Progress of
  Theoretical Physics}\ }\textbf {\bibinfo {volume} {72}},\ \bibinfo {pages}
  {33--37} (\bibinfo {year} {1984})},\ \Eprint
  {http://arxiv.org/abs/https://academic.oup.com/ptp/article-pdf/72/1/33/19572888/72-1-33.pdf}
  {https://academic.oup.com/ptp/article-pdf/72/1/33/19572888/72-1-33.pdf}
  \BibitemShut {NoStop}%
\bibitem [{\citenamefont {Kawahara}\ \emph {et~al.}(1992)\citenamefont
  {Kawahara}, \citenamefont {Araki},\ and\ \citenamefont
  {Toh}}]{KAWAHARA199279}%
  \BibitemOpen
  \bibfield  {author} {\bibinfo {author} {\bibfnamefont {Takuji}\ \bibnamefont
  {Kawahara}}, \bibinfo {author} {\bibfnamefont {Keisuke}\ \bibnamefont
  {Araki}}, \ and\ \bibinfo {author} {\bibfnamefont {Sadayoshi}\ \bibnamefont
  {Toh}},\ }\bibfield  {title} {\enquote {\bibinfo {title} {Interactions of
  two-dimensionally localized pulses of the regularized-long-wave equation},}\
  }\href {\doibase https://doi.org/10.1016/0167-2789(92)90207-4} {\bibfield
  {journal} {\bibinfo  {journal} {Physica D: Nonlinear Phenomena}\ }\textbf
  {\bibinfo {volume} {59}},\ \bibinfo {pages} {79--89} (\bibinfo {year}
  {1992})}\BibitemShut {NoStop}%
\bibitem [{\citenamefont {Gao}\ \emph {et~al.}(2022)\citenamefont {Gao},
  \citenamefont {Guo},\ and\ \citenamefont {Shan}}]{GAO2022112293}%
  \BibitemOpen
  \bibfield  {author} {\bibinfo {author} {\bibfnamefont {Xin-Yi}\ \bibnamefont
  {Gao}}, \bibinfo {author} {\bibfnamefont {Yong-Jiang}\ \bibnamefont {Guo}}, \
  and\ \bibinfo {author} {\bibfnamefont {Wen-Rui}\ \bibnamefont {Shan}},\
  }\bibfield  {title} {\enquote {\bibinfo {title} {Taking into consideration an
  extended coupled (2+1)-dimensional burgers system in oceanography, acoustics
  and hydrodynamics},}\ }\href {\doibase
  https://doi.org/10.1016/j.chaos.2022.112293} {\bibfield  {journal} {\bibinfo
  {journal} {Chaos, Solitons and Fractals}\ }\textbf {\bibinfo {volume}
  {161}},\ \bibinfo {pages} {112293} (\bibinfo {year} {2022})}\BibitemShut
  {NoStop}%
\bibitem [{\citenamefont {Kaup}(1981)}]{Kaup1981TheLS}%
  \BibitemOpen
  \bibfield  {author} {\bibinfo {author} {\bibfnamefont {D.~J.}\ \bibnamefont
  {Kaup}},\ }\bibfield  {title} {\enquote {\bibinfo {title} {The lump solutions
  and the b{\"a}cklund transformation for the three-dimensional three-wave
  resonant interaction},}\ }\href@noop {} {\bibfield  {journal} {\bibinfo
  {journal} {Journal of Mathematical Physics}\ }\textbf {\bibinfo {volume}
  {22}},\ \bibinfo {pages} {1176--1181} (\bibinfo {year} {1981})}\BibitemShut
  {NoStop}%
\bibitem [{\citenamefont {Imai}\ and\ \citenamefont {Nozaki}(1996)}]{Imai1996}%
  \BibitemOpen
  \bibfield  {author} {\bibinfo {author} {\bibfnamefont {Kenji}\ \bibnamefont
  {Imai}}\ and\ \bibinfo {author} {\bibfnamefont {Kazuhiro}\ \bibnamefont
  {Nozaki}},\ }\bibfield  {title} {\enquote {\bibinfo {title} {{Lump Solutions
  of the Ishimori-II Equation}},}\ }\href {\doibase 10.1143/PTP.96.521}
  {\bibfield  {journal} {\bibinfo  {journal} {Progress of Theoretical Physics}\
  }\textbf {\bibinfo {volume} {96}},\ \bibinfo {pages} {521--526} (\bibinfo
  {year} {1996})},\ \Eprint
  {http://arxiv.org/abs/https://academic.oup.com/ptp/article-pdf/96/3/521/5414710/96-3-521.pdf}
  {https://academic.oup.com/ptp/article-pdf/96/3/521/5414710/96-3-521.pdf}
  \BibitemShut {NoStop}%
\bibitem [{\citenamefont {Zhang}\ and\ \citenamefont {Ma}(2017)}]{Zhang2017}%
  \BibitemOpen
  \bibfield  {author} {\bibinfo {author} {\bibfnamefont {Hai-Qiang}\
  \bibnamefont {Zhang}}\ and\ \bibinfo {author} {\bibfnamefont {Wen-Xiu}\
  \bibnamefont {Ma}},\ }\bibfield  {title} {\enquote {\bibinfo {title} {Lump
  solutions to the 2+1-dimensional sawada-kotera equation},}\ }\href {\doibase
  10.1007/s11071-016-3190-6} {\bibfield  {journal} {\bibinfo  {journal}
  {Nonlinear Dynamics}\ }\textbf {\bibinfo {volume} {87}},\ \bibinfo {pages}
  {2305--2310} (\bibinfo {year} {2017})}\BibitemShut {NoStop}%
\bibitem [{\citenamefont {Zou}\ \emph {et~al.}(2018)\citenamefont {Zou},
  \citenamefont {Yu}, \citenamefont {Tian}, \citenamefont {Feng},\ and\
  \citenamefont {Li}}]{Zou2018}%
  \BibitemOpen
  \bibfield  {author} {\bibinfo {author} {\bibfnamefont {Li}~\bibnamefont
  {Zou}}, \bibinfo {author} {\bibfnamefont {Zong-Bing}\ \bibnamefont {Yu}},
  \bibinfo {author} {\bibfnamefont {Shou-Fu}\ \bibnamefont {Tian}}, \bibinfo
  {author} {\bibfnamefont {Lian-Li}\ \bibnamefont {Feng}}, \ and\ \bibinfo
  {author} {\bibfnamefont {Jin}\ \bibnamefont {Li}},\ }\bibfield  {title}
  {\enquote {\bibinfo {title} {Lump solutions with interaction phenomena in the
  (2+1)-dimensional ito equation},}\ }\href {\doibase
  10.1142/S021798491850104X} {\bibfield  {journal} {\bibinfo  {journal} {Modern
  Physics Letters B}\ }\textbf {\bibinfo {volume} {32}},\ \bibinfo {pages}
  {1850104} (\bibinfo {year} {2018})},\ \Eprint
  {http://arxiv.org/abs/https://doi.org/10.1142/S021798491850104X}
  {https://doi.org/10.1142/S021798491850104X} \BibitemShut {NoStop}%
\bibitem [{\citenamefont {Anker}\ \emph {et~al.}(1978)\citenamefont {Anker},
  \citenamefont {Freeman},\ and\ \citenamefont {Stewartson}}]{Anker78}%
  \BibitemOpen
  \bibfield  {author} {\bibinfo {author} {\bibfnamefont {D.}~\bibnamefont
  {Anker}}, \bibinfo {author} {\bibfnamefont {N.~C.}\ \bibnamefont {Freeman}},
  \ and\ \bibinfo {author} {\bibfnamefont {Keith}\ \bibnamefont {Stewartson}},\
  }\bibfield  {title} {\enquote {\bibinfo {title} {{On the soliton solutions of
  the Davey-Stewartson equation for long waves}},}\ }\href {\doibase
  10.1098/rspa.1978.0083} {\bibfield  {journal} {\bibinfo  {journal}
  {Proceedings of the Royal Society of London. A. Mathematical and Physical
  Sciences}\ }\textbf {\bibinfo {volume} {360}},\ \bibinfo {pages} {529--540}
  (\bibinfo {year} {1978})}\BibitemShut {NoStop}%
\bibitem [{\citenamefont {Boiti}\ \emph {et~al.}(1995)\citenamefont {Boiti},
  \citenamefont {Martina},\ and\ \citenamefont {Pempinelli}}]{BOITI19952377}%
  \BibitemOpen
  \bibfield  {author} {\bibinfo {author} {\bibfnamefont {M.}~\bibnamefont
  {Boiti}}, \bibinfo {author} {\bibfnamefont {L.}~\bibnamefont {Martina}}, \
  and\ \bibinfo {author} {\bibfnamefont {F.}~\bibnamefont {Pempinelli}},\
  }\bibfield  {title} {\enquote {\bibinfo {title} {{Multidimensional localized
  solitons}},}\ }\href {\doibase https://doi.org/10.1016/0960-0779(94)E0106-Y}
  {\bibfield  {journal} {\bibinfo  {journal} {Chaos, Solitons and Fractals}\
  }\textbf {\bibinfo {volume} {5}},\ \bibinfo {pages} {2377--2417} (\bibinfo
  {year} {1995})}\BibitemShut {NoStop}%
\bibitem [{\citenamefont {Larichev}\ and\ \citenamefont
  {Reznik}(1976)}]{Larichev76}%
  \BibitemOpen
  \bibfield  {author} {\bibinfo {author} {\bibfnamefont {V.}~\bibnamefont
  {Larichev}}\ and\ \bibinfo {author} {\bibfnamefont {G.}~\bibnamefont
  {Reznik}},\ }\bibfield  {title} {\enquote {\bibinfo {title} {{Two-dimensional
  solitary Rossby waves}},}\ }\href@noop {} {\bibfield  {journal} {\bibinfo
  {journal} {Doklady Akademii nauk SSSR}\ }\textbf {\bibinfo {volume} {231}},\
  \bibinfo {pages} {1077--1079} (\bibinfo {year} {1976})}\BibitemShut {NoStop}%
\bibitem [{\citenamefont {Hasegawa}\ and\ \citenamefont
  {Mima}(1978)}]{Hasegawa78}%
  \BibitemOpen
  \bibfield  {author} {\bibinfo {author} {\bibfnamefont {Akira}\ \bibnamefont
  {Hasegawa}}\ and\ \bibinfo {author} {\bibfnamefont {Kunioki}\ \bibnamefont
  {Mima}},\ }\bibfield  {title} {\enquote {\bibinfo {title}
  {{Pseudo-three-dimensional turbulence in magnetized nonuniform plasma}},}\
  }\href {\doibase 10.1063/1.862083} {\bibfield  {journal} {\bibinfo  {journal}
  {The Physics of Fluids}\ }\textbf {\bibinfo {volume} {21}},\ \bibinfo {pages}
  {87--92} (\bibinfo {year} {1978})}\BibitemShut {NoStop}%
\bibitem [{\citenamefont {Hasegawa}\ \emph {et~al.}(1979)\citenamefont
  {Hasegawa}, \citenamefont {Maclennan},\ and\ \citenamefont
  {Kodama}}]{Hasegawa79}%
  \BibitemOpen
  \bibfield  {author} {\bibinfo {author} {\bibfnamefont {Akira}\ \bibnamefont
  {Hasegawa}}, \bibinfo {author} {\bibfnamefont {Carol~G.}\ \bibnamefont
  {Maclennan}}, \ and\ \bibinfo {author} {\bibfnamefont {Yuji}\ \bibnamefont
  {Kodama}},\ }\bibfield  {title} {\enquote {\bibinfo {title} {{Nonlinear
  behavior and turbulence spectra of drift waves and Rossby waves}},}\ }\href
  {\doibase 10.1063/1.862504} {\bibfield  {journal} {\bibinfo  {journal} {The
  Physics of Fluids}\ }\textbf {\bibinfo {volume} {22}},\ \bibinfo {pages}
  {2122--2129} (\bibinfo {year} {1979})}\BibitemShut {NoStop}%
\bibitem [{\citenamefont {Zakharov}\ and\ \citenamefont
  {Kuznetsov}(1974)}]{Zakharov74}%
  \BibitemOpen
  \bibfield  {author} {\bibinfo {author} {\bibfnamefont {V.}~\bibnamefont
  {Zakharov}}\ and\ \bibinfo {author} {\bibfnamefont {E~A.}\ \bibnamefont
  {Kuznetsov}},\ }\bibfield  {title} {\enquote {\bibinfo {title}
  {{Three-dimensional solitons}},}\ }\href@noop {} {\bibfield  {journal}
  {\bibinfo  {journal} {Soviet Physics JETP}\ }\textbf {\bibinfo {volume}
  {29}},\ \bibinfo {pages} {594--597} (\bibinfo {year} {1974})}\BibitemShut
  {NoStop}%
\bibitem [{\citenamefont {Petviashvili}(1981)}]{PETVIASHVILI1981329}%
  \BibitemOpen
  \bibfield  {author} {\bibinfo {author} {\bibfnamefont {Vladimir~I.}\
  \bibnamefont {Petviashvili}},\ }\bibfield  {title} {\enquote {\bibinfo
  {title} {Multidimensional and dissipative solitons},}\ }\href {\doibase
  https://doi.org/10.1016/0167-2789(81)90136-6} {\bibfield  {journal} {\bibinfo
   {journal} {Physica D: Nonlinear Phenomena}\ }\textbf {\bibinfo {volume}
  {3}},\ \bibinfo {pages} {329--334} (\bibinfo {year} {1981})}\BibitemShut
  {NoStop}%
\bibitem [{\citenamefont {Toh}\ \emph {et~al.}(1989)\citenamefont {Toh},
  \citenamefont {Iwasaki},\ and\ \citenamefont {Kawahara}}]{Toh89}%
  \BibitemOpen
  \bibfield  {author} {\bibinfo {author} {\bibfnamefont {Sadayoshi}\
  \bibnamefont {Toh}}, \bibinfo {author} {\bibfnamefont {Hiroshi}\ \bibnamefont
  {Iwasaki}}, \ and\ \bibinfo {author} {\bibfnamefont {Takuji}\ \bibnamefont
  {Kawahara}},\ }\bibfield  {title} {\enquote {\bibinfo {title}
  {Two-dimensionally localized pulses of a nonlinear equation with dissipation
  and dispersion},}\ }\href {\doibase 10.1103/PhysRevA.40.5472} {\bibfield
  {journal} {\bibinfo  {journal} {Phys. Rev. A}\ }\textbf {\bibinfo {volume}
  {40}},\ \bibinfo {pages} {5472--5475} (\bibinfo {year} {1989})}\BibitemShut
  {NoStop}%
\bibitem [{\citenamefont {Melkonian}\ and\ \citenamefont
  {Maslowe}(1989)}]{MELKONIAN1989255}%
  \BibitemOpen
  \bibfield  {author} {\bibinfo {author} {\bibfnamefont {S.}~\bibnamefont
  {Melkonian}}\ and\ \bibinfo {author} {\bibfnamefont {S.A.}\ \bibnamefont
  {Maslowe}},\ }\bibfield  {title} {\enquote {\bibinfo {title} {Two-dimensional
  amplitude evolution equations for nonlinear dispersive waves on thin
  films},}\ }\href {\doibase https://doi.org/10.1016/0167-2789(89)90238-8}
  {\bibfield  {journal} {\bibinfo  {journal} {Physica D: Nonlinear Phenomena}\
  }\textbf {\bibinfo {volume} {34}},\ \bibinfo {pages} {255--269} (\bibinfo
  {year} {1989})}\BibitemShut {NoStop}%
\bibitem [{\citenamefont {V.I.Petviashvili}(1980)}]{Petviashvilli80}%
  \BibitemOpen
  \bibfield  {author} {\bibinfo {author} {\bibnamefont {V.I.Petviashvili}},\
  }\bibfield  {title} {\enquote {\bibinfo {title} {{Red spot of Jupiter and the
  drift soliton in a plasma}},}\ }\href@noop {} {\bibfield  {journal} {\bibinfo
   {journal} {JETP Lett.}\ }\textbf {\bibinfo {volume} {32}},\ \bibinfo {pages}
  {619--621} (\bibinfo {year} {1980})}\BibitemShut {NoStop}%
\bibitem [{\citenamefont {Nozaki}(1981)}]{Nozaki81}%
  \BibitemOpen
  \bibfield  {author} {\bibinfo {author} {\bibfnamefont {K.}~\bibnamefont
  {Nozaki}},\ }\bibfield  {title} {\enquote {\bibinfo {title} {Vortex solitons
  of drift waves and anomalous diffusion},}\ }\href {\doibase
  10.1103/PhysRevLett.46.184} {\bibfield  {journal} {\bibinfo  {journal} {Phys.
  Rev. Lett.}\ }\textbf {\bibinfo {volume} {46}},\ \bibinfo {pages} {184--187}
  (\bibinfo {year} {1981})}\BibitemShut {NoStop}%
\bibitem [{\citenamefont {Xu}\ and\ \citenamefont {Shu}(2005)}]{XU200521}%
  \BibitemOpen
  \bibfield  {author} {\bibinfo {author} {\bibfnamefont {Yan}\ \bibnamefont
  {Xu}}\ and\ \bibinfo {author} {\bibfnamefont {Chi-Wang}\ \bibnamefont
  {Shu}},\ }\bibfield  {title} {\enquote {\bibinfo {title} {Local discontinuous
  galerkin methods for two classes of two-dimensional nonlinear wave
  equations},}\ }\href {\doibase https://doi.org/10.1016/j.physd.2005.06.007}
  {\bibfield  {journal} {\bibinfo  {journal} {Physica D: Nonlinear Phenomena}\
  }\textbf {\bibinfo {volume} {208}},\ \bibinfo {pages} {21--58} (\bibinfo
  {year} {2005})}\BibitemShut {NoStop}%
\bibitem [{\citenamefont {Lannes}\ \emph {et~al.}(2013)\citenamefont {Lannes},
  \citenamefont {Linares},\ and\ \citenamefont {Saut}}]{Lannes2013}%
  \BibitemOpen
  \bibfield  {author} {\bibinfo {author} {\bibfnamefont {David}\ \bibnamefont
  {Lannes}}, \bibinfo {author} {\bibfnamefont {Felipe}\ \bibnamefont
  {Linares}}, \ and\ \bibinfo {author} {\bibfnamefont {Jean-Claude}\
  \bibnamefont {Saut}},\ }\enquote {\bibinfo {title} {The cauchy problem for
  the euler--poisson system and derivation of the zakharov--kuznetsov
  equation},}\ in\ \href {\doibase 10.1007/978-1-4614-6348-1_10} {\emph
  {\bibinfo {booktitle} {Studies in Phase Space Analysis with Applications to
  PDEs}}},\ \bibinfo {editor} {edited by\ \bibinfo {editor} {\bibfnamefont
  {Massimo}\ \bibnamefont {Cicognani}}, \bibinfo {editor} {\bibfnamefont
  {Ferruccio}\ \bibnamefont {Colombini}}, \ and\ \bibinfo {editor}
  {\bibfnamefont {Daniele}\ \bibnamefont {Del~Santo}}}\ (\bibinfo  {publisher}
  {Springer New York},\ \bibinfo {address} {New York, NY},\ \bibinfo {year}
  {2013})\ pp.\ \bibinfo {pages} {181--213}\BibitemShut {NoStop}%
\bibitem [{\citenamefont {Klein}\ \emph {et~al.}(2021)\citenamefont {Klein},
  \citenamefont {Roudenko},\ and\ \citenamefont {Stoilov}}]{Klein21}%
  \BibitemOpen
  \bibfield  {author} {\bibinfo {author} {\bibfnamefont {Christian}\
  \bibnamefont {Klein}}, \bibinfo {author} {\bibfnamefont {Svetlana}\
  \bibnamefont {Roudenko}}, \ and\ \bibinfo {author} {\bibfnamefont {Nikola}\
  \bibnamefont {Stoilov}},\ }\bibfield  {title} {\enquote {\bibinfo {title}
  {{Numerical study of Zakhavor-Kuznetsov equations in two dimensions}},}\
  }\href@noop {} {\bibfield  {journal} {\bibinfo  {journal} {Journal of
  Nonlinear Science}\ }\textbf {\bibinfo {volume} {31}},\ \bibinfo {pages}
  {1--28} (\bibinfo {year} {2021})}\BibitemShut {NoStop}%
\bibitem [{\citenamefont {Iwasaki}\ \emph {et~al.}(1990)\citenamefont
  {Iwasaki}, \citenamefont {Toh},\ and\ \citenamefont
  {Kawahara}}]{IWASAKI1990293}%
  \BibitemOpen
  \bibfield  {author} {\bibinfo {author} {\bibfnamefont {Hiroshi}\ \bibnamefont
  {Iwasaki}}, \bibinfo {author} {\bibfnamefont {Sadayoshi}\ \bibnamefont
  {Toh}}, \ and\ \bibinfo {author} {\bibfnamefont {Takuji}\ \bibnamefont
  {Kawahara}},\ }\bibfield  {title} {\enquote {\bibinfo {title} {{Cylindrical
  quasi-solitons of the Zakharov-Kuznetsov equation}},}\ }\href {\doibase
  https://doi.org/10.1016/0167-2789(90)90138-F} {\bibfield  {journal} {\bibinfo
   {journal} {Physica D: Nonlinear Phenomena}\ }\textbf {\bibinfo {volume}
  {43}},\ \bibinfo {pages} {293--303} (\bibinfo {year} {1990})}\BibitemShut
  {NoStop}%
\bibitem [{\citenamefont {Charney}\ and\ \citenamefont
  {Flierl}(1981)}]{charney1981oceanic}%
  \BibitemOpen
  \bibfield  {author} {\bibinfo {author} {\bibfnamefont {Jule~G.}\ \bibnamefont
  {Charney}}\ and\ \bibinfo {author} {\bibfnamefont {Glenn~R.}\ \bibnamefont
  {Flierl}},\ }\bibfield  {title} {\enquote {\bibinfo {title} {Oceanic
  analogues of large-scale atmospheric motions},}\ }\href@noop {} {\bibfield
  {journal} {\bibinfo  {journal} {Evolution of physical oceanography}\ }\textbf
  {\bibinfo {volume} {504}},\ \bibinfo {pages} {548} (\bibinfo {year}
  {1981})}\BibitemShut {NoStop}%
\bibitem [{\citenamefont {Bouchet}\ and\ \citenamefont
  {Venaille}(2012)}]{bouchet2012statistical}%
  \BibitemOpen
  \bibfield  {author} {\bibinfo {author} {\bibfnamefont {Freddy}\ \bibnamefont
  {Bouchet}}\ and\ \bibinfo {author} {\bibfnamefont {Antoine}\ \bibnamefont
  {Venaille}},\ }\bibfield  {title} {\enquote {\bibinfo {title} {Statistical
  mechanics of two-dimensional and geophysical flows},}\ }\href@noop {}
  {\bibfield  {journal} {\bibinfo  {journal} {Physics reports}\ }\textbf
  {\bibinfo {volume} {515}},\ \bibinfo {pages} {227--295} (\bibinfo {year}
  {2012})}\BibitemShut {NoStop}%
\bibitem [{\citenamefont {Constantin}(2009)}]{constantin2009relevance}%
  \BibitemOpen
  \bibfield  {author} {\bibinfo {author} {\bibfnamefont {Adrian}\ \bibnamefont
  {Constantin}},\ }\bibfield  {title} {\enquote {\bibinfo {title} {On the
  relevance of soliton theory to tsunami modelling},}\ }\href@noop {}
  {\bibfield  {journal} {\bibinfo  {journal} {Wave Motion}\ }\textbf {\bibinfo
  {volume} {46}},\ \bibinfo {pages} {420--426} (\bibinfo {year}
  {2009})}\BibitemShut {NoStop}%
\bibitem [{\citenamefont {Constantin}\ and\ \citenamefont
  {Henry}(2009)}]{constantin2009solitons}%
  \BibitemOpen
  \bibfield  {author} {\bibinfo {author} {\bibfnamefont {Adrian}\ \bibnamefont
  {Constantin}}\ and\ \bibinfo {author} {\bibfnamefont {David}\ \bibnamefont
  {Henry}},\ }\bibfield  {title} {\enquote {\bibinfo {title} {Solitons and
  tsunamis},}\ }\href@noop {} {\bibfield  {journal} {\bibinfo  {journal}
  {Zeitschrift f{\"u}r Naturforschung A}\ }\textbf {\bibinfo {volume} {64}},\
  \bibinfo {pages} {65--68} (\bibinfo {year} {2009})}\BibitemShut {NoStop}%
\bibitem [{\citenamefont {Constantin}(2011)}]{constantin2011nonlinear}%
  \BibitemOpen
  \bibfield  {author} {\bibinfo {author} {\bibfnamefont {Adrian}\ \bibnamefont
  {Constantin}},\ }\href@noop {} {\emph {\bibinfo {title} {Nonlinear water
  waves with applications to wave-current interactions and tsunamis}}}\
  (\bibinfo  {publisher} {SIAM},\ \bibinfo {year} {2011})\BibitemShut {NoStop}%
\bibitem [{\citenamefont {Ibragimov}\ \emph {et~al.}(2013)\citenamefont
  {Ibragimov}, \citenamefont {Jefferson},\ and\ \citenamefont
  {Carminati}}]{ibragimov2013invariant}%
  \BibitemOpen
  \bibfield  {author} {\bibinfo {author} {\bibfnamefont {Ranis}\ \bibnamefont
  {Ibragimov}}, \bibinfo {author} {\bibfnamefont {Grace}\ \bibnamefont
  {Jefferson}}, \ and\ \bibinfo {author} {\bibfnamefont {John}\ \bibnamefont
  {Carminati}},\ }\bibfield  {title} {\enquote {\bibinfo {title} {Invariant and
  approximately invariant solutions of non-linear internal gravity waves
  forming a column of stratified fluid affected by the earth's rotation},}\
  }\href@noop {} {\bibfield  {journal} {\bibinfo  {journal} {International
  Journal of Non-Linear Mechanics}\ }\textbf {\bibinfo {volume} {51}},\
  \bibinfo {pages} {28--44} (\bibinfo {year} {2013})}\BibitemShut {NoStop}%
\bibitem [{\citenamefont {Ibragimov}\ and\ \citenamefont
  {Ibragimov}(2013)}]{ibragimov2013nonlinear}%
  \BibitemOpen
  \bibfield  {author} {\bibinfo {author} {\bibfnamefont {Nail~H.}\ \bibnamefont
  {Ibragimov}}\ and\ \bibinfo {author} {\bibfnamefont {Ranis~N.}\ \bibnamefont
  {Ibragimov}},\ }\bibfield  {title} {\enquote {\bibinfo {title} {Nonlinear
  whirlpools versus harmonic waves in a rotating column of stratified fluid},}\
  }\href@noop {} {\bibfield  {journal} {\bibinfo  {journal} {Mathematical
  Modelling of Natural Phenomena}\ }\textbf {\bibinfo {volume} {8}},\ \bibinfo
  {pages} {122--131} (\bibinfo {year} {2013})}\BibitemShut {NoStop}%
\bibitem [{\citenamefont {Ibragimov}\ and\ \citenamefont
  {Lin}(2017)}]{ibragimov2017nonlinear}%
  \BibitemOpen
  \bibfield  {author} {\bibinfo {author} {\bibfnamefont {R.N.}\ \bibnamefont
  {Ibragimov}}\ and\ \bibinfo {author} {\bibfnamefont {G.}~\bibnamefont
  {Lin}},\ }\bibfield  {title} {\enquote {\bibinfo {title} {Nonlinear analysis
  of perturbed rotating whirlpools in the ocean and atmosphere},}\ }\href@noop
  {} {\bibfield  {journal} {\bibinfo  {journal} {Mathematical Modelling of
  Natural Phenomena}\ }\textbf {\bibinfo {volume} {12}},\ \bibinfo {pages}
  {94--114} (\bibinfo {year} {2017})}\BibitemShut {NoStop}%
\bibitem [{\citenamefont {Nezlin}\ and\ \citenamefont
  {Sutyrin}(1989)}]{nezlin1989long}%
  \BibitemOpen
  \bibfield  {author} {\bibinfo {author} {\bibfnamefont {M.V.}\ \bibnamefont
  {Nezlin}}\ and\ \bibinfo {author} {\bibfnamefont {G.G.}\ \bibnamefont
  {Sutyrin}},\ }\bibfield  {title} {\enquote {\bibinfo {title} {Long-lived
  solitary anticyclones in the planetary atmospheres and oceans, in laboratory
  experiments and in theory},}\ }in\ \href@noop {} {\emph {\bibinfo {booktitle}
  {Elsevier oceanography series}}},\ Vol.~\bibinfo {volume} {50}\ (\bibinfo
  {publisher} {Elsevier},\ \bibinfo {year} {1989})\ pp.\ \bibinfo {pages}
  {701--719}\BibitemShut {NoStop}%
\bibitem [{\citenamefont {Nezlin}\ \emph {et~al.}(1993)\citenamefont {Nezlin},
  \citenamefont {Snezhkin}, \citenamefont {Dobroslavsky},\ and\ \citenamefont
  {Pletnev}}]{BA2125604X}%
  \BibitemOpen
  \bibfield  {author} {\bibinfo {author} {\bibfnamefont {M.~V.}\ \bibnamefont
  {Nezlin}}, \bibinfo {author} {\bibfnamefont {E.~N.}\ \bibnamefont
  {Snezhkin}}, \bibinfo {author} {\bibfnamefont {A.}~\bibnamefont
  {Dobroslavsky}}, \ and\ \bibinfo {author} {\bibfnamefont {A.}~\bibnamefont
  {Pletnev}},\ }\href {https://ci.nii.ac.jp/ncid/BA2125604X} {\emph {\bibinfo
  {title} {Rossby vortices, spiral structures, solitons : astrophysics and
  plasma physics in shallow water experiments}}},\ Springer series in nonlinear
  dynamics\ (\bibinfo  {publisher} {Springer-Verlag},\ \bibinfo {year}
  {1993})\BibitemShut {NoStop}%
\bibitem [{\citenamefont {Williams}\ and\ \citenamefont
  {Yamagata}(1984)}]{Williams84}%
  \BibitemOpen
  \bibfield  {author} {\bibinfo {author} {\bibfnamefont {Gareth}\ \bibnamefont
  {Williams}}\ and\ \bibinfo {author} {\bibfnamefont {Toshio}\ \bibnamefont
  {Yamagata}},\ }\bibfield  {title} {\enquote {\bibinfo {title} {{Geostrophic
  Regimes, Intermediate Solitary Vortices and Jovian Eddies}},}\ }\href
  {\doibase 10.1175/1520-0469(1984)041<0453:GRISVA>2.0.CO;2} {\bibfield
  {journal} {\bibinfo  {journal} {Journal of The Atmospheric Sciences - J ATMOS
  SCI}\ }\textbf {\bibinfo {volume} {41}},\ \bibinfo {pages} {453--478}
  (\bibinfo {year} {1984})}\BibitemShut {NoStop}%
\bibitem [{\citenamefont {Charney}(1963)}]{Charney63}%
  \BibitemOpen
  \bibfield  {author} {\bibinfo {author} {\bibfnamefont {Jule~G.}\ \bibnamefont
  {Charney}},\ }\bibfield  {title} {\enquote {\bibinfo {title} {A note on
  large-scale motions in the tropics},}\ }\href {\doibase
  10.1175/1520-0469(1963)020<0607:ANOLSM>2.0.CO;2} {\bibfield  {journal}
  {\bibinfo  {journal} {Journal of Atmospheric Sciences}\ }\textbf {\bibinfo
  {volume} {20}},\ \bibinfo {pages} {607 -- 609} (\bibinfo {year}
  {1963})}\BibitemShut {NoStop}%
\bibitem [{\citenamefont {Calogero}\ and\ \citenamefont
  {Degasperis}(1978)}]{calogero1978exact}%
  \BibitemOpen
  \bibfield  {author} {\bibinfo {author} {\bibfnamefont {F.}~\bibnamefont
  {Calogero}}\ and\ \bibinfo {author} {\bibfnamefont {Ao}~\bibnamefont
  {Degasperis}},\ }\bibfield  {title} {\enquote {\bibinfo {title} {Exact
  solution via the spectral transform of a generalization with linearly
  x-dependent coefficients of the modified korteweg-de-vries equation},}\
  }\href@noop {} {\bibfield  {journal} {\bibinfo  {journal} {Lett. Nuovo Cim}\
  }\textbf {\bibinfo {volume} {22}},\ \bibinfo {pages} {270--273} (\bibinfo
  {year} {1978})}\BibitemShut {NoStop}%
\bibitem [{\citenamefont {Brugarino}\ and\ \citenamefont
  {Pantano}(1980)}]{brugarino1980integration}%
  \BibitemOpen
  \bibfield  {author} {\bibinfo {author} {\bibfnamefont {T.}~\bibnamefont
  {Brugarino}}\ and\ \bibinfo {author} {\bibfnamefont {P.}~\bibnamefont
  {Pantano}},\ }\bibfield  {title} {\enquote {\bibinfo {title} {The integration
  of burgers and korteweg-de vries equations with nonuniformities},}\
  }\href@noop {} {\bibfield  {journal} {\bibinfo  {journal} {Physics Letters
  A}\ }\textbf {\bibinfo {volume} {80}},\ \bibinfo {pages} {223--224} (\bibinfo
  {year} {1980})}\BibitemShut {NoStop}%
\bibitem [{\citenamefont {Joshi}(1987)}]{joshi1987painleve}%
  \BibitemOpen
  \bibfield  {author} {\bibinfo {author} {\bibfnamefont {Nalini}\ \bibnamefont
  {Joshi}},\ }\bibfield  {title} {\enquote {\bibinfo {title} {Painlev{\'e}
  property of general variable-coefficient versions of the korteweg-de vries
  and non-linear schr{\"o}dinger equations},}\ }\href@noop {} {\bibfield
  {journal} {\bibinfo  {journal} {Physics Letters A}\ }\textbf {\bibinfo
  {volume} {125}},\ \bibinfo {pages} {456--460} (\bibinfo {year}
  {1987})}\BibitemShut {NoStop}%
\bibitem [{\citenamefont {Hlavat{\`y}}(1988)}]{hlavaty1988painleve}%
  \BibitemOpen
  \bibfield  {author} {\bibinfo {author} {\bibfnamefont {Ladislav}\
  \bibnamefont {Hlavat{\`y}}},\ }\bibfield  {title} {\enquote {\bibinfo {title}
  {Painlev{\'e} analysis of nonautonomous evolution equations},}\ }\href@noop
  {} {\bibfield  {journal} {\bibinfo  {journal} {Physics Letters A}\ }\textbf
  {\bibinfo {volume} {128}},\ \bibinfo {pages} {335--338} (\bibinfo {year}
  {1988})}\BibitemShut {NoStop}%
\bibitem [{\citenamefont {Brugarino}\ and\ \citenamefont
  {Greco}(1991)}]{brugarino1991painleve}%
  \BibitemOpen
  \bibfield  {author} {\bibinfo {author} {\bibfnamefont {Tommaso}\ \bibnamefont
  {Brugarino}}\ and\ \bibinfo {author} {\bibfnamefont {Antonio~M.}\
  \bibnamefont {Greco}},\ }\bibfield  {title} {\enquote {\bibinfo {title}
  {Painlev{\'e} analysis and reducibility to the canonical form for the
  generalized kadomtsev--petviashvili equation},}\ }\href@noop {} {\bibfield
  {journal} {\bibinfo  {journal} {Journal of mathematical physics}\ }\textbf
  {\bibinfo {volume} {32}},\ \bibinfo {pages} {69--71} (\bibinfo {year}
  {1991})}\BibitemShut {NoStop}%
\bibitem [{\citenamefont {Gao}\ and\ \citenamefont
  {Tian}(2001)}]{gao2001variable}%
  \BibitemOpen
  \bibfield  {author} {\bibinfo {author} {\bibfnamefont {Yi-Tian}\ \bibnamefont
  {Gao}}\ and\ \bibinfo {author} {\bibfnamefont {Bo}~\bibnamefont {Tian}},\
  }\bibfield  {title} {\enquote {\bibinfo {title} {Variable-coefficient
  balancing-act algorithm extended to a variable-coefficient mkp model for the
  rotating fluids},}\ }\href@noop {} {\bibfield  {journal} {\bibinfo  {journal}
  {International Journal of Modern Physics C}\ }\textbf {\bibinfo {volume}
  {12}},\ \bibinfo {pages} {1383--1389} (\bibinfo {year} {2001})}\BibitemShut
  {NoStop}%
\bibitem [{\citenamefont {Kobayashi}\ and\ \citenamefont
  {Toda}(2005)}]{kobayashi2005generalized}%
  \BibitemOpen
  \bibfield  {author} {\bibinfo {author} {\bibfnamefont {Tadashi}\ \bibnamefont
  {Kobayashi}}\ and\ \bibinfo {author} {\bibfnamefont {Kouichi}\ \bibnamefont
  {Toda}},\ }\bibfield  {title} {\enquote {\bibinfo {title} {A generalized
  kdv-family with variable coefficients in (2+ 1) dimensions},}\ }\href@noop {}
  {\bibfield  {journal} {\bibinfo  {journal} {IEICE Transactions on
  Fundamentals of Electronics, Communications and Computer Sciences}\ }\textbf
  {\bibinfo {volume} {88}},\ \bibinfo {pages} {2548--2553} (\bibinfo {year}
  {2005})}\BibitemShut {NoStop}%
\bibitem [{\citenamefont {Kobayashi}\ and\ \citenamefont
  {Toda}(2006)}]{kobayashi2006painleve}%
  \BibitemOpen
  \bibfield  {author} {\bibinfo {author} {\bibfnamefont {Tadashi}\ \bibnamefont
  {Kobayashi}}\ and\ \bibinfo {author} {\bibfnamefont {Kouichi}\ \bibnamefont
  {Toda}},\ }\bibfield  {title} {\enquote {\bibinfo {title} {The painlev{\'e}
  test and reducibility to the canonical forms for higher-dimensional soliton
  equations with variable-coefficients},}\ }\href@noop {} {\bibfield  {journal}
  {\bibinfo  {journal} {SIGMA. Symmetry, Integrability and Geometry: Methods
  and Applications}\ }\textbf {\bibinfo {volume} {2}},\ \bibinfo {pages} {063}
  (\bibinfo {year} {2006})}\BibitemShut {NoStop}%
\bibitem [{\citenamefont {Honda}\ and\ \citenamefont
  {Choptuik}(2002)}]{Honda:2001xg}%
  \BibitemOpen
  \bibfield  {author} {\bibinfo {author} {\bibfnamefont {Ethan~P.}\
  \bibnamefont {Honda}}\ and\ \bibinfo {author} {\bibfnamefont {Matthew~W.}\
  \bibnamefont {Choptuik}},\ }\bibfield  {title} {\enquote {\bibinfo {title}
  {{Fine structure of oscillons in the spherically symmetric phi**4
  Klein-Gordon model}},}\ }\href {\doibase 10.1103/PhysRevD.65.084037}
  {\bibfield  {journal} {\bibinfo  {journal} {Phys. Rev. D}\ }\textbf {\bibinfo
  {volume} {65}},\ \bibinfo {pages} {084037} (\bibinfo {year} {2002})},\
  \Eprint {http://arxiv.org/abs/hep-ph/0110065} {arXiv:hep-ph/0110065}
  \BibitemShut {NoStop}%
\bibitem [{\citenamefont {Bogolyubsky}\ and\ \citenamefont
  {Makhankov}(1976)}]{Bogolyubsky:1976yu}%
  \BibitemOpen
  \bibfield  {author} {\bibinfo {author} {\bibfnamefont {I.~L.}\ \bibnamefont
  {Bogolyubsky}}\ and\ \bibinfo {author} {\bibfnamefont {V.~G.}\ \bibnamefont
  {Makhankov}},\ }\bibfield  {title} {\enquote {\bibinfo {title} {{Lifetime of
  Pulsating Solitons in Some Classical Models}},}\ }\href@noop {} {\bibfield
  {journal} {\bibinfo  {journal} {Pisma Zh. Eksp. Teor. Fiz.}\ }\textbf
  {\bibinfo {volume} {24}},\ \bibinfo {pages} {15--18} (\bibinfo {year}
  {1976})}\BibitemShut {NoStop}%
\bibitem [{\citenamefont {Gleiser}(1994)}]{Gleiser:1993pt}%
  \BibitemOpen
  \bibfield  {author} {\bibinfo {author} {\bibfnamefont {Marcelo}\ \bibnamefont
  {Gleiser}},\ }\bibfield  {title} {\enquote {\bibinfo {title} {{Pseudostable
  bubbles}},}\ }\href {\doibase 10.1103/PhysRevD.49.2978} {\bibfield  {journal}
  {\bibinfo  {journal} {Phys. Rev. D}\ }\textbf {\bibinfo {volume} {49}},\
  \bibinfo {pages} {2978--2981} (\bibinfo {year} {1994})},\ \Eprint
  {http://arxiv.org/abs/hep-ph/9308279} {arXiv:hep-ph/9308279} \BibitemShut
  {NoStop}%
\bibitem [{\citenamefont {Copeland}\ \emph {et~al.}(1995)\citenamefont
  {Copeland}, \citenamefont {Gleiser},\ and\ \citenamefont
  {Muller}}]{Copeland:1995fq}%
  \BibitemOpen
  \bibfield  {author} {\bibinfo {author} {\bibfnamefont {Edmund~J.}\
  \bibnamefont {Copeland}}, \bibinfo {author} {\bibfnamefont {M.}~\bibnamefont
  {Gleiser}}, \ and\ \bibinfo {author} {\bibfnamefont {H.~R.}\ \bibnamefont
  {Muller}},\ }\bibfield  {title} {\enquote {\bibinfo {title} {{Oscillons:
  Resonant configurations during bubble collapse}},}\ }\href {\doibase
  10.1103/PhysRevD.52.1920} {\bibfield  {journal} {\bibinfo  {journal} {Phys.
  Rev. D}\ }\textbf {\bibinfo {volume} {52}},\ \bibinfo {pages} {1920--1933}
  (\bibinfo {year} {1995})},\ \Eprint {http://arxiv.org/abs/hep-ph/9503217}
  {arXiv:hep-ph/9503217} \BibitemShut {NoStop}%
\bibitem [{\citenamefont {Fodor}\ \emph {et~al.}(2008)\citenamefont {Fodor},
  \citenamefont {Forgacs}, \citenamefont {Horvath},\ and\ \citenamefont
  {Lukacs}}]{Fodor:2008es}%
  \BibitemOpen
  \bibfield  {author} {\bibinfo {author} {\bibfnamefont {Gyula}\ \bibnamefont
  {Fodor}}, \bibinfo {author} {\bibfnamefont {Peter}\ \bibnamefont {Forgacs}},
  \bibinfo {author} {\bibfnamefont {Zalan}\ \bibnamefont {Horvath}}, \ and\
  \bibinfo {author} {\bibfnamefont {Arpad}\ \bibnamefont {Lukacs}},\ }\bibfield
   {title} {\enquote {\bibinfo {title} {{Small amplitude quasi-breathers and
  oscillons}},}\ }\href {\doibase 10.1103/PhysRevD.78.025003} {\bibfield
  {journal} {\bibinfo  {journal} {Phys. Rev. D}\ }\textbf {\bibinfo {volume}
  {78}},\ \bibinfo {pages} {025003} (\bibinfo {year} {2008})},\ \Eprint
  {http://arxiv.org/abs/0802.3525} {arXiv:0802.3525 [hep-th]} \BibitemShut
  {NoStop}%
\bibitem [{\citenamefont {Fodor}\ \emph {et~al.}(2009)\citenamefont {Fodor},
  \citenamefont {Forgacs}, \citenamefont {Horvath},\ and\ \citenamefont
  {Mezei}}]{Fodor:2009xw}%
  \BibitemOpen
  \bibfield  {author} {\bibinfo {author} {\bibfnamefont {Gyula}\ \bibnamefont
  {Fodor}}, \bibinfo {author} {\bibfnamefont {Peter}\ \bibnamefont {Forgacs}},
  \bibinfo {author} {\bibfnamefont {Zalan}\ \bibnamefont {Horvath}}, \ and\
  \bibinfo {author} {\bibfnamefont {Mark}\ \bibnamefont {Mezei}},\ }\bibfield
  {title} {\enquote {\bibinfo {title} {{Oscillons in dilaton-scalar
  theories}},}\ }\href {\doibase 10.1088/1126-6708/2009/08/106} {\bibfield
  {journal} {\bibinfo  {journal} {JHEP}\ }\textbf {\bibinfo {volume} {08}},\
  \bibinfo {pages} {106} (\bibinfo {year} {2009})},\ \Eprint
  {http://arxiv.org/abs/0906.4160} {arXiv:0906.4160 [hep-th]} \BibitemShut
  {NoStop}%
\bibitem [{\citenamefont {Gleiser}\ and\ \citenamefont
  {Stamatopoulos}(2012)}]{Gleiser:2011di}%
  \BibitemOpen
  \bibfield  {author} {\bibinfo {author} {\bibfnamefont {Marcelo}\ \bibnamefont
  {Gleiser}}\ and\ \bibinfo {author} {\bibfnamefont {Nikitas}\ \bibnamefont
  {Stamatopoulos}},\ }\bibfield  {title} {\enquote {\bibinfo {title} {{Entropic
  Measure for Localized Energy Configurations: Kinks, Bounces, and Bubbles}},}\
  }\href {\doibase 10.1016/j.physletb.2012.05.064} {\bibfield  {journal}
  {\bibinfo  {journal} {Phys. Lett. B}\ }\textbf {\bibinfo {volume} {713}},\
  \bibinfo {pages} {304--307} (\bibinfo {year} {2012})},\ \Eprint
  {http://arxiv.org/abs/1111.5597} {arXiv:1111.5597 [hep-th]} \BibitemShut
  {NoStop}%
\bibitem [{\citenamefont {Petviashvili}\ and\ \citenamefont
  {Yan'kov}(1982)}]{PetYan82}%
  \BibitemOpen
  \bibfield  {author} {\bibinfo {author} {\bibfnamefont {V.~I.}\ \bibnamefont
  {Petviashvili}}\ and\ \bibinfo {author} {\bibfnamefont {V.~V.}\ \bibnamefont
  {Yan'kov}},\ }\bibfield  {title} {\enquote {\bibinfo {title} {{Bilayer
  vortices in rotating stratified fluid}},}\ }\href@noop {} {\bibfield
  {journal} {\bibinfo  {journal} {Dokl. Akad. Nauk SSSR}\ }\textbf {\bibinfo
  {volume} {267}},\ \bibinfo {pages} {825--828} (\bibinfo {year}
  {1982})}\BibitemShut {NoStop}%
\bibitem [{\citenamefont {Kuznetsov}\ \emph {et~al.}(1986)\citenamefont
  {Kuznetsov}, \citenamefont {Rubenchik},\ and\ \citenamefont
  {Zakharov}}]{KUZNETSOV1986103}%
  \BibitemOpen
  \bibfield  {author} {\bibinfo {author} {\bibfnamefont {E.A.}\ \bibnamefont
  {Kuznetsov}}, \bibinfo {author} {\bibfnamefont {A.M.}\ \bibnamefont
  {Rubenchik}}, \ and\ \bibinfo {author} {\bibfnamefont {V.E.}\ \bibnamefont
  {Zakharov}},\ }\bibfield  {title} {\enquote {\bibinfo {title} {{Soliton
  stability in plasmas and hydrodynamics}},}\ }\href@noop {} {\bibfield
  {journal} {\bibinfo  {journal} {Physics Reports}\ }\textbf {\bibinfo {volume}
  {142}},\ \bibinfo {pages} {103--165} (\bibinfo {year} {1986})}\BibitemShut
  {NoStop}%
\bibitem [{\citenamefont {Kaup}(1993)}]{Kaup_1993}%
  \BibitemOpen
  \bibfield  {author} {\bibinfo {author} {\bibfnamefont {D.~J.}\ \bibnamefont
  {Kaup}},\ }\bibfield  {title} {\enquote {\bibinfo {title} {Asymptotic
  expansions for the transmission coefficients of the so-called
  davey-stewartson i system},}\ }\href {\doibase 10.1088/0266-5611/9/3/004}
  {\bibfield  {journal} {\bibinfo  {journal} {Inverse Problems}\ }\textbf
  {\bibinfo {volume} {9}},\ \bibinfo {pages} {417--432} (\bibinfo {year}
  {1993})}\BibitemShut {NoStop}%
\bibitem [{\citenamefont {Zabusky}(1981)}]{ZABUSKY1981195}%
  \BibitemOpen
  \bibfield  {author} {\bibinfo {author} {\bibfnamefont {Norman~J.}\
  \bibnamefont {Zabusky}},\ }\bibfield  {title} {\enquote {\bibinfo {title}
  {Computational synergetics and mathematical innovation},}\ }\href {\doibase
  https://doi.org/10.1016/0021-9991(81)90120-0} {\bibfield  {journal} {\bibinfo
   {journal} {Journal of Computational Physics}\ }\textbf {\bibinfo {volume}
  {43}},\ \bibinfo {pages} {195--249} (\bibinfo {year} {1981})}\BibitemShut
  {NoStop}%
\bibitem [{\citenamefont {Taha}\ and\ \citenamefont
  {Ablowitz}(1984{\natexlab{a}})}]{TAHA1984203}%
  \BibitemOpen
  \bibfield  {author} {\bibinfo {author} {\bibfnamefont {Thiab~R}\ \bibnamefont
  {Taha}}\ and\ \bibinfo {author} {\bibfnamefont {Mark~I}\ \bibnamefont
  {Ablowitz}},\ }\bibfield  {title} {\enquote {\bibinfo {title} {Analytical and
  numerical aspects of certain nonlinear evolution equations. ii. numerical,
  nonlinear schroedinger equation},}\ }\href {\doibase
  https://doi.org/10.1016/0021-9991(84)90003-2} {\bibfield  {journal} {\bibinfo
   {journal} {Journal of Computational Physics}\ }\textbf {\bibinfo {volume}
  {55}},\ \bibinfo {pages} {203--230} (\bibinfo {year}
  {1984}{\natexlab{a}})}\BibitemShut {NoStop}%
\bibitem [{\citenamefont {Taha}\ and\ \citenamefont
  {Ablowitz}(1984{\natexlab{b}})}]{TAHA1984231}%
  \BibitemOpen
  \bibfield  {author} {\bibinfo {author} {\bibfnamefont {Thiab~R.}\
  \bibnamefont {Taha}}\ and\ \bibinfo {author} {\bibfnamefont {Mark~I.}\
  \bibnamefont {Ablowitz}},\ }\bibfield  {title} {\enquote {\bibinfo {title}
  {Analytical and numerical aspects of certain nonlinear evolution equations.
  iii. numerical, korteweg-de vries equation},}\ }\href {\doibase
  https://doi.org/10.1016/0021-9991(84)90004-4} {\bibfield  {journal} {\bibinfo
   {journal} {Journal of Computational Physics}\ }\textbf {\bibinfo {volume}
  {55}},\ \bibinfo {pages} {231--253} (\bibinfo {year}
  {1984}{\natexlab{b}})}\BibitemShut {NoStop}%
\bibitem [{\citenamefont {Li}(1995)}]{LI1995121}%
  \BibitemOpen
  \bibfield  {author} {\bibinfo {author} {\bibfnamefont {Ping-Wah}\
  \bibnamefont {Li}},\ }\bibfield  {title} {\enquote {\bibinfo {title} {On the
  numerical study of the kdv equation by the semi-implicit and leap-frog
  method},}\ }\href {\doibase https://doi.org/10.1016/0010-4655(95)00060-S}
  {\bibfield  {journal} {\bibinfo  {journal} {Computer Physics Communications}\
  }\textbf {\bibinfo {volume} {88}},\ \bibinfo {pages} {121--127} (\bibinfo
  {year} {1995})}\BibitemShut {NoStop}%
\bibitem [{\citenamefont {Fornberg}\ and\ \citenamefont
  {Whitham}(1978)}]{Fornberg78}%
  \BibitemOpen
  \bibfield  {author} {\bibinfo {author} {\bibfnamefont {B.}~\bibnamefont
  {Fornberg}}\ and\ \bibinfo {author} {\bibfnamefont {Gerald~Beresford}\
  \bibnamefont {Whitham}},\ }\bibfield  {title} {\enquote {\bibinfo {title} {A
  numerical and theoretical study of certain nonlinear wave phenomena},}\
  }\href {\doibase 10.1098/rsta.1978.0064} {\bibfield  {journal} {\bibinfo
  {journal} {Philosophical Transactions of the Royal Society of London. Series
  A, Mathematical and Physical Sciences}\ }\textbf {\bibinfo {volume} {289}},\
  \bibinfo {pages} {373--404} (\bibinfo {year} {1978})},\ \Eprint
  {http://arxiv.org/abs/https://royalsocietypublishing.org/doi/pdf/10.1098/rsta.1978.0064}
  {https://royalsocietypublishing.org/doi/pdf/10.1098/rsta.1978.0064}
  \BibitemShut {NoStop}%
\bibitem [{\citenamefont {Jain}\ \emph {et~al.}(1997)\citenamefont {Jain},
  \citenamefont {Shankar},\ and\ \citenamefont {Bhardwaj}}]{JAIN1997943}%
  \BibitemOpen
  \bibfield  {author} {\bibinfo {author} {\bibfnamefont {P.C.}\ \bibnamefont
  {Jain}}, \bibinfo {author} {\bibfnamefont {Rama}\ \bibnamefont {Shankar}}, \
  and\ \bibinfo {author} {\bibfnamefont {Dheeraj}\ \bibnamefont {Bhardwaj}},\
  }\bibfield  {title} {\enquote {\bibinfo {title} {Numerical solution of the
  korteweg-de vries (kdv) equation},}\ }\href {\doibase
  https://doi.org/10.1016/S0960-0779(96)00135-X} {\bibfield  {journal}
  {\bibinfo  {journal} {Chaos, Solitons and Fractals}\ }\textbf {\bibinfo
  {volume} {8}},\ \bibinfo {pages} {943--951} (\bibinfo {year}
  {1997})}\BibitemShut {NoStop}%
\bibitem [{\citenamefont {Muslu}\ and\ \citenamefont
  {Erbay}(2003)}]{MUSLU2003503}%
  \BibitemOpen
  \bibfield  {author} {\bibinfo {author} {\bibfnamefont {G.M.}\ \bibnamefont
  {Muslu}}\ and\ \bibinfo {author} {\bibfnamefont {H.A.}\ \bibnamefont
  {Erbay}},\ }\bibfield  {title} {\enquote {\bibinfo {title} {A split-step
  fourier method for the complex modified korteweg-de vries equation},}\ }\href
  {\doibase https://doi.org/10.1016/S0898-1221(03)80033-0} {\bibfield
  {journal} {\bibinfo  {journal} {Computers and Mathematics with Applications}\
  }\textbf {\bibinfo {volume} {45}},\ \bibinfo {pages} {503--514} (\bibinfo
  {year} {2003})}\BibitemShut {NoStop}%
\bibitem [{\citenamefont {Bona}\ \emph {et~al.}(1986)\citenamefont {Bona},
  \citenamefont {Dougalis},\ and\ \citenamefont {Karakashian}}]{BONA1986859}%
  \BibitemOpen
  \bibfield  {author} {\bibinfo {author} {\bibfnamefont {Jerry~L.}\
  \bibnamefont {Bona}}, \bibinfo {author} {\bibfnamefont {Vassilios~A.}\
  \bibnamefont {Dougalis}}, \ and\ \bibinfo {author} {\bibfnamefont
  {Ohannes~A.}\ \bibnamefont {Karakashian}},\ }\bibfield  {title} {\enquote
  {\bibinfo {title} {Fully discrete galerkin methods for the korteweg-de vries
  equation},}\ }\href {\doibase https://doi.org/10.1016/0898-1221(86)90031-3}
  {\bibfield  {journal} {\bibinfo  {journal} {Computers and Mathematics with
  Applications}\ }\textbf {\bibinfo {volume} {12}},\ \bibinfo {pages}
  {859--884} (\bibinfo {year} {1986})}\BibitemShut {NoStop}%
\bibitem [{\citenamefont {Smith}(1985)}]{smith1985numerical}%
  \BibitemOpen
  \bibfield  {author} {\bibinfo {author} {\bibfnamefont {G.D.}\ \bibnamefont
  {Smith}},\ }\href {https://books.google.co.jp/books?id=hDpvljaHOrMC} {\emph
  {\bibinfo {title} {Numerical Solution of Partial Differential Equations:
  Finite Difference Methods}}},\ Oxford applied mathematics and computing
  science series\ (\bibinfo  {publisher} {Clarendon Press},\ \bibinfo {year}
  {1985})\BibitemShut {NoStop}%
\bibitem [{\citenamefont {Singer}\ and\ \citenamefont
  {Turkel}(1998)}]{Singer98}%
  \BibitemOpen
  \bibfield  {author} {\bibinfo {author} {\bibfnamefont {I.}~\bibnamefont
  {Singer}}\ and\ \bibinfo {author} {\bibfnamefont {Eli}\ \bibnamefont
  {Turkel}},\ }\bibfield  {title} {\enquote {\bibinfo {title} {High order
  finite difference methods for the helmholtz equation},}\ }\href@noop {}
  {\bibfield  {journal} {\bibinfo  {journal} {Comput. Meth. Appl. Mech. Eng.}\
  }\textbf {\bibinfo {volume} {163}} (\bibinfo {year} {1998})}\BibitemShut
  {NoStop}%
\bibitem [{\citenamefont {Hassanien}\ \emph {et~al.}(2005)\citenamefont
  {Hassanien}, \citenamefont {Salama},\ and\ \citenamefont
  {Hosham}}]{HASSANIEN2005781}%
  \BibitemOpen
  \bibfield  {author} {\bibinfo {author} {\bibfnamefont {I.A.}\ \bibnamefont
  {Hassanien}}, \bibinfo {author} {\bibfnamefont {A.A.}\ \bibnamefont
  {Salama}}, \ and\ \bibinfo {author} {\bibfnamefont {H.A.}\ \bibnamefont
  {Hosham}},\ }\bibfield  {title} {\enquote {\bibinfo {title} {Fourth-order
  finite difference method for solving burgers?equation},}\ }\href {\doibase
  https://doi.org/10.1016/j.amc.2004.12.052} {\bibfield  {journal} {\bibinfo
  {journal} {Applied Mathematics and Computation}\ }\textbf {\bibinfo {volume}
  {170}},\ \bibinfo {pages} {781--800} (\bibinfo {year} {2005})}\BibitemShut
  {NoStop}%
\bibitem [{\citenamefont {Lee}\ \emph {et~al.}(2014)\citenamefont {Lee},
  \citenamefont {Jeong}, \citenamefont {Shin}, \citenamefont {Li},\ and\
  \citenamefont {Kim}}]{LEE201417}%
  \BibitemOpen
  \bibfield  {author} {\bibinfo {author} {\bibfnamefont {Chaeyoung}\
  \bibnamefont {Lee}}, \bibinfo {author} {\bibfnamefont {Darae}\ \bibnamefont
  {Jeong}}, \bibinfo {author} {\bibfnamefont {Jaemin}\ \bibnamefont {Shin}},
  \bibinfo {author} {\bibfnamefont {Yibao}\ \bibnamefont {Li}}, \ and\ \bibinfo
  {author} {\bibfnamefont {Junseok}\ \bibnamefont {Kim}},\ }\bibfield  {title}
  {\enquote {\bibinfo {title} {A fourth-order spatial accurate and practically
  stable compact scheme for the cahn-hilliard equation},}\ }\href {\doibase
  https://doi.org/10.1016/j.physa.2014.04.038} {\bibfield  {journal} {\bibinfo
  {journal} {Physica A: Statistical Mechanics and its Applications}\ }\textbf
  {\bibinfo {volume} {409}},\ \bibinfo {pages} {17--28} (\bibinfo {year}
  {2014})}\BibitemShut {NoStop}%
\bibitem [{\citenamefont {Wang}\ and\ \citenamefont {Dai}(2019)}]{WANG2019310}%
  \BibitemOpen
  \bibfield  {author} {\bibinfo {author} {\bibfnamefont {Xiaofeng}\
  \bibnamefont {Wang}}\ and\ \bibinfo {author} {\bibfnamefont {Weizhong}\
  \bibnamefont {Dai}},\ }\bibfield  {title} {\enquote {\bibinfo {title} {A
  conservative fourth-order stable finite difference scheme for the generalized
  rosenau-kdv equation in both 1d and 2d},}\ }\href {\doibase
  https://doi.org/10.1016/j.cam.2019.01.041} {\bibfield  {journal} {\bibinfo
  {journal} {Journal of Computational and Applied Mathematics}\ }\textbf
  {\bibinfo {volume} {355}},\ \bibinfo {pages} {310--331} (\bibinfo {year}
  {2019})}\BibitemShut {NoStop}%
\bibitem [{\citenamefont {Arakawa}(1966)}]{ARAKAWA1966119}%
  \BibitemOpen
  \bibfield  {author} {\bibinfo {author} {\bibfnamefont {Akio}\ \bibnamefont
  {Arakawa}},\ }\bibfield  {title} {\enquote {\bibinfo {title} {{Computational
  design for long-term numerical integration of the equations of fluid motion:
  Two-dimensional incompressible flow. Part I}},}\ }\href {\doibase
  https://doi.org/10.1016/0021-9991(66)90015-5} {\bibfield  {journal} {\bibinfo
   {journal} {Journal of Computational Physics}\ }\textbf {\bibinfo {volume}
  {1}},\ \bibinfo {pages} {119--143} (\bibinfo {year} {1966})}\BibitemShut
  {NoStop}%
\bibitem [{\citenamefont {Gleiser}\ \emph {et~al.}(2018)\citenamefont
  {Gleiser}, \citenamefont {Stephens},\ and\ \citenamefont
  {Sowinski}}]{Gleiser:2018kbq}%
  \BibitemOpen
  \bibfield  {author} {\bibinfo {author} {\bibfnamefont {Marcelo}\ \bibnamefont
  {Gleiser}}, \bibinfo {author} {\bibfnamefont {Michelle}\ \bibnamefont
  {Stephens}}, \ and\ \bibinfo {author} {\bibfnamefont {Damian}\ \bibnamefont
  {Sowinski}},\ }\bibfield  {title} {\enquote {\bibinfo {title}
  {{Configurational entropy as a lifetime predictor and pattern discriminator
  for oscillons}},}\ }\href {\doibase 10.1103/PhysRevD.97.096007} {\bibfield
  {journal} {\bibinfo  {journal} {Phys. Rev. D}\ }\textbf {\bibinfo {volume}
  {97}},\ \bibinfo {pages} {096007} (\bibinfo {year} {2018})},\ \Eprint
  {http://arxiv.org/abs/1803.08550} {arXiv:1803.08550 [hep-th]} \BibitemShut
  {NoStop}%
\bibitem [{\citenamefont {Gleiser}\ and\ \citenamefont
  {Krackow}(2019)}]{Gleiser:2019rvw}%
  \BibitemOpen
  \bibfield  {author} {\bibinfo {author} {\bibfnamefont {Marcelo}\ \bibnamefont
  {Gleiser}}\ and\ \bibinfo {author} {\bibfnamefont {Max}\ \bibnamefont
  {Krackow}},\ }\bibfield  {title} {\enquote {\bibinfo {title} {{Resonant
  configurations in scalar field theories: Can some oscillons live forever?}}}\
  }\href {\doibase 10.1103/PhysRevD.100.116005} {\bibfield  {journal} {\bibinfo
   {journal} {Phys. Rev. D}\ }\textbf {\bibinfo {volume} {100}},\ \bibinfo
  {pages} {116005} (\bibinfo {year} {2019})},\ \Eprint
  {http://arxiv.org/abs/1906.04070} {arXiv:1906.04070 [hep-th]} \BibitemShut
  {NoStop}%
\bibitem [{\citenamefont {Gleiser}\ and\ \citenamefont
  {Krackow}(2020)}]{Gleiser:2020zaj}%
  \BibitemOpen
  \bibfield  {author} {\bibinfo {author} {\bibfnamefont {Marcelo}\ \bibnamefont
  {Gleiser}}\ and\ \bibinfo {author} {\bibfnamefont {Max}\ \bibnamefont
  {Krackow}},\ }\bibfield  {title} {\enquote {\bibinfo {title}
  {{Configurational entropic study of the enhanced longevity in resonant
  oscillons}},}\ }\href {\doibase 10.1016/j.physletb.2020.135450} {\bibfield
  {journal} {\bibinfo  {journal} {Phys. Lett. B}\ }\textbf {\bibinfo {volume}
  {805}},\ \bibinfo {pages} {135450} (\bibinfo {year} {2020})},\ \Eprint
  {http://arxiv.org/abs/2003.10899} {arXiv:2003.10899 [hep-th]} \BibitemShut
  {NoStop}%
\bibitem [{\citenamefont {Gleiser}\ and\ \citenamefont
  {Sowinski}(2013)}]{Gleiser:2013mga}%
  \BibitemOpen
  \bibfield  {author} {\bibinfo {author} {\bibfnamefont {Marcelo}\ \bibnamefont
  {Gleiser}}\ and\ \bibinfo {author} {\bibfnamefont {Damian}\ \bibnamefont
  {Sowinski}},\ }\bibfield  {title} {\enquote {\bibinfo {title}
  {{Information-Entropic Stability Bound for Compact Objects: Application to
  Q-Balls and the Chandrasekhar Limit of Polytropes}},}\ }\href {\doibase
  10.1016/j.physletb.2013.10.005} {\bibfield  {journal} {\bibinfo  {journal}
  {Phys. Lett. B}\ }\textbf {\bibinfo {volume} {727}},\ \bibinfo {pages}
  {272--275} (\bibinfo {year} {2013})},\ \Eprint
  {http://arxiv.org/abs/1307.0530} {arXiv:1307.0530 [hep-th]} \BibitemShut
  {NoStop}%
\bibitem [{\citenamefont {Gleiser}\ and\ \citenamefont
  {Jiang}(2015)}]{Gleiser:2015rwa}%
  \BibitemOpen
  \bibfield  {author} {\bibinfo {author} {\bibfnamefont {Marcelo}\ \bibnamefont
  {Gleiser}}\ and\ \bibinfo {author} {\bibfnamefont {Nan}\ \bibnamefont
  {Jiang}},\ }\bibfield  {title} {\enquote {\bibinfo {title} {{Stability Bounds
  on Compact Astrophysical Objects from Information-Entropic Measure}},}\
  }\href {\doibase 10.1103/PhysRevD.92.044046} {\bibfield  {journal} {\bibinfo
  {journal} {Phys. Rev. D}\ }\textbf {\bibinfo {volume} {92}},\ \bibinfo
  {pages} {044046} (\bibinfo {year} {2015})},\ \Eprint
  {http://arxiv.org/abs/1506.05722} {arXiv:1506.05722 [gr-qc]} \BibitemShut
  {NoStop}%
\end{thebibliography}%

\end{document}